\documentclass[twocolumn]{aastex63} 
\hypersetup{linkcolor=red,citecolor=blue,filecolor=cyan, urlcolor=black}
\usepackage{multirow}
\usepackage{amsmath}
\usepackage{amssymb}
\usepackage{balance}
\usepackage{lmodern}
\usepackage{changepage}

\received{November 30, 2022}
\revised{December 13, 2022}
\accepted{December 14, 2022}
\submitjournal{ApJS}

\shorttitle{A three-dimensional map of the Milky Way using 66,000 Mira variable stars}
\shortauthors{Iwanek et al.}

\begin{document}

\title{A three-dimensional map of the Milky Way using 66,000 Mira variable stars}

\correspondingauthor{Patryk Iwanek}
\email{piwanek@astrouw.edu.pl}

\author[0000-0002-6212-7221]{Patryk Iwanek}
\affiliation{Astronomical Observatory, University of Warsaw, Al. Ujazdowskie 4, 00-478 Warsaw, Poland}

\author[0000-0002-9245-6368]{Radosław Poleski}
\affiliation{Astronomical Observatory, University of Warsaw, Al. Ujazdowskie 4, 00-478 Warsaw, Poland}

\author[0000-0003-4084-880X]{Szymon Koz{\l}owski}
\affiliation{Astronomical Observatory, University of Warsaw, Al. Ujazdowskie 4, 00-478 Warsaw, Poland}

\author[0000-0002-7777-0842]{Igor Soszy{\'n}ski}
\affiliation{Astronomical Observatory, University of Warsaw, Al. Ujazdowskie 4, 00-478 Warsaw, Poland}

\author[0000-0002-2339-5899]{Paweł Pietrukowicz}
\affiliation{Astronomical Observatory, University of Warsaw, Al. Ujazdowskie 4, 00-478 Warsaw, Poland}

\author[0000-0002-6079-3335]{Makiko Ban}
\affiliation{Astronomical Observatory, University of Warsaw, Al. Ujazdowskie 4, 00-478 Warsaw, Poland}

\author[0000-0002-2335-1730]{Jan Skowron}
\affiliation{Astronomical Observatory, University of Warsaw, Al. Ujazdowskie 4, 00-478 Warsaw, Poland}

\author[0000-0001-7016-1692]{Przemysław Mróz}
\affiliation{Astronomical Observatory, University of Warsaw, Al. Ujazdowskie 4, 00-478 Warsaw, Poland}

\author[0000-0002-3051-274X]{Marcin Wrona}
\affiliation{Astronomical Observatory, University of Warsaw, Al. Ujazdowskie 4, 00-478 Warsaw, Poland}

\author[0000-0001-5207-5619]{Andrzej Udalski}
\affiliation{Astronomical Observatory, University of Warsaw, Al. Ujazdowskie 4, 00-478 Warsaw, Poland}

\author[0000-0002-0548-8995]{Michał K. Szymański}
\affiliation{Astronomical Observatory, University of Warsaw, Al. Ujazdowskie 4, 00-478 Warsaw, Poland}

\author[0000-0001-9439-604X]{Dorota M. Skowron}
\affiliation{Astronomical Observatory, University of Warsaw, Al. Ujazdowskie 4, 00-478 Warsaw, Poland}

\author[0000-0001-6364-408X]{Krzysztof Ulaczyk}
\affiliation{Department of Physics, University of Warwick, Coventry CV4 7 AL, UK}
\affiliation{Astronomical Observatory, University of Warsaw, Al. Ujazdowskie 4, 00-478 Warsaw, Poland}

\author[0000-0002-1650-1518]{Mariusz Gromadzki}
\affiliation{Astronomical Observatory, University of Warsaw, Al. Ujazdowskie 4, 00-478 Warsaw, Poland}

\author[0000-0002-9326-9329]{Krzysztof Rybicki}
\affiliation{Astronomical Observatory, University of Warsaw, Al. Ujazdowskie 4, 00-478 Warsaw, Poland}

\author[0000-0002-3218-2684]{Milena Ratajczak}
\affiliation{Astronomical Observatory, University of Warsaw, Al. Ujazdowskie 4, 00-478 Warsaw, Poland}

\begin{abstract}

\noindent  We study the three-dimensional structure of the Milky Way using 65,981 Mira variable stars discovered by the Optical Gravitational Lensing Experiment (OGLE) survey. The spatial distribution of the Mira stars is analyzed with a model containing three barred components that include the X-shaped boxy component in the Galactic center (GC), and an axisymmetric disk. We take into account the distance uncertainties by implementing the Bayesian hierarchical inference method. The distance to the GC is $R_0 = 7.66 \pm 0.01 \mathrm{(stat.)} \pm 0.39 \mathrm{(sys.)}$ kpc, while the inclination of the major axis of the bulge to the Sun-GC line-of-sight is $\theta = 20.2^\circ \pm 0.6^\circ \mathrm{(stat.)} \pm 0.7^\circ \mathrm{(sys.)} $. We present, for the first time, a detailed three-dimensional map of the Milky Way composed of young and intermediate-age stellar populations. Our analysis provides independent evidence for both the X-shaped bulge component and the flaring disk (being plausibly warped). We provide the complete dataset of properties of Miras that were used for calculations in this work. The table includes: mean brightness and amplitudes in nine photometric bands (covering a range of wavelength from 0.5 to 12 $\mu$m), photometric chemical type, estimated extinction, and calculated distance with its uncertainty for each Mira variable. The median distance accuracy to a Mira star is at the level of $6.6\%$.

\end{abstract}

\keywords{Galaxy: bulge -- Galaxy: center -- Galaxy: disk -- Galaxy: structure -- stars: AGB and post-AGB -- astronomical databases: catalogs -- astronomical databases: surveys}

\section{Introduction} \label{sec:intro}

The exploration of our backyard, the Milky Way, is hampered by the unfortunate location of the Solar System in the plane of the Galactic disk (GD), and line-of-sight dependent extinction toward the center of the Galaxy. The first serious scientific attempt to map the shape of the Milky Way was made in 1785 by William Herschel, who counted stars in different regions of the visible sky \citep{1785RSPT...75..213H, 2011arXiv1112.3635T}. At that time, it was assumed that stars are uniformly distributed in the Galaxy, there are no stars beyond the boundaries of the Milky Way, and the Sun is located near the center of the Galaxy. William's son, John Herschel, continued his father's work and stated that the central part of the Galaxy is surrounded by stars clustered in twisting arms, a shape that he could not define. The question of whether the Milky Way has a spiral structure arose when William Parsons in M51 (the Whirlpool Galaxy) discovered the spiral arms in 1845. The first drawing of the Milky Way as a spiral-shaped structure was presented in 1900 by Cornelis Easton.

In 1904, Jacobus Kapteyn, while studying the proper motions of nearby stars, noticed that these were not random. Kapteyn reported that stars could be divided into two streams that move in opposite directions. This was the first (but unconscious) evidence of the rotation of the Milky Way. At the time, the accepted view on the structure of the Galaxy was that it was a lens-shaped system of stars with the Sun near its middle and a diameter of about 3 kpc \citep{1999ApJ...525C.135G}.

It was not until the work of Harlow Shapley that he made a milestone in studies of the structure of the Milky Way. He calibrated the newly-discovered period-luminosity relation (PLR) for Cepheids in the Small Magellanic Cloud \citep[][]{1912HarCi.173....1L} and used it to determine distances to 69 Galactic globular clusters \citep{1918ApJ....48...89S, 1918ApJ....48..154S, 1918PASP...30...42S}. 
These distances significantly exceed any cosmic distances measured so far, and Shapley's analysis ended with the estimation of the realistic size of the Galaxy and the conclusion that the Sun does not lie close to its center.

Despite all the difficulties and challenges, the structure of the Milky Way has been extensively studied using various tracers. \citet{1988gera.book..295B, 2003PASJ...55..191N} used neutral gas, while \citet{2001ApJ...547..792D, 2006PASJ...58..847N, 2020MNRAS.498.1710P}
applied molecular gas for these studies. The other authors based their research on, e.g., star-forming regions \citep{2014ApJ...797...39B, 2014ApJ...783..130R}, red clump stars \citep{1994ApJ...425L..81W, 1994ApJ...429L..73S, 1997ApJ...477..163S, 1997MNRAS.292L..15L, 2010ApJ...721L..28N, 2010ApJ...724.1491M, 2013MNRAS.435.1874W, 2019A&A...627A...3L, 2020ApJ...897..119W}, halo and local metal-poor stars \citep{2007Natur.450.1020C, 2022ApJ...927..145S}, K-type giants \citep{2020ApJ...899..110T}, or finally pulsating stars of different types.

The latter tracers, pulsating stars, thanks to well-defined PLRs, are used as distance indicators in different stellar environments. Large-scale variability surveys such as the Optical Gravitational Lensing Experiment \citep[OGLE;][]{2015AcA....65....1U}, the All-Sky Automated Survey for Supernovae \citep[ASAS-SN;][]{2014ApJ...788...48S}, the VISTA Variables in The Via Lactea \citep[VVV;][]{2010NewA...15..433M}, the Zwicky Transient Facility \citep[ZTF;][]{2019PASP..131a8002B}, or Gaia \citep{2016A&A...595A...1G, 2018A&A...616A...1G} play first fiddle in this area of interests, as they allow us to discover thousands of pulsating stars, which enable us to directly study the three-dimensional structure of the Milky Way and other galaxies.

The rich sample of RR Lyrae-type stars is commonly used to study the Galactic bulge \citep[BLG;][]{2015ApJ...811..113P, 2020AcA....70..121P, 2020AJ....159..270K, 2020A&A...641A..96S, 2022MNRAS.509.4532S}, while spiral arms and GD are studied using young stars -- classical Cepheids \citep{2019Sci...365..478S, 2019AcA....69..305S, 2019NatAs...3..320C, 2019ApJ...870L..10M}. Past research showed that Mira variables, thanks to their wide range of ages, abundance throughout the Milky Way, and relatively easy detection, can be used to study the structure of the BLG, as well as the spiral arms \citep{1980MNRAS.190..227F, 1982MNRAS.198..199G, 2005A&A...443..143G, 2005MNRAS.364..117M, 2009MNRAS.399.1709M, 2016MNRAS.455.2216C, 2017ApJ...836..218L, 2020MNRAS.492.3128G, 2020ApJ...891...50U, 2022MNRAS.517.6060S, 2022MNRAS.517..257S}.

The Mira-type stars have been studied for centuries and it is a well-known subgroup of long period variables (LPVs). They are asymptotic giant branch (AGB) stars pulsating in the fundamental mode, with pulsation periods ranging from about 80 days to over 1000 days \citep{2022ApJS..260...46I}. Miras have large bolometric luminosities (of 100-10,000 $L_\odot$) and a large brightness amplitude \citep[above \mbox{0.8 mag} in the {\it I}-band;][]{2009AcA....59..239S, 2013AcA....63...21S, 2022ApJS..260...46I} which decreases with increasing wavelength \citep{2021MNRAS.500...82I, 2021ApJS..257...23I}. 

Mira-type variables (as other LPVs) can be divided into oxygen-rich (O-rich; those containing molecules such as H$_2$O, TiO, SiO) and carbon-rich (C-rich; those containing molecules C$_2$, CN) based on their surface chemical compositions and the C/O ratio \citep{2010ApJ...723.1195R, 2005AcA....55..331S, 2009AcA....59..239S}. The abundance of both types in galaxies depends on the metallicity of the environment \citep{2019MNRAS.482.5567M}. In the Milky Way, it is expected to observe mostly O-rich Miras, with only a small number of C-rich ones \citep{2006MNRAS.369..751W, 2017MNRAS.469.4949M}. This ratio is reversed for lower-metallicity galaxies, where C-rich Miras predominate \citep[e.g., in the Magellanic Clouds;][]{2009AcA....59..239S}.

Almost a century of research on the Mira PLRs \citep[][]{1928PNAS...14..963G, 1942ApJ....95..248W, 1981Natur.291..303G, 1989MNRAS.241..375F, 2000PASA...17...18W, 2008MNRAS.386..313W, 2010ApJ...723.1195R, 2011MNRAS.412.2345I, 2017AJ....153..170Y, 2017AJ....154..149Y, 2018AJ....156..112Y, 2019ApJ...884...20B, 2021ApJ...919...99I} showed, that these stars are attractive distance indicators. Recently, \citet{2021ApJS..257...23I} reported that the scatter of Miras PLRs decreases with the increasing wavelength. Observations at longer wavelengths also reduce the influence of interstellar extinction on stellar light. Therefore, using mid-infrared (mid-IR) PLRs makes Miras extremely valuable tracers of the Milky Way structure.

In this paper, we combine the recently discovered 66,000 Miras in the BLG and GD fields \citep[][]{2022ApJS..260...46I} based on the OGLE observations, with mid-IR observations of these objects obtained by the Wide Field Infrared Survey Explorer \citep[WISE;][]{2010AJ....140.1868W}
and Spitzer \citep{2004ApJS..154....1W} space telescopes, and with the most accurate to date mid-IR PLRs for Miras \citep{2021ApJ...919...99I} to derive the precise distance to each Mira star and study in detail the three-dimensional structure of the Milky Way.

The paper is organized as follows. In Section \ref{sec:data}, we briefly describe the sample of Miras and mid-IR data from space telescopes. Section \ref{sec:mean_brightness_and_extinction} is devoted to a description of the measurement of mean brightness and amplitudes in the mid-IR bands, as well as the method of interstellar extinction correction. The distance measurement method is presented in Section \ref{sec:distances}, while in Section \ref{sec:OCdivision} we discuss the division of Miras into O- and C-rich types. In Section \ref{sec:3dmap}, we describe the Galactic bar model, the sample selection used for modeling, and the fitting procedure of the model to this sample of Miras. In this section, we also explore potential biases of the fitting method and possible systematic errors that could affect the final solution. The final results are discussed in Section \ref{sec:discussion}, while in Section \ref{sec:conclusions} we conclude the paper.

\section{Data} \label{sec:data}

\subsection{Sample of Miras} \label{subsection:sample_of_miras}
In this study, we used 65,981 Miras discovered in the Milky Way based on the third \citep[OGLE-III;][]{2003AcA....53..291U, 2013AcA....63...21S, 2013AcA....63..379P} and the fourth phases of the OGLE project \citep[OGLE-IV;][]{2015AcA....65....1U, 2022ApJS..260...46I}. The recently published collection of Mira-type variables \citep{2022ApJS..260...46I} contains 40,356 stars from BLG fields and 25,625 Miras discovered in GD fields. The authors provided light curves in the {\it I}- and {\it V}-band from the Johnson-Cousins photometric system collected from December 1996 to March 2020. The collection covers $\sim$3000 deg$^2$ of the sky. A full description of the Mira collection can be found in \citet{2022ApJS..260...46I}. The catalog containing equatorial coordinates, pulsation periods, {\it I}-band brightness amplitudes, mean magnitudes in {\it V} and {\it I} bands, light curves, and finding charts is publicly available and can be accessed through the OGLE Internet Archive\footnote{\url{https://www.astrouw.edu.pl/ogle/ogle4/OCVS/
blg/lpv/} \newline \url{https://www.astrouw.edu.pl/ogle/ogle4/OCVS/
gd/lpv/}}.

\begin{figure*}
\centering
\includegraphics[scale=0.25]{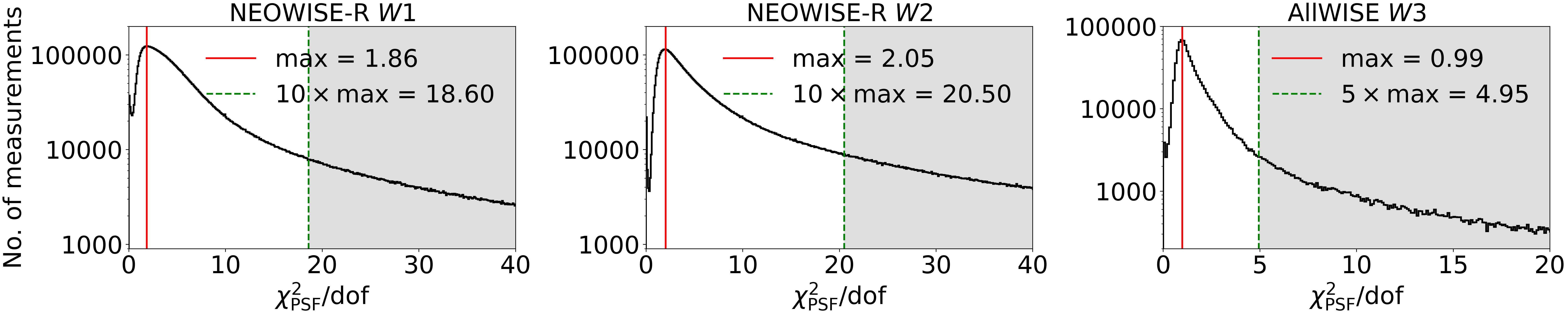}
\caption{Distributions of $\chi^2_{\rm PSF}/\mathrm{dof}$ from fitting the point-spread function (PSF) to images collected during WISE observations. In the first step of cleaning the WISE light curves from the low-quality measurements, data points with the value of $\chi^2_{\rm PSF}/\mathrm{dof}$ from the gray area were removed.}
\label{fig:chi2}
\end{figure*}

\subsection{Mid-IR data} \label{subsec:selection}

Mid-IR observations are important in studies of environments strongly obscured by interstellar matter, as the influence of interstellar extinction decreases with the increasing wavelength. Both studied regions of the Galaxy (BLG and GD) are obscured by a significant amount of dust that absorbs stellar light. Therefore, extinction in the Milky Way could change significantly in a small angular range. Another asset of \mbox{mid-IR} observations is the fact that the amplitudes of the Mira brightness variation decrease with increasing wavelength, which has been thoroughly investigated and explained in \citet{2021ApJS..257...23I}. Therefore, the use of mid-IR data minimizes the uncertainty associated with determining mean brightness, even for poorly-covered targets.
A detailed discussion on the mean brightness determination of the Miras analyzed in the mid-IR bands is presented in Section \ref{subsec:mean_brightness}.

\subsubsection{WISE Data}

WISE \citep{2010AJ....140.1868W} is a 40-cm diameter infrared space telescope that observed the sky in four bands: {\it W}1 ($\lambda_{\rm eff} = 3.4$ $\mu$m), {\it W}2 ($\lambda_{\rm eff} = 4.6$ $\mu$m), {\it W}3 ($\lambda_{\rm eff} = 12$ $\mu$m), {\it W}4 ($\lambda_{\rm eff} = 22$ $\mu$m). The main mission of the WISE telescope was completed in 2010, while at the beginning of 2011 due to the depletion of the solid hydrogen cryostat, the telescope was placed into hibernation mode. Since then, the observations in {\it W}3- and {\it W}4-bands have ended, however, the observations in the {\it W}1- and {\it W}2-bands were resumed as a Near Earth Object WISE Reactivation Mission \citep[NEOWISE-R;][]{2011ApJ...731...53M, 2014ApJ...792...30M}, and continue to this day.

We cross-matched the sample of the Milky Way Miras detected by OGLE \citep{2022ApJS..260...46I} with the AllWISE Multiepoch Photometry Table \citep[data collected before the telescope hibernation;][]{https://doi.org/10.26131/irsa134}\footnote{\url{https://wise2.ipac.caltech.edu/docs/release/allwise/}}, and with the NEOWISE-R Single Exposure (L1b) Source Table \citep{https://doi.org/10.26131/irsa144} using NASA/IPAC Infrared Science Archive\footnote{\url{https://irsa.ipac.caltech.edu/applications/Gator/}}. Our previous work on Miras variability in the mid-IR bands \citep{2021ApJS..257...23I} showed two important characteristics of the WISE data:
\begin{itemize}
\item Internal scatter of the light curves collected before the telescope hibernation (AllWISE) is larger than scatter of light curves collected during \mbox{NEOWISE-R}. Moreover, the time span of the NEOWISE-R light curves is much longer.

\item During the spectral energy distributions (SEDs) analysis, it turned out that most measurements in the {\it W}4-band significantly deviate from the model, which makes it an outlier. Furthermore, studies on PLR in this band showed that {\it W}4 measurements are not reliable \citep{2021ApJ...919...99I}, and should not be used.
\end{itemize}

\noindent Due to the above-mentioned reasons, in this analysis we used the light curves in the {\it W}1- and {\it W}2-bands from the NEOWISE-R only, while from the AllWISE table we used the {\it W}3 light curves.

We searched the AllWISE and NEOWISE-R tables for objects within $1''$ around Mira positions. 
We downloaded all available measurements in the {\it W}1, {\it W}2, and {\it W}3 bands. In the NEOWISE-R data, we found counterparts to 65,562 and 63,661 Miras (out of 65,981) in the {\it W}1 and {\it W}2 bands, respectively, while in the AllWISE table, we found counterparts to 59,936 Miras.

Each WISE measurement from both the AllWISE and NEOWISE-R tables can be examined for quality using the reduced $\chi^2_{\rm PSF}$ which is calculated by the fitting of the point-spread function (PSF) to objects detected in the collected images. We used a similar procedure as described in \citet{2021ApJS..257...23I}, and we cleaned the WISE light curves from the low-quality points assuming that good-quality measurements in the {\it W}1 and {\it W}2 bands had \mbox{$\chi^2_{\rm PSF}/\mathrm{dof} < 10 \times \chi^2_{\rm PSF, max}/\mathrm{dof}$}, while {\it W}3 best-quality measurements had \mbox{$\chi^2_{\rm PSF}/\mathrm{dof} < 5 \times \chi^2_{\rm PSF, max}/\mathrm{dof}$}, where $\chi^2_{\rm PSF, max}/\mathrm{dof}$ is the mode of the $\chi^2_{\rm PSF}/\mathrm{dof}$ distributions, separately for each WISE band. The $\chi^2_{\rm PSF}/\mathrm{dof}$ distributions are presented in Figure \ref{fig:chi2}. As a result, we rejected $24.42\%$, $37.03\%$, and $19.21\%$ observations in the {\it W}1, {\it W}2, and {\it W}3 bands, respectively.

The WISE telescope observes a given field for a short time (called a 'batch'; typically 2 days) and returns to the field half a year later. A single batch contains typically several dozen measurements, but for slowly-varying Miras, each batch should be treated as a single epoch.
The densely-covered WISE light curves (in the {\it W}1 and {\it W}2 bands) contained up to 13 such epochs, while the {\it W}3 light curves had a maximum of two epochs. Throughout the paper, we use distinct terms: 'epochs' and 'measurements/data points' regarding the WISE data. We use 'epochs' to refer to the batches of observations described above, while we use 'measurements/data points' to refer to the individual measurements that make up the 'epochs'. The maximum and the median number of data points per light curve were 360, 368, 144, and 115, 125, 12, for {\it W}1, {\it W}2, and {\it W}3, respectively. The first observations in the \mbox{NEOWISE-R} light curves were taken in 2013 December 13, while the newest observations come from 2020 December 11. Data in the AllWISE database were collected between 2010 January~6 and 2010 August~5. 

\newpage
\subsubsection{Spitzer Data}

The Spitzer Space Telescope was an 85-cm diameter telescope with three infrared instruments onboard: Infrared Spectrograph (IRS), Multiband Imaging Photometry for Spitzer (MIPS), and Infrared Array Camera \citep[IRAC;][]{2004ApJS..154....1W}. The latter instrument, IRAC, was equipped with four channels: $[3.6]$ ($\lambda_{\mathrm{eff}} = 3.6$ $\mu$m), $[4.5]$ ($\lambda_{\mathrm{eff}} = 4.5$ $\mu$m), $[5.8]$ ($\lambda_{\mathrm{eff}} = 5.8$ $\mu$m), $[8.0]$ ($\lambda_{\mathrm{eff}} = 8.0$ $\mu$m). In this paper, we used channels $[3.6]$, $[4.5]$, $[5.8]$, $[8.0]$, later also referred to as {\it I}1, {\it I}2, {\it I}3, {\it I}4. Most of the Milky Way observations were carried out during the Galactic Legacy Infrared Mid-Plane Survey Extraordinaire \citep[GLIMPSE;][]{2003PASP..115..953B, 2009PASP..121..213C}\footnote{\url{http://www.astro.wisc.edu/glimpse/all\_GLIMPSE-data-AAS2013.pdf}}.

We searched the databases: GLIMPSE~I \citep{https://doi.org/10.26131/irsa223}, GLIMPSE II \citep{https://doi.org/10.26131/irsa200}, GLIMPSE 3D \citep{2009PASP..121..213C, https://doi.org/10.26131/irsa203}, GLIMPSE 360 \citep{2008sptz.prop60020W, https://doi.org/10.26131/irsa214}, Vela-Carina \citep{2009ApJ...707..510Z, https://doi.org/10.26131/irsa213}, Deep GLIMPSE \citep{2011sptz.prop80074W, https://doi.org/10.26131/irsa217}, Spitzer Mapping of the Outer Galaxy \citep[SMOG;][]{2008sptz.prop50398C, https://doi.org/10.26131/irsa226}, A Spitzer Legacy Survey of the Cygnus-X Complex \citep[Cygnus-X;][]{2007sptz.prop40184H, https://doi.org/10.26131/irsa225} for objects within $1''$ around Mira positions using the NASA/IPAC Infrared Science Archive\footnote{\url{https://irsa.ipac.caltech.edu/applications/Gator/}}. In total, we found counterparts to 15,741, 18,005, 25,595, and 24,867 Miras in the {\it I}1, {\it I}2, {\it I}3, and {\it I}4 bands, respectively. Most Spitzer light curves have one or two measurements. A few dozen Miras had four observations per light curve in each band. The observation time-stamps are not provided in the aforementioned databases. Due to the small number of observations per star, we did not remove any outlying measurements.

\newpage
\section{Mean brightness and extinction} \label{sec:mean_brightness_and_extinction}

The main aim of this paper is to measure the distance to each detected Mira star and study the three-dimensional structure of our Galaxy. Therefore, first, we calculated the mean brightness in various bands and tried to evaluate the extinction toward every individual star.

\subsection{Mean brightness and amplitudes in mid-IR bands} \label{subsec:mean_brightness}

The method of measuring the mean magnitude was straightforward and assumed that the mean brightness was represented by the weighted mean with the weight of each data point taken as an inverse square of the uncertainty reported by the survey ($\sigma_i^{-2}$). We searched the light curves for outlying measurements that deviated more than $3\sigma$ from the median, and we removed such measurements. The mean brightness was estimated in the flux scale and then transformed to the magnitude scale.
The final mean brightness for each Mira star is referred to as $m_{\lambda}$ throughout the paper, where $\lambda$ indicates each mid-IR band. When we write about the color index, for example ({\it W}1$-${\it W}2), we mean the difference between mean magnitudes, i.e., $(m_{W1}-m_{W2})$. 

The most accurate measurement of mean brightness would be achieved by making pulsating light-curve templates based on the phase of the optical light curves from the OGLE survey and fitting them at each measured epoch of the WISE and Spitzer data, as was done for Miras in the Large Magellanic Cloud \citep[LMC;][]{2021ApJS..257...23I}. In the case of the Galactic Miras, this approach is not possible because of two reasons: most Mira-type variables do not have light curves covered well enough to make reliable and accurate templates, and Spitzer observations time-stamp are not available, so it is not even possible to determine in what phase of pulsation the observations were made. Therefore, it is extremely important to carefully assign proper uncertainties to the mean brightness calculated in this approximate fashion, as later these uncertainties will be propagated to the final distance uncertainties.

We calculated brightness amplitude using the relations between the variability amplitude ratio and the wavelength obtained by \citet[][see Figure 2 and Table 1 therein]{2021ApJS..257...23I}. These relations allow us to estimate the amplitude in any photometric band, knowing the amplitude in the OGLE {\it I}-band, separately for \mbox{O-rich} and C-rich Miras. The OGLE {\it I}-band amplitude for each Mira star was taken from the OGLE Miras catalog \citep{2022ApJS..260...46I}. Furthermore, for Miras without any measurements in the OGLE {\it V}-band, we estimated the {\it V}-band brightness variability amplitude using the same relation. The division of Galactic Miras into \mbox{O-rich} and C-rich is discussed in detail in Section \ref{sec:OCdivision}. 

A subset of stars had enough measurements in the two WISE bands ({\it W}1 and {\it W}2) that the direct measurement of the brightness amplitude was possible. If the WISE light curve had more datapoints than the OGLE one, we used the direct measurement from the WISE light curve, instead of rescaling the optical amplitude. For such light curves, we calculated brightness amplitude as the difference between the 95th and 5th percentiles of the brightness distribution, calculated after \mbox{3$\sigma$} clipping procedure, and we divided this difference by 0.9, i.e., the range between those two percentiles.

When determining the mean brightness uncertainties, we took into account two factors: the standard error of the weighted mean calculated from the weights $(\sum_i \sigma_i^{-2})^{-1/2}$ and the expected root mean square (RMS) scatter of brightness caused by pulsations, and we combined them in the quadrature. Assuming that mid-IR light curves are nearly sinusoidal, we estimated {\it W}1, {\it W}2, and {\it W}3 RMS scatter as 0.7 of the brightness variability amplitude divided by the square root of the number of epochs in the analyzed light curve. In the case of the Spitzer data, we estimated the RMS scatter as 0.7 of the brightness variability amplitude in the case of only one measurement or as 0.5 of the brightness variability amplitude for two or more data points. The final mean brightness uncertainty for each Mira star is referred to as $\sigma_{\lambda}$ throughout the paper, where $\lambda$ indicates each mid-IR band.

\subsection{Correction for the interstellar extinction} \label{subsec:extinction}

The interstellar extinction in the Milky Way, caused by irregularly distributed clouds of dust and gas, is highly line-of-sight dependent. The strongest interstellar extinction is observed toward the BLG and near the Galactic plane. The influence of interstellar extinction on stellar light is the strongest in the optical bands and weakens with wavelength to be much smaller in the infrared bands.

We derived the interstellar extinction using the three-dimensional map
\mbox{"mwdust"}\footnote{\url{https://github.com/jobovy/mwdust}} \citep{2016ApJ...818..130B} which provides a value in a given direction in the Galactic coordinates $(l, b)$, for the distance $d$ from the Sun, and in many bands (in our case, we chose the $K_s$ band). The "mwdust" code combined several interstellar extinction maps, i.e., \citet{2003A&A...409..205D, 2006A&A...453..635M, 2019ApJ...887...93G}. \citet{2019Sci...365..478S} tested $A_{Ks}$

\clearpage
\begin{turnpage}
\begin{table}
\begin{adjustwidth}{-5cm}{}
\caption{Mean magnitudes, distances, extinction values, and photometric chemical types for 65,981 Galactic Miras.}
{\fontsize{5}{6} \selectfont
\begin{tabular}{lccccccccccccccccccc}
\hline \hline
ID & P & R.A. & Decl. & $l$ & $b$ & Loc. & Type & $m_I$ & \ldots & $m_{I4}$ & $\sigma_{I4}$ & $\Delta m_{I4}$ &
$A_{Ks}$ & $d$ & $\sigma_d$ & $\mu$ & $\sigma_\mu$ & n & mask \\

 & (d) & (h:m:s) & ($^\circ$:m:s) & ($^\circ$) & ($^\circ$) &  &  & (mag) & \ldots & (mag) & (mag) & (mag) &
(mag) & (pc) & (pc) & (mag) & (mag) & &  \\ \hline

OGLE-BLG-LPV-000009 & 275.30 & 17:05:28.47 & $-$32:44:22.4 & 352.019277 & 5.008994 & BLG & O & 13.008 & \ldots & $-$99.999 & $-$99.999 & $-$99.999 & 0.165 & 13473 & 815 & 15.647 & 0.131 & 3 & $-$99.999 \\
OGLE-BLG-LPV-000018 & 363.90 & 17:05:37.00 & $-$33:02:36.6 & 351.793338 & 4.802601 & BLG & O & 13.568 & \ldots & $-$99.999 & $-$99.999 & $-$99.999 & 0.198 & 8658 & 545 & 14.687 & 0.137 & 3 & 1 \\
OGLE-BLG-LPV-000024 & 245.70 & 17:05:43.89 & $-$33:02:32.6 & 351.808730 & 4.784023 & BLG & O & 12.718 & \ldots & $-$99.999 & $-$99.999 & $-$99.999 & 0.198 & 11759 & 603 & 15.352 & 0.111 & 3 & 1 \\
OGLE-BLG-LPV-000028 & 386.80 & 17:05:52.29 & $-$32:39:49.4 & 352.130480 & 4.987692 & BLG & O & 12.868 & \ldots & $-$99.999 & $-$99.999 & $-$99.999 & 0.165 & 8224 & 440 & 14.575 & 0.116 & 3 & 1 \\
OGLE-BLG-LPV-000030 & 172.01 & 17:05:55.21 & $-$32:50:24.6 & 351.994917 & 4.873692 & BLG & O & 12.133 & \ldots & $-$99.999 & $-$99.999 & $-$99.999 & 0.165 & 10681 & 726 & 15.143 & 0.148 & 3 & 1 \\
OGLE-BLG-LPV-000092 & 348.10 & 17:07:36.82 & $-$32:48:40.3 & 352.231693 & 4.605862 & BLG & O & 12.980 & \ldots & $-$99.999 & $-$99.999 & $-$99.999 & 0.216 & 12952 & 943 & 15.562 & 0.158 & 3 & $-$99.999 \\
OGLE-BLG-LPV-000106 & 363.10 & 17:07:44.70 & $-$32:30:20.8 & 352.494298 & 4.765683 & BLG & O & 13.051 & \ldots & $-$99.999 & $-$99.999 & $-$99.999 & 0.193 & 12486 & 1049 & 15.482 & 0.182 & 3 & $-$99.999 \\
OGLE-BLG-LPV-000113 & 342.10 & 17:07:50.52 & $-$32:43:04.1 & 352.335630 & 4.622973 & BLG & O & 13.315 & \ldots & $-$99.999 & $-$99.999 & $-$99.999 & 0.216 & 7693 & 490 & 14.431 & 0.138 & 3 & 1 \\
OGLE-BLG-LPV-000142 & 268.10 & 17:08:39.85 & $-$32:58:09.9 & 352.235834 & 4.334476 & BLG & O & 13.190 & \ldots & $-$99.999 & $-$99.999 & $-$99.999 & 0.239 & 15180 & 1063 & 15.906 & 0.152 & 3 & $-$99.999 \\
OGLE-BLG-LPV-000147 & 208.10 & 17:08:40.98 & $-$32:54:57.2 & 352.281366 & 4.363104 & BLG & O & 12.493 & \ldots & $-$99.999 & $-$99.999 & $-$99.999 & 0.243 & 9241 & 562 & 14.829 & 0.132 & 3 & 1 \\
OGLE-BLG-LPV-000161 & 337.00 & 17:08:45.88 & $-$32:03:07.8 & 352.988923 & 4.862020 & BLG & O & 13.117 & \ldots & $-$99.999 & $-$99.999 & $-$99.999 & 0.221 & 8838 & 669 & 14.732 & 0.164 & 3 & 1 \\
OGLE-BLG-LPV-000190 & 544.90 & 17:08:59.33 & $-$33:16:12.7 & 352.033709 & 4.101120 & BLG & O & 14.424 & \ldots & $-$99.999 & $-$99.999 & $-$99.999 & 0.265 & 9616 & 834 & 14.915 & 0.188 & 3 & 1 \\
OGLE-BLG-LPV-000218 & 233.90 & 17:09:09.16 & $-$32:07:03.4 & 352.984927 & 4.757037 & BLG & O & 14.125 & \ldots & $-$99.999 & $-$99.999 & $-$99.999 & 0.221 & 10234 & 685 & 15.050 & 0.145 & 3 & 1 \\
OGLE-BLG-LPV-000257 & 328.80 & 17:09:25.43 & $-$31:54:59.6 & 353.181672 & 4.829763 & BLG & O & 13.544 & \ldots & $-$99.999 & $-$99.999 & $-$99.999 & 0.225 & 8559 & 453 & 14.662 & 0.115 & 3 & 1 \\
OGLE-BLG-LPV-000284 & 373.90 & 17:09:38.56 & $-$32:58:57.0 & 352.347350 & 4.161511 & BLG & O & 13.089 & \ldots & $-$99.999 & $-$99.999 & $-$99.999 & 0.239 & 8976 & 779 & 14.765 & 0.189 & 3 & 1 \\
OGLE-BLG-LPV-000294 & 332.40 & 17:09:41.82 & $-$32:45:03.3 & 352.541258 & 4.289436 & BLG & O & 12.913 & \ldots & $-$99.999 & $-$99.999 & $-$99.999 & 0.210 & 8205 & 598 & 14.570 & 0.158 & 3 & 1 \\
OGLE-BLG-LPV-000318 & 509.90 & 17:09:46.49 & $-$31:56:10.3 & 353.209979 & 4.758093 & BLG & C & 17.663 & \ldots & $-$99.999 & $-$99.999 & $-$99.999 & 0.225 & 11559 & 1312 & 15.315 & 0.246 & 3 & 1 \\
OGLE-BLG-LPV-000333 & 265.30 & 17:09:51.77 & $-$31:53:39.4 & 353.254985 & 4.767798 & BLG & O & 11.520 & \ldots & $-$99.999 & $-$99.999 & $-$99.999 & 0.225 & 8423 & 572 & 14.627 & 0.148 & 3 & 1 \\
OGLE-BLG-LPV-000334 & 333.70 & 17:09:51.89 & $-$32:53:50.3 & 352.443862 & 4.174377 & BLG & O & 12.333 & \ldots & $-$99.999 & $-$99.999 & $-$99.999 & 0.230 & 6202 & 509 & 13.962 & 0.178 & 3 & 1 \\
OGLE-BLG-LPV-000362 & 351.10 & 17:10:00.28 & $-$32:10:58.4 & 353.039177 & 4.573028 & BLG & O & 13.816 & \ldots & $-$99.999 & $-$99.999 & $-$99.999 & 0.221 & 16471 & 1302 & 16.084 & 0.172 & 3 & $-$99.999 \\
\vdots & \vdots & \vdots & \vdots & \vdots & \vdots & \vdots & \vdots & \vdots & \ldots & \vdots & \vdots & \vdots & \vdots & \vdots & \vdots & \vdots & \vdots & \vdots & \vdots \\
OGLE-GD-LPV-025617 & 306.50 & 19:15:50.43 & +07:06:39.0 & 42.044152 & $-$2.143960 & GD & O & 14.303 & \ldots & $-$99.999 & $-$99.999 & $-$99.999 & 0.526 & 12163 & 738 & 15.425 & 0.132 & 3 & $-$99.999 \\
OGLE-GD-LPV-025618 & 262.70 & 19:15:53.20 & $-$03:04:55.5 & 32.968241 & $-$6.842566 & GD & O & 12.908 & \ldots & $-$99.999 & $-$99.999 & $-$99.999 & 0.186 & 11280 & 1343 & 15.261 & 0.259 & 3 & $-$99.999 \\
OGLE-GD-LPV-025619 & 490.10 & 19:15:53.96 & +09:36:55.7 & 44.269102 & $-$0.992049 & GD & O & 12.770 & \ldots & $-$99.999 & $-$99.999 & $-$99.999 & $-$99.999 & $-$99.999 & $-$99.999 & $-$99.999 & $-$99.999 & 0 & $-$99.999 \\
OGLE-GD-LPV-025620 & 364.90 & 19:15:58.21 & +13:10:10.8 & 47.422567 & 0.647777 & GD & O & 17.518 & \ldots & 6.154 & 0.654 & 0.933 & 0.954 & 9829 & 680 & 14.963 & 0.150 & 7 & $-$99.999 \\
OGLE-GD-LPV-025621 & 414.70 & 19:15:58.84 & +09:17:30.4 & 43.991899 & $-$1.160472 & GD & O & 16.414 & \ldots & $-$99.999 & $-$99.999 & $-$99.999 & 0.929 & 8439 & 629 & 14.632 & 0.162 & 3 & $-$99.999 \\
OGLE-GD-LPV-025622 & 418.60 & 19:15:59.32 & +16:46:00.5 & 50.610078 & 2.317125 & GD & O & 15.887 & \ldots & $-$99.999 & $-$99.999 & $-$99.999 & 0.662 & 9066 & 682 & 14.787 & 0.163 & 5 & $-$99.999 \\
OGLE-GD-LPV-025623 & 451.80 & 19:16:00.03 & +08:04:29.8 & 42.916727 & $-$1.730961 & GD & O & 12.167 & \ldots & $-$99.999 & $-$99.999 & $-$99.999 & 0.518 & 5924 & 744 & 13.863 & 0.273 & 3 & $-$99.999 \\
OGLE-GD-LPV-025624 & 218.20 & 19:16:01.19 & $-$04:03:03.4 & 32.112832 & $-$7.310733 & GD & O & 12.127 & \ldots & $-$99.999 & $-$99.999 & $-$99.999 & 0.203 & 6740 & 395 & 14.143 & 0.127 & 3 & 1 \\
OGLE-GD-LPV-025625 & 207.50 & 19:16:01.78 & $-$07:14:13.5 & 29.238370 & $-$8.742704 & GD & O & 10.560 & \ldots & $-$99.999 & $-$99.999 & $-$99.999 & 0.124 & 4295 & 266 & 13.165 & 0.134 & 3 & $-$99.999 \\
OGLE-GD-LPV-025626 & 246.30 & 19:16:02.02 & +14:16:08.5 & 48.402796 & 1.145997 & GD & O & 13.045 & \ldots & $-$99.999 & $-$99.999 & $-$99.999 & 0.313 & 3550 & 193 & 12.751 & 0.118 & 4 & $-$99.999 \\
OGLE-GD-LPV-025627 & 365.50 & 19:16:02.25 & +09:32:12.1 & 44.215228 & $-$1.058867 & GD & O & 15.775 & \ldots & 4.831 & 0.548 & 0.782 & 0.881 & 4162 & 284 & 13.096 & 0.148 & 5 & $-$99.999 \\
OGLE-GD-LPV-025628 & 552.10 & 19:16:03.38 & +10:55:03.7 & 45.439597 & $-$0.419830 & GD & C & 16.860 & \ldots & $-$99.999 & $-$99.999 & $-$99.999 & 0.937 & 3658 & 1588 & 12.817 & 0.943 & 1 & $-$99.999 \\
OGLE-GD-LPV-025629 & 338.40 & 19:16:03.62 & +00:45:43.2 & 36.427432 & $-$5.126788 & GD & O & 13.670 & \ldots & $-$99.999 & $-$99.999 & $-$99.999 & 0.162 & 13744 & 1003 & 15.690 & 0.158 & 3 & $-$99.999 \\
OGLE-GD-LPV-025630 & 407.80 & 19:16:05.40 & $-$04:09:09.2 & 32.029396 & $-$7.372240 & GD & C & 10.804 & \ldots & $-$99.999 & $-$99.999 & $-$99.999 & 0.155 & 2140 & 854 & 11.652 & 0.866 & 1 & $-$99.999 \\
OGLE-GD-LPV-025631 & 170.18 & 19:16:05.53 & $-$03:59:31.0 & 32.174114 & $-$7.300167 & GD & O & 11.540 & \ldots & $-$99.999 & $-$99.999 & $-$99.999 & 0.203 & 9265 & 437 & 14.834 & 0.102 & 3 & $-$99.999 \\
OGLE-GD-LPV-025632 & 267.70 & 19:16:05.84 & +23:17:28.7 & 56.422200 & 5.308147 & GD & O & 13.044 & \ldots & $-$99.999 & $-$99.999 & $-$99.999 & 0.223 & 10835 & 771 & 15.174 & 0.155 & 3 & $-$99.999 \\
OGLE-GD-LPV-025633 & 290.10 & 19:16:06.44 & $-$03:01:03.6 & 33.051229 & $-$6.862450 & GD & O & 13.017 & \ldots & $-$99.999 & $-$99.999 & $-$99.999 & 0.238 & 14920 & 1124 & 15.869 & 0.164 & 3 & $-$99.999 \\
OGLE-GD-LPV-025634 & 335.40 & 19:16:06.64 & +14:17:31.9 & 48.431990 & 1.140272 & GD & O & 13.421 & \ldots & $-$99.999 & $-$99.999 & $-$99.999 & 0.523 & 7124 & 568 & 14.263 & 0.173 & 5 & $-$99.999 \\
OGLE-GD-LPV-025635 & 443.00 & 19:16:06.76 & +14:37:24.3 & 48.725425 & 1.294035 & GD & O & 13.221 & \ldots & $-$99.999 & $-$99.999 & $-$99.999 & 0.369 & 4345 & 446 & 13.190 & 0.223 & 2 & $-$99.999 \\
OGLE-GD-LPV-025636 & 255.50 & 19:16:07.04 & $-$00:06:14.3 & 35.661065 & $-$5.536504 & GD & O & 12.207 & \ldots & $-$99.999 & $-$99.999 & $-$99.999 & 0.162 & 13278 & 854 & 15.616 & 0.140 & 3 & $-$99.999 \\
\hline

\label{tab:all_information}
\end{tabular}
}
\end{adjustwidth}

\begin{adjustwidth}{-1.5cm}{}
\tablecomments{For each star, we provide the original ID from the catalog \citet{2022ApJS..260...46I}, pulsation period $P$, equatorial coordinates (R.A. and Decl.), Galactic coordinates (l, b), region of the Milky Way toward which the star is observed (BLG for the Galactic bulge or GD for the Galactic disk), chemical type of Mira (O or C), OGLE {\it I}-band and {\it V}-band mean brightness ($m_I$ and $m_V$) and amplitudes ($\Delta m_I$ and $\Delta m_V$), mid-IR mean magnitudes ($m_\lambda$) with uncertainties ($\sigma_\lambda$) and amplitudes ($\Delta m_\lambda$), for WISE {\it W}1-{\it W}3 bands, and Spitzer {\it I}1-{\it I}4 bands, extinction ($A_{Ks}$), distance ($d$) and its uncertainty ($\sigma_d$), distance modulus ($\mu$) and its uncertainty ($\sigma_\mu$), number of mid-IR measurements ($n$) used for distance and extinction calculations, and finally mask flag, which indicates if the star is inside the mask (mask value $0$) or outside the mask (mask value $1$). Stars that have a $-99.999$ value in the mask column are not located in the analyzed cuboid. More information about the mask flag and analyzed cuboid can be found in Section \ref{subsec:sample_selection}. The table rows are sorted by the star ID and location in the sky. A value $-99.999$ in the columns $m_V$, $m_\lambda$, $A_{Ks}$, $d$, $\sigma_d$, $\mu$ or $\sigma_\mu$ means that the star has no OGLE {\it V}-band observations, was not found in the mid-IR databases (for chosen bands, or all of them), or measurements were unavailable. This table is available in its entirety in a machine-readable form in the online journal. Here we present the first and last twenty rows for guidance regarding its form and content.}
\end{adjustwidth}

\end{table}
\end{turnpage}

\clearpage

\noindent extinction obtained with the "mwdust" code against extinction obtained using mid-IR photometry and $K_s$ photometry from the VISTA Variables in the Via Lactea \citep{2012A&A...537A.107S} survey data for 273 Galactic Cepheids. The authors conclude that due to the strong dependence of the accuracy of the extinction determination on the individual star's $K_s$ photometry, the use of \mbox{"mwdust"} extinction yields more reliable and homogeneous results. We follow their approach and use \mbox{"mwdust"} $A_{Ks}$ extinction estimation to correct the mid-IR brightness. The $A_{Ks}$ values were transformed into extinctions $A_\lambda$ in the mid-IR bands using the mid-IR extinction curve based on the results from the "Apache Point Observatory Galactic Evolution Experiment" (APOGEE) survey \citep{2016ApJS..224...23X}: $A_{W1}/A_{Ks} = 0.591$, $A_{W2}/A_{Ks} = 0.463$, $A_{W3}/A_{Ks} = 0.537$, $A_{I1}/A_{Ks} = 0.553$, $A_{I2}/A_{Ks} = 0.461$, $A_{I3}/A_{Ks} = 0.389$, $A_{I4}/A_{Ks} = 0.463$.

\begin{figure*}
\centering
\includegraphics[scale=0.35]{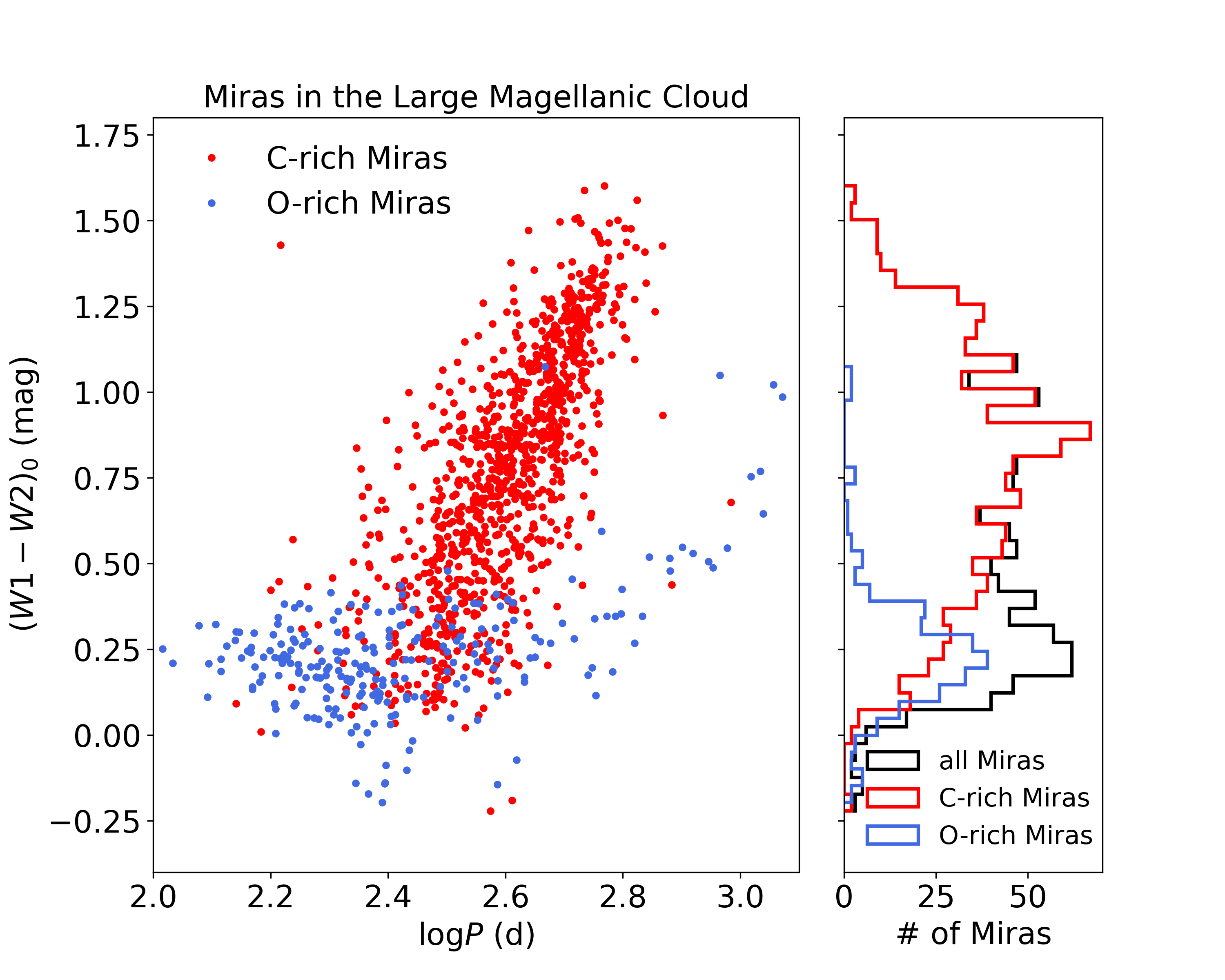}
\includegraphics[scale=0.35]{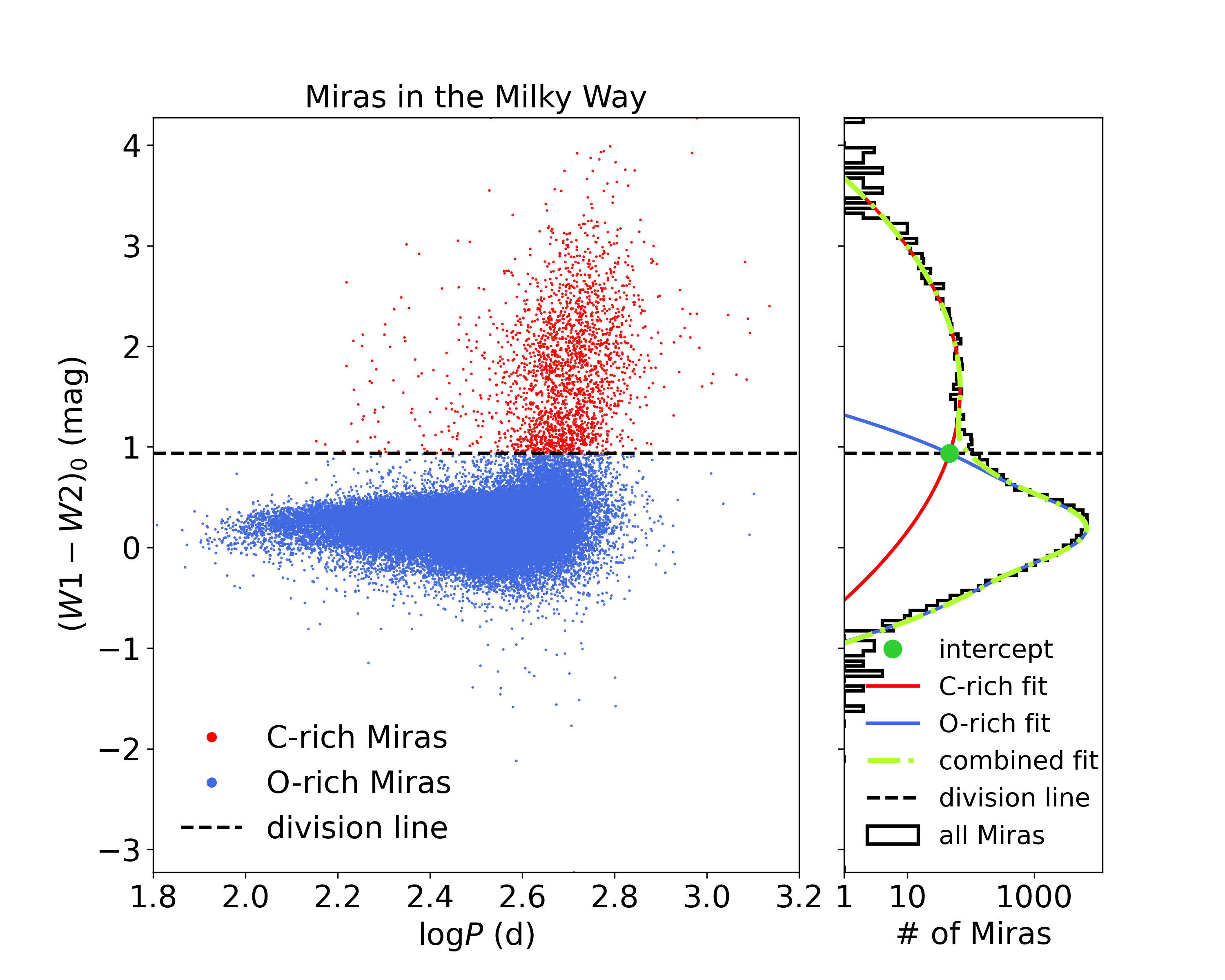}
\caption{Mid-IR dereddened color index $(W1-W2)_0$ vs. $\log P$ diagram. Left plot: Diagram for Miras located in the LMC, colored using the classification into O- and C-rich as assigned by \citet{2009AcA....59..239S}. Right plot: Division into O-rich and C-rich Galactic Miras made in this paper. O-rich Miras are marked with blue points, whereas the C-rich Miras are marked with red points. The total distribution of $(W1-W2)_0$ is shown by a black histogram. The combined model fitted to the $(W1-W2)_0$ Milky Way Mira distribution is plotted with a dashed-dotted green line, while solid blue and red lines present separate models for the O-rich and C-rich Mira distributions. The intercept points are marked by a green dot, whereas the division line is plotted in black.}
\label{fig:OCdivision}
\end{figure*}

\section{Distances to individual Miras} \label{sec:distances}

The distance $d_\lambda$ in a given band $\lambda$ can be calculated using relation:

\begin{equation}
    d_\lambda = 10^{0.2\mu_{\lambda,0} + 1} \hspace{0.1cm} \mathrm{(pc)},
\label{eqn:distance}
\end{equation}

\noindent where $\mu_{\lambda, 0}$ is the extinction-corrected distance modulus given by the equation:

\begin{equation}
    \mu_{\lambda, 0} = m_\lambda - M_\lambda - A_\lambda \hspace{0.1cm} \mathrm{(mag)},
\label{eqn:distance_modulus}
\end{equation}

\noindent where $m_\lambda$ is the mean (observed) brightness calculated as described in Section \ref{subsec:mean_brightness}, $M_\lambda$ is the absolute magnitude, and $A_\lambda$ is the extinction in a given mid-IR band, calculated from  $A_{Ks}$ band as described in Section \ref{subsec:extinction}.

Absolute magnitudes $M_\lambda$ in each WISE and Spitzer band can be calculated separately for O-rich and C-rich Miras using PLRs obtained for the LMC Miras \citep[][]{2021ApJ...919...99I}. These PLRs are the most accurate to date and allow for the measurement of distances with precision at the level of $5\%$ for O-rich Miras and $12\%$ for C-rich Miras. Before the distance calculation, we divided the Miras into O- and C-rich. A detailed description of this procedure can be found in Section~\ref{sec:OCdivision}. For O-rich Miras, we used PLRs in quadratic forms, as presented in \citet[][]{2021ApJ...919...99I}. The uncertainties of the distance moduli $\sigma_{\mu_{\lambda, 0}}$ were calculated using the mean brightness uncertainties $\sigma_{\lambda}$ and the PLR scatter \citep{2021ApJ...919...99I} combined in the quadrature. 

The distance to each Mira star can be measured separately for each available WISE and Spitzer band (in total, up to seven measurements per star). The estimation of the interstellar extinction from the "mwdust" model depends on the distance to the given target. Therefore, we measured the distance and the extinction for each band iteratively. In the first iteration, we calculated the distance modulus in each band, assuming no extinction. Then we calculated the weighted mean distance modulus while at the same time removing the distance modulus that deviated more than 3$\sigma$ from the mean. As weights, we took an inverse square of the uncertainties of the distance moduli ($\sigma_{\mu_{\lambda, 0}}^{-2}$). From the resulting mean distance modulus we calculated distance $d$, thus, we could estimate extinction $A_{Ks}$ from "mwdust" and transform it to the extinction in each of the mid-IR bands as described in Section \ref{subsec:extinction}. In the second and several subsequent iterations, we calculated the distance moduli $\mu_{\lambda, 0}$ again, taking into account extinction values from the previous iteration. In each iteration, we calculated the weighted mean distance modulus and the updated distance $d$. After a few iterations, both the distance and the extinction converge. Then, we obtained the final distance $d$ and its uncertainty given as:

\begin{equation}
    \sigma_d = d \times \ln(10) \times 0.2\sigma_\mu,
\label{eqn:distance_uncertainty}
\end{equation}

\noindent where $\sigma_\mu$ is the standard error of the weighted distance modulus $\mu$ calculated from the weights, i.e.,  $(\sum_i \sigma_{\mu_{\lambda, 0}}^{-2})^{-1/2}$. In Table \ref{tab:all_information}, we present all parameters derived in this paper of 65,981 Mira variables from the catalog by \citet{2022ApJS..260...46I}. This table is also available through the OGLE website\footnote{\url{https://www.astrouw.edu.pl/ogle/ogle4/MILKY_WAY_3D_MAP/}} and Zenodo at \dataset[DOI: 10.5281/zenodo.7472598]{http://doi.org/10.5281/zenodo.7472598}.

The saturation limits for Spitzer bands are 6.0 mag, 5.5 mag, 3.0 mag, 3.0 mag for {\it I}1, {\it I}2, {\it I}3, {\it I}4 bands, respectively. The WISE detectors saturate below \mbox{8.0 mag}, 7.0 mag, 3.8 mag for {\it W}1, {\it W}2, and {\it W}3 bands, respectively. However, the WISE profile-fitting photometry allows us to extract reliable measurements for saturated sources using non-saturated wings of their profiles. As a result, the NEOWISE-R and AllWISE databases contain useful information for sources up to the brightness of \mbox{{\it W}1~$=2$~mag}, {\it W}2~$=0$~mag, and \mbox{{\it W}3~$=-3$~mag}\footnote{\url{https://wise2.ipac.caltech.edu/docs/release/neowise/expsup/sec2\_1c.html} \\ and \\ \url{https://wise2.ipac.caltech.edu/docs/release/allsky/expsup/sec6\_3d.html}}. All of our Miras are fainter than the above-mentioned saturation limits. However, to avoid contaminating distance measurements with perhaps inaccurately measured brightness during the profile-fitting photometry using non-saturated wings, we removed from the distance calculations {\it W}1, {\it W}2, and {\it W}3 measurements brighter than 4~mag, 3~mag, and 0~mag, respectively. This was the case for 2172 Miras. We marked these measurements as missing with $-99.999$ value in \mbox{Table \ref{tab:all_information}}.

We measured distances to 65,385 Miras (out of 65,981). Most stars (59,902) have distances measured with accuracy better than $10\%$. The median distance accuracy is at a level of $6.6\%$.  

\section{Division into O-rich and C-rich Miras} \label{sec:OCdivision}

Miras (like other AGB stars) can be divided into \mbox{O- and C-rich} types. In the LMC, \citet{2005AcA....55..331S, 2009AcA....59..239S} separated O-rich and C-rich LPVs using optical and near-infrared Wesenheit indices. However, the analysis of LPVs toward the BLG showed that this method cannot be used for Milky Way stars due to the considerable depth of the bulge along the line-of-sight \citep{2013AcA....63...21S}. Recently, we found that such a division can be made based on the mid-IR color index and the pulsation period of stars \citep{2021ApJ...919...99I}.

In Figure \ref{fig:OCdivision} (left plot), we show $(W1-W2)_0$-$\log P$ plane for the LMC Miras. We used dereddened mid-IR brightness and periods from \citet{2021ApJ...919...99I}, and  the O/C classification made by \citet{2009AcA....59..239S}. It is clearly seen that C-rich Miras have redder colors $(W1-W2)_0$ than O-rich ones, and the spread of the O-rich Miras colors is much smaller, and they clump between $(W1-W2)_0 = 0$ mag and $(W1-W2)_0 = 0.5$ mag. There is a partial overlap between both populations.

We constructed a similar plot (Figure \ref{fig:OCdivision}, right plot) for the Milky Way Miras using pulsation periods from \citet{2022ApJS..260...46I} and mid-IR mean brightness measured in this paper (see Section \ref{subsec:mean_brightness}). We noticed that similar deflection toward the redder colors of some Miras is present in the Milky Way, with a comparable clump consisting of a much larger number of stars around $(W1-W2) = 0$. Therefore, we used $(W1-W2)_0$ vs. $\log P$ plane to divide Milky Way Miras into \mbox{O- and C-rich}.

We used 63,577 Miras that have a mid-IR mean brightness in both {\it W}1 and {\it W}2 WISE bands. The O/C division was performed in two separate iterations. In the first iteration, we assumed that the extinction toward each star was absent. We fitted the Gaussian Mixture Model (GMM) with three components to the $(W1-W2)$ distribution.

We searched the parameter space using the Monte Carlo Markov Chain (MCMC) and Python library \texttt{emcee}\footnote{\url{https://github.com/dfm/emcee}} \citep{2013PASP..125..306F}. The O-rich Mira distribution is characterized by two components of the GMM, while the C-rich Mira distribution is characterized by one component. We found a first approximation of the intercept point of these groups is at $(W1-W2) = 1.046$ mag and assumed this as a classification boundary between O-rich and C-rich stars before the next steps (i.e., stars with $(W1-W2) < 1.046$ mag are O-rich).

\begin{figure*}
\centering
\includegraphics[scale=0.48]{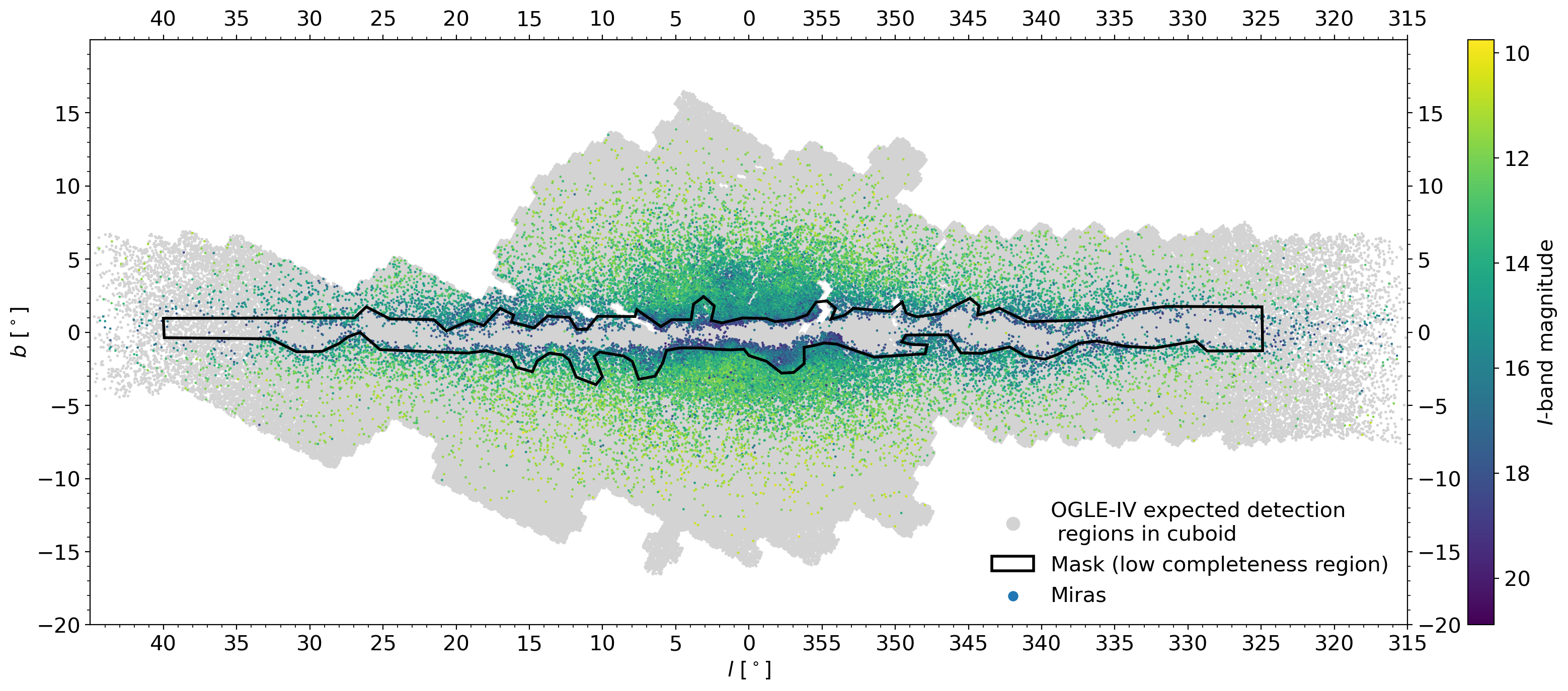}
\caption{Distribution of Miras in Galactic coordinates. The sample of Miras shown is limited to the cuboid in three-dimensional $(X, Y, Z)$ space defined by: $4$ kpc $\leq X \leq 12$ kpc, $-4$ kpc $\leq Y \leq 4$ kpc, and $-2$ kpc $\leq Z \leq 2$ kpc. Miras are presented by points color-coded with the {\it I}-band magnitude. Gray points represent areas of the OGLE fields, limited to the above-mentioned cuboid, where we expect any detection. More about expected detections in the OGLE fields can be found in Section \ref{subsec:fitting_procedure}. The black polygon represents an area that we excluded from the analysis because of the low completeness of the Mira sample caused by the high extinction near the Galactic plane.}
\label{fig:mask}
\end{figure*}

During the second iteration, we iteratively calculated the distance and extinction toward each Mira star as described in Sections \ref{subsec:extinction} and \ref{sec:distances}, taking into account the Mira type from the previous iteration to choose an appropriate PLR for obtaining absolute brightness. Then we calculated the final extinction-corrected mean brightness in {\it W}1 and {\it W}2 bands and the dereddened color index $(W1-W2)_0$. We repeated the fitting procedure from the previous iteration and found the dereddened intercept point at $0.936$ mag. The final division of the Milky Way Miras is presented in Figure \ref{fig:OCdivision} (right plot).

We divided the Milky Way Miras into O-rich (\mbox{$(W1-W2)_0 < 0.936$ mag}; 61,333 objects), and C-rich (\mbox{$(W1-W2)_0 \geq 0.936$ mag}; 2244 objects). However, in Figure~\ref{fig:OCdivision} (right plot) it is clearly seen that both O-rich and C-rich Mira distributions partly overlap. We estimated possible contamination in both the O-rich and C-rich groups by Miras from the other group by integrating the area under the model curve. The possible contamination of the C-rich Miras by O-rich Miras is at a level of $4.34\%$, while the contamination of the O-rich Miras by the C-rich Miras is at the level $0.72\%$. In other words, 442 of the 61,333 O-rich Miras could be C-rich, while 98 of the 2244 C-rich Miras could be O-rich ones. Such small contamination should not significantly affect our analysis and the final result. Since this division is made solely based on the photometric properties of Mira stars, spectroscopic observations are needed to confirm their chemical type.

For stars without $W1$ or $W2$ mean brightness measurements, we set their type as O-rich (2404 such cases), since this type predominates in the Milky Way (no more than $4\%$ is C-rich). Finally, we divided the collection of 65,981 Milky Way Miras into 63,737 O-rich and 2244 C-rich stars.

\section{Three-dimensional structure of the Milky Way} \label{sec:3dmap}

Mira stars are characterized by a very wide range of ages; therefore, they can be found both in the structures represented by the old- and intermediate-age populations, i.e., bulge/bar and a long bar, as well as near the structures typical for the young population, i.e., spiral arms  \citep{2020ApJ...891...50U, 2022ApJS..260...46I}. Most Miras, however, belong to the intermediate-age stellar population, and they clump around the center of the Milky Way. 
\citet{2022MNRAS.517.6060S} showed an anticorrelation between the location of Miras along the Galactic longitude $l$ and their pulsation periods (which is related to the age; the longer the pulsation period the younger the Mira star) with respect to the GC. The authors explained that such an anticorrelation can be explained by an age-morphology dependence of the boxy/peanut bulge.

The BLG is an extremely interesting region to study in three dimensions based on such a large sample of Miras. A number of previous studies, often based on the analysis of red clump stars, indicated the existence of the X-shaped structure \citep[e.g.,][]{2010ApJ...721L..28N, 2010ApJ...724.1491M, 2013MNRAS.435.1874W, 2019MNRAS.489.3519C}. On the other hand, other studies argue that the X-shaped structure is not visible in the Mira stars distribution, and the boxy model is more appropriate to describe the BLG's shape \citep[e.g.,][]{2017ApJ...836..218L, 2022A&A...666L..13C}.

Another puzzle of the Milky Way structure is related to the long bar -- a thin and long stellar overdensity extending central bar. This component has been discovered in the near-infrared data in 1990s \citep[e.g.,][]{1995MNRAS.276..301C}, and later studied by several groups \citep[e.g., ][]{2007AJ....133..154L, 2017MNRAS.471.3988C}. The long bar appears to be a very thin structure, about $200$ pc wide, with a different orientation angle than the bar \citep[e.g.,][]{2007AJ....133..154L, 2008A&A...491..781C}. It turns out that the determination of the length and orientation angle of the long bar is crucial for the development of models of the inner part of the Milky Way. Mira variables, due to their intermediate ages, seem to be the best tool for conducting such an analysis.

\subsection{Cartesian coordinates} \label{subsec:cartesian_coordinates}
We studied the three-dimensional distribution of Miras in the Milky Way in the Cartesian coordinate system with the origin at the Sun. We transform the galactic $(l, b, d)$ coordinates to $(X, Y, Z)$ coordinates using transformations:

\begin{align}
\begin{split}
    X &= d \times \cos{l}\cos{b}, \\
    Y &= d \times \sin{l}\cos{b}, \\
    Z &= d \times \sin{b}. 
\end{split}
\label{eqn:cartesian_coordinates}    
\end{align}

\subsection{Sample selection} \label{subsec:sample_selection}

We limited our exploration space to a three-dimensional cuboid: \mbox{$4$ kpc $\leq X \leq 12$ kpc}, \mbox{$-4$ kpc $\leq Y \leq 4$ kpc} and \mbox{$-2$ kpc $\leq Z \leq 2$ kpc} covering the central parts of the Milky Way. This limit left us with 39,619 Miras.

The completeness of the sample is essential when studying three-dimensional structures. Our Galactic Miras collection \citep{2022ApJS..260...46I} does not contain stars located very close to the Galactic plane (in the range of about $-1^\circ \leq b \leq 1^\circ$). This is caused by the very large amount of dust inside this region that nearly completely obscures the stars in the OGLE {\it I} and {\it V} bands. Additionally, at slightly larger $|b|$, the clouds of dust are less optically thick but still obscure the stars. This effect is clearly shown in Figure \ref{fig:mask}, where Miras near the Galactic plane are fainter. In this figure, we present Miras located in the above-mentioned cuboid only. Therefore, we completely removed the region near the Galactic plane from the further analysis, i.e., we removed regions with no Miras detection, or with significantly reduced completeness of our catalog. Admittedly, we significantly reduced the impact of interstellar extinction by the use of mid-IR data, where its value is noticeably smaller. However, this does not increase the completeness of the sample close to the Galactic plane. We removed the stars from the incomplete area using the mask shown with the black polygon in Figure \ref{fig:mask}. With this mask, we removed 7180 Miras from the analysis, leaving us with 32,439 Miras. Stars located inside the mask were assigned the value of $0$ in the column "mask" in Table \ref{tab:all_information}, while stars located outside the mask (so the ones used in further analyses) were assigned the value of $1$. A similar cut was made by \citet[][]{2015ApJ...811..113P} during their analysis of the three-dimensional structure of the BLG using RR Lyrae stars.

The last cut we made was based on the distance accuracy and the chemical type of Miras. We removed all C-rich Miras from the analysis (2244 objects), as these stars usually change their mean brightness over time due to the significant mass-loss phenomenon \citep{2021ApJS..257...23I}. This means that their distances are likely to be affected by inaccurate mean magnitude measurements. Moreover, we limited our sample to stars with distances measured with accuracy better than $20\%$ only. These two conditions left us with 31,992 Miras, which is our final sample used to study their three-dimensional distribution in the Milky Way.

\subsection{Galactic bar model}

Recently, \citet{2022MNRAS.514L...1S} proposed the \mbox{39-parameter} analytical model of the Milky Way bar, which reproduces in detail the three-dimensional distribution of the N-body bar including the X-shaped structure. The authors obtained this model by fitting a multi-component parametric density distribution to a made-to-measure N-body model of \citet{2017MNRAS.465.1621P}. The latter model was proposed because it accurately represents observations, both the density data and the kinematics of stars in the center of the Galaxy.

\citet{2022MNRAS.514L...1S} model describes a three-dimensional density in the Cartesian coordinates, and is composed of three barred components and an axisymmetric disk:

\begin{equation}
    \rho (X_\mathrm{S},Y_\mathrm{S},Z_\mathrm{S}) = \rho_{\mathrm{bar, 1}} + \rho_{\mathrm{bar, 2}} + \rho_{\mathrm{bar, 3}} + \rho_{\mathrm{disk}},
\label{eqn:model}    
\end{equation}

\noindent where $\rho_{\mathrm{bar, 1}}$ and $\rho_{\mathrm{bar, 2}}$ represent the bar component, including the X-shaped structure in the center, $\rho_{\mathrm{bar, 3}}$ represents the long bar, i.e., vertically flat extension of the bar, and $\rho_{\mathrm{disk}}$ gives the contribution from the Milky Way disk to the total bar density. For the detailed equations of this model, we refer to \citet{2022MNRAS.514L...1S}.

The coordinate system of \citet{2022MNRAS.514L...1S} model has the origin at the GC. It does not take into account the orientation angle of the bar, and the density $\rho(X_\mathrm{S}, Y_\mathrm{S}, Z_\mathrm{S})$ is defined in units $10^{10} M_\odot$ kpc$^{-3}$. We slightly modified this model by shifting and rotating the coordinate system and modifying the density unit, as we did not have prior information on the ratio of the total mass of O-rich Miras to the total stellar mass. We defined a modified model as:

\begin{equation}
    \rho'(X',Y',Z') = \beta(\rho_{\mathrm{bar, 1}} + \rho_{\mathrm{bar, 2}} + \rho_{\mathrm{bar, 3}} + \rho_{\mathrm{disk}}),
\label{eqn:modified_model}    
\end{equation}

\noindent where coordinates $(X', Y', Z')$ are shifted with parameters $X_{\mathrm{GC}}$, $Y_{\mathrm{GC}}$, $Z_{\mathrm{GC}}$ (these parameters indicate the Galaxy center in the coordinates system with the origin at the Sun), and rotated by the angle $\theta$, and they are given by equations:

\begin{align}
\begin{split}
    X' &= (Y - Y_{\mathrm{GC}}) \times \sin{\theta} + (X - X_{\mathrm{GC}}) \times \cos{\theta}, \\
    Y' &= (Y - Y_{\mathrm{GC}}) \times \cos{\theta} - (X - X_{\mathrm{GC}}) \times \sin{\theta}, \\
    Z' &= Z - Z_{\mathrm{GC}},
\end{split}
\label{eqn:modified_cartesian_coordinates}    
\end{align}

\noindent while $\beta$ is the scaling parameter that has a unit of the number of Miras per $10^{10} M_\odot$. Finally, we fitted the 44-parameter model to the Milky Way bar.

\subsection{Fitting procedure} \label{subsec:fitting_procedure}

Although the positions in the sky of each Mira variable are known with great accuracy, the distances are not. There is significant uncertainty in this coordinate, therefore, it is necessary to take it into account to reproduce a meaningful fit to the three-dimensional distribution of Miras.

We took into account the distance uncertainties by implementing Bayesian hierarchical inference in our method. The theory of hierarchical Bayesian methods is beyond the scope of this paper, and we refer to, e.g., \citet{Gelman2004} and \citet{Gelman2007}. The use of Bayesian hierarchical inference in astronomical research is becoming increasingly common, and it is used in many branches of modern astronomy, as shown by, e.g., \citet{2014ApJ...795...64F, 2018A&A...616A...9L, 2019A&A...623A.156D, 2021AcA....71....1P}.

Assuming that our collection of Mira stars \citep{2022ApJS..260...46I} contains $N$ Mira variables located in the Milky Way, we can define the likelihood function as:

\begin{equation}
    \mathcal{L}({\bf p}) = e^{-N_\mathrm{exp}} \prod_{i=1}^{N} \rho'(X'_i, Y'_i, Z'_i;\,\, {\bf p}),
    \vspace{0.1cm}
\label{eqn:likelihood1}
\end{equation}

\noindent where $N_\mathrm{exp}$ in the normalization term is the expected number of Miras observed in the OGLE fields in the analyzed cuboid (Section \ref{subsec:sample_selection}) and is given by:

\begin{equation}
    N_\mathrm{exp} = \int \rho'(X', Y', Z';\,\, {\bf p}) d\mathrm{log}X' d\mathrm{log}Y' d\mathrm{log}Z'.
\label{eqn:n_expected1}
\end{equation}

\noindent In both equations, ${\bf p}$ denotes a vector of 44 model parameters. The assumption that we observe all the stars is not true in any sky survey, as the observed number of stars is affected by many factors, e.g. instrumental ones. Therefore, in the likelihood function, a term taking into account detection efficiency is necessary. Since the number of all Miras is given by the Poisson distribution, under the assumption of negligible uncertainties of the distances (and thus  $(X', Y', Z')$ coordinates), we can define the likelihood function as:

\begin{equation}
    \mathcal{L}({\bf p}) = e^{-N_\mathrm{exp}} \prod_{i=1}^{N_\mathrm{obs}} \rho'(X'_i, Y'_i, Z'_i;\, {\bf p}),
    \vspace{0.1cm}
\label{eqn:likelihood_2}
\end{equation}

\noindent where $N_\mathrm{obs}$ is the number of Miras observed/analyzed. In our case $N_\mathrm{obs} = 31,992$, as described in Section \ref{subsec:sample_selection}.

The likelihood $\mathcal{L}$ given by Equation \ref{eqn:likelihood_2} can be modified into hierarchical form by including the posterior distribution of distance to each Mira in Equation \ref{eqn:likelihood_2} and marginalizing over these distributions. Therefore, for each star $i$ we drew $K=10$ samples of distance according to the normal distribution with a mean equal to the estimated distance $d_i$ and a standard deviation equal to the distance uncertainty $\sigma_{d,i}$. We then combined the drawn distances and on-sky coordinates to calculate the Cartesian coordinates of these samples $(X_{i,k}', Y_{i,k}', Z_{i,k}')$.  Finally, the hierarchical likelihood could be defined as:

\begin{equation}
    \mathcal{L}({\bf p}) = e^{-N_\mathrm{exp}}\prod_{i=1}^{N_\mathrm{obs}} \left(\frac{1}{K}\sum_{k=1}^{K}\rho'(X'_{i,k}, Y'_{i,k}, Z'_{i,k};\, {\bf p})\right).
\vspace{0.1cm}
\label{eqn:likelihood_hierarchical}
\end{equation}

During the sample drawing, it could happen that for Miras lying near the boundaries of the analyzed cuboid, the randomly selected samples could fall outside the analyzed area. In such cases, we checked what percentage of the drawn samples fell outside the cuboid, and if it was greater than $50\%$, we removed such stars from the analysis. It was the case for 112 Miras. On the other hand, if more than $50\%$ of the samples at the first draw were inside the cuboid, we repeated the draw until we had full 10 samples for each Mira inside the cuboid.

When calculating $N_{\mathrm{exp}}$, we took into account the success rate of finding Miras in the cuboid analyzed in the OGLE survey fields. We randomly selected from the uniform distribution 1,000,000 points from the defined cuboid. We calculated the Galactic coordinates $(l,b)$ for each point and checked how many points could be observed by the OGLE survey. In the OGLE fields, we would be able to find $67\%$ points out of the whole sample. All these points are presented in Figure \ref{fig:mask}.

We fitted the model given by Equation~\ref{eqn:modified_model} to the Galactic Mira distribution using the MCMC implemented in the Python package \texttt{emcee}\footnote{\url{https://github.com/dfm/emcee}}\citep{2013PASP..125..306F}. In all fitting runs, i.e., during bias exploration (see Section \ref{subsec:bias_exploration}) and fitting the final model to the Mira distribution (see Section \ref{subsec:final_fit}), we used 1000 walkers and 5000 steps for a single run. We adopted the flat prior distributions for most parameters, limiting their values to the physically reasonable range (as summarized in \mbox{Table \ref{tab:final_solution}}). The only parameter with the informative prior defined by the normal distribution was $\theta$, i.e, the inclination of the bar to the line of sight. As a $\theta$ prior, we used $p(\theta) = \mathcal{N}(\theta; 20, 3)$, where the mean and standard deviation values were taken from \citet{2015ApJ...811..113P}. Priors for all parameters that we used during model fitting are presented in Table \ref{tab:final_solution}.

\begin{table*}
\centering

\caption{Final solution for the Milky Way bar with statistical and systematic uncertainties. We also provide the parameter values as presented in the original paper about the model, i.e., \citet{2022MNRAS.514L...1S}, and priors used during model fitting. The only parameter with the informative prior defined by the normal distribution was $\theta$. For other parameters, we adopted flat prior distributions between indicated values.}

\footnotesize
\begin{tabular}{cccccccc}

\hline \hline

parameter & \citet{2022MNRAS.514L...1S} value  & this paper value & $\sigma_{\mathrm{stat},+}$ & $\sigma_{\mathrm{stat},-}$ & $\sigma_{\mathrm{sys}}$ & unit & this paper prior \\ \hline \hline
\multicolumn{8}{c}{coordinate system parameters}      \\
\hline \hline 
$\beta$   &          &  21101     &     2046       &    1940      &    & $\mathrm{\# \; of \; Miras} \; 10^{-10} M_\odot^{-1}$ &  (10000, 25000)     \\
$X_{\mathrm{GC}}$ &    & 7.66    &      0.01      &       0.01   & 0.39  &   kpc  & (4, 12)  \\
$Y_{\mathrm{GC}}$ &    & $-$0.01    &   0.01         &       0.01   & 0.01    &   kpc  &  ($-$4, 4)  \\
$Z_{\mathrm{GC}}$ &    & 0.01    &     0.01        &     0.01    &  0.01   &   kpc  & ($-$2, 2)   \\
$\theta$ &            & 20.2     &     0.6        &     0.5    &  0.7 &   $^\circ$ & $\mathcal{N}(20, 3)$   \\ \hline \hline
\multicolumn{8}{c}{bar 1 component}      \\
\hline \hline 
$\rho_1$   & 0.316          &  0.682    &   0.069       &     0.062   &       & $10^{10} M_\odot \mathrm{kpc}^{-3}$ & (0, 10) \\
$x_1$ &  0.490   &  0.339     &   0.019         &      0.017        &  &  kpc & (0, 10)   \\
$y_1$ &  0.392  &   0.205    &    0.017         &      0.015        &  &  kpc & (0, 10)     \\
$z_1$ &  0.229  &    0.103   &     0.018        &       0.018       &  &  kpc & (0, 10)    \\
$c_\parallel$  & 1.991         &  2.686     &    0.574         &     0.574     &     &   & (0, 10)    \\
$c_\bot$    & 2.232       &  2.636     &     0.308        &        0.290    &   &  & (0, 10)    \\
$m$  & 0.873        &    0.742   &     0.054        &      0.050      &   &    & (0, 10) \\
$\alpha$  & 0.626        &   0.780    &     0.065        &    0.065    &     &  & (0, 10)  \\
$n$  & 1.940       &   126.914    &       52.005      &      42.649     &   &    & (0, 400)  \\
$c$ & 1.342          &    3.598   &      0.205       &       0.202    &   &   & (0, 10) \\
$x_c$  & 0.751       &   0.236    &       0.001      &      0.001   &     &   kpc  & (0, 10) \\
$y_c$ & 0.469       &   0.334    &    0.004         &      0.004    &    &   kpc & (0, 10)   \\
$r_\mathrm{cut}$  & 4.370         &  16.806     &    4.972        &       3.877   &      &  kpc & (0, 100)  \\ \hline \hline
\multicolumn{8}{c}{bar 2 component}      \\
\hline \hline 
$\rho_2$  &    0.050        &  0.055     &       0.020     &       0.016   &     & $10^{10} M_\odot \mathrm{kpc}^{-3}$ & (0, 10) \\
$x_2$ & 5.364   &    3.388   &     1.224        &    0.665      &     &   kpc & (0, 10) \\
$y_2$ & 0.959   &    0.463   &     0.063        &     0.065     &     &   kpc   & (0, 10)  \\
$z_2$ & 0.611  &   0.557    &       0.023      &      0.021   &      &   kpc  & (0, 10) \\
$n_2$ &  3.051         &   1.506    &       0.269     &        0.223  &     &  & (0, 10)    \\
$c_{\bot,2}$ & 0.970        &  4.045     &   1.406          &      1.016 &      &   & (0, 10)  \\
$R_{2,\mathrm{out}}$ & 3.190        & 3.422      &    0.359         &   0.318  &          & kpc & (0, 10)  \\
$R_{2,\mathrm{in}}$  & 0.558       &  0.608     &      0.034       &       0.030  &      &  kpc  & (0, 10) \\
$n_{2,\mathrm{out}}$ & 16.731        &  3.571     &       1.050      &     0.783   &       &   & ($-$100, 100)   \\
$n_{2,\mathrm{in}}$ & 3.196         &   9.954    &     5.422       &      3.203  &       &  & ($-$100, 100)  \\ \hline \hline
\multicolumn{8}{c}{bar 3 component}      \\
\hline \hline 
$\rho_3$ & 1743.049       &  1962.209     &   936.854          &     771.640    &    & $10^{10} M_\odot \mathrm{kpc}^{-3}$ & (500, 4000) \\
$x_3$ & 0.478   &   0.231    &    0.070         &       0.062    &    &   kpc  & (0, 10)\\
$y_3$ & 0.267   &    0.173   &     0.053        &      0.047    &     &   kpc   & (0, 10)  \\
$z_3$ & 0.252  &   0.181    &     0.014        &   0.014    &     &   kpc  & (0, 10) \\
$n_3$ &  0.980           &   0.733    &      0.071       &      0.067    &     &  & (0, 10)   \\
$c_{\bot,3}$ & 1.879        & 1.606     &       0.093      &       0.090  &      &   & (0, 10)   \\
$R_{3,\mathrm{out}}$ & 2.204       &   0.764    &   0.507          &      0.529 &        & kpc & (0, 10)  \\
$R_{3,\mathrm{in}}$ & 7.607        &  12.289    &   3.554         &    3.231   &      &  kpc & (0, 100) \\
$n_{3,\mathrm{out}}$ & $-$27.291       &   $-$98.157    &    65.511 &      68.029  &       &    & ($-$200, 200)  \\
$n_{3,\mathrm{in}}$ & 1.630         &   0.935    &       0.145      &  0.087   &      &  & (0, 10) \\ \hline \hline
\multicolumn{8}{c}{disk component}      \\
\hline \hline 
$\Sigma_0$ & 0.103            &   0.158    &     0.052        &       0.046   &     & $10^{10} M_\odot \mathrm{kpc}^{-2}$ & (0, 10) \\
$R_d$ & 4.754  &   4.015    &    0.270         &    0.291     &      &   kpc  & (0, 10) \\
$z_d$ & 0.151   &    0.121   &        0.046     &      0.038   &      &   kpc  & (0, 10)   \\
$R_{\mathrm{cut}}$ & 4.688  &  8.968    &     0.900        &    0.903     &      &   kpc & (0, 100)  \\
$n_d$ & 1.536            &   2.663    &    0.485         &       0.437  &      &     & (0, 10) \\
$m_d$  & 0.716      &    0.359  &     0.132       &        0.112  &     &     & (0, 10) \\ \hline \hline

\end{tabular}
\label{tab:final_solution}
\end{table*}

\subsection{Bias exploration with simulations} \label{subsec:bias_exploration}

The model that we fit to the data has 44 parameters. Such a large parameter space may be plagued by biases and degeneracies that in turn may affect the final parameters of the model. Therefore, we explored potential biases by assuming the model parameters and simulating mock distributions of Miras, fitting the model, and comparing the assumed parameters with the recovered values.

We generated two mock datasets, each with different model parameters. To best reproduce the true data and its uncertainties, we first counted how many Miras are expected to be observed in the examined cuboid (for more information, see Section \ref{subsec:cartesian_coordinates}). Then, we drew a number of stars from the Poisson distribution with an expected value equal to the expected number of Miras calculated a step earlier. 
For each star from this sample, we drew position, i.e., $(X', Y', Z')$ from the $\rho'(X', Y', Z')$ distribution (Equation \ref{eqn:modified_model}), and we transformed it to the Galactic coordinates and distances $(l, b, d)$. Since the best accuracy of the distance measurement in our sample of real Miras was about $4\%$, and at the same time, we have limited the sample to the distance accuracy not worse than $20\%$ (for more information, see Section \ref{subsec:sample_selection}), this is the range we have assigned to the mock sample.
Therefore, we calculated the distance uncertainties in the mock datasets by multiplying the distance by randomly selected distance accuracy in the range between $4\%$ and $20\%$. As the mock distance is accurate, we added noise by randomizing the new distance from the normal distribution with the mock distance taken as a mean and calculated distance uncertainty taken as a standard deviation.

We performed 10 fitting runs for each mock dataset (i.e., 20 fitting runs in total). The MCMC fitting procedure retrieved almost all original parameters within the $3\sigma$ regions, in at least 15 fits.
The only parameter, $\theta$, has not been reproduced in any fit. Therefore, we investigated the $\theta$ bias in detail.

We simulated and fitted other mock datasets with the same model parameters as previously, changing only the inclination of the bar, i.e., $\theta$. In this test, the bar was inclined at angles of $\theta_\mathrm{TRUE}$ equal to $15^\circ$, $20^\circ$, $25^\circ$, $30^\circ$, $35^\circ$, $40^\circ$, $45^\circ$, $50^\circ$, $55^\circ$, and the MCMC procedure returned values for $\theta_\mathrm{MCMC}$ equal to $8.6^\circ$, $12.1^\circ$, $16.1^\circ$, $19.1^\circ$, $24.2^\circ$, $27.6^\circ$, $32.3^\circ$, $36.2^\circ$, $39.7^\circ$, respectively. We noticed that there is a linear relation between $\theta_\mathrm{MCMC}$ and $\theta_\mathrm{TRUE}$:

\begin{equation}
    \theta_{\mathrm{MCMC}} = 0.792(\pm 0.012) \times \theta_{\mathrm{TRUE}} - 3.731(\pm 0.448).
\label{eqn:theta_bias}
\end{equation}

\noindent The scatter of this relation is equal to $0.4^\circ$. The scatter, as well as parameter uncertainties of the relation given by Equation \ref{eqn:theta_bias}, will be taken into account in the subsequent estimation of $\theta$.

Knowing that our procedure returned an underestimated $\theta$, we performed another test. We repeated the fitting of both mock datasets, however, we fixed $\theta$ at the assumed value, i.e., $20^\circ$. We again performed 10 fitting runs for each mock dataset, this time fitting a 43-parameter model (with fixed $\theta$). This test revealed that the posterior uncertainties of the model are underestimated. We compensated for this underestimation by multiplying the measured (statistics) uncertainties by factors derived from fits to simulated data by calculating an average ratio of the expected and measured parameters.

\newpage
\subsection{Final bar fit} \label{subsec:final_fit}

As $\theta$ is biased in our fitting procedure, we performed the final fit during two iterations. In the first iteration, we fitted the full 44-parameter model. This run returned the inclination of the bar $\theta_{\mathrm{MCMC}} = 12.3^{\circ +0.6^\circ} _ { \hspace{0.15cm} -0.5^\circ} \mathrm{(stat.)}$. We calculated $\theta_{\mathrm{TRUE}}$ using Equation \ref{eqn:theta_bias}. Therefore, the final slope of the bar is $\theta_{\mathrm{TRUE}} = \theta = 20.2^{\circ +0.6^\circ} _ { \hspace{0.15cm} -0.5^\circ} \mathrm{(stat.)}$.
In the second iteration, we fixed $\theta$, and we fitted the 43-parameter model. The final solution with statistical uncertainties for the Milky Way bar is presented in Table \ref{tab:final_solution}. The biases exploration showed that the posterior uncertainties of the model are underestimated.  The statistics uncertainties of the final fit reported in Table \ref{tab:final_solution} are already corrected as described in Section \ref{subsec:bias_exploration}.

\begin{figure*}
\centering
\includegraphics[scale=0.5]{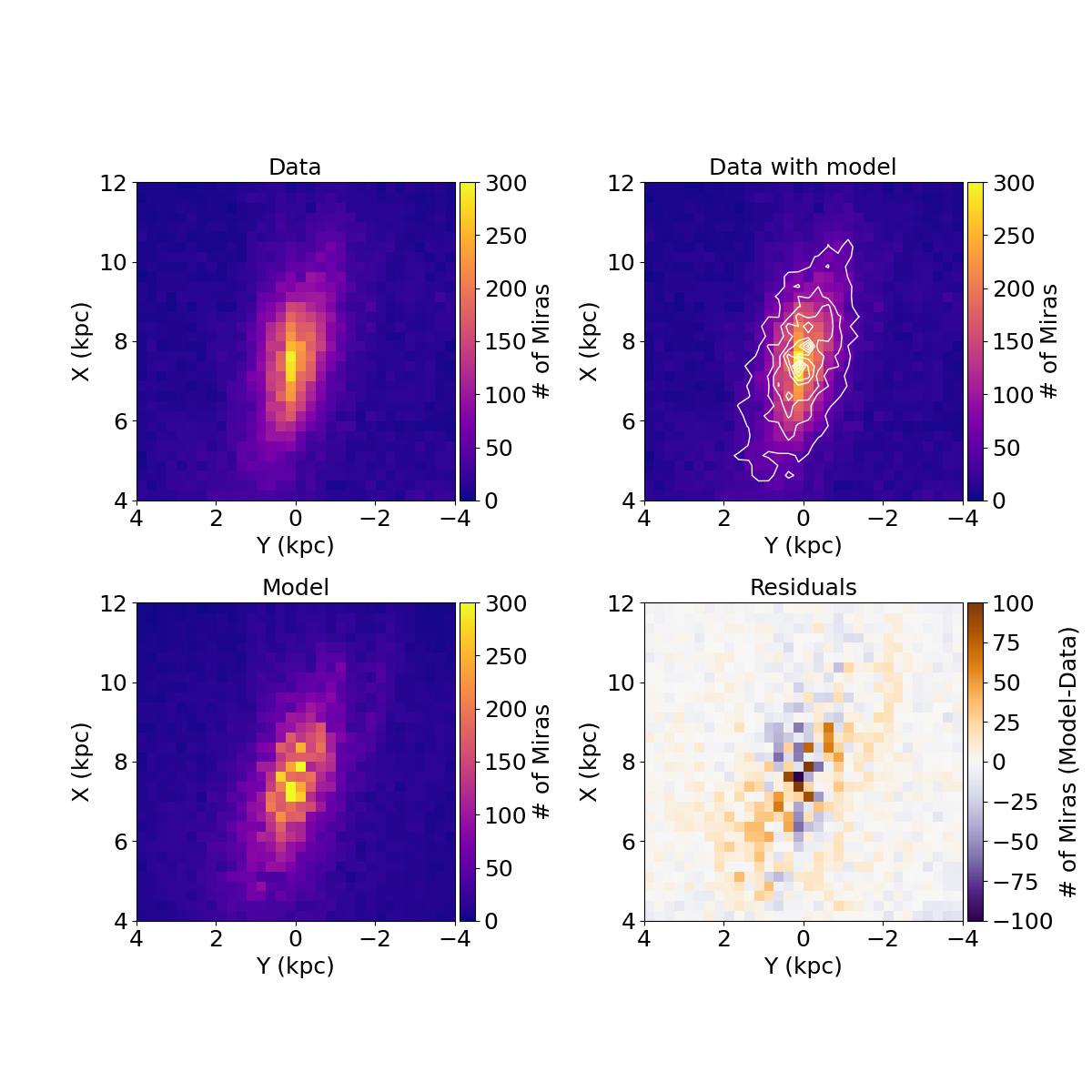}
\caption{The {\it Y--X} projection of the three-dimensional distribution of Miras and the fitted model. Miras are summed along the $Z$ coordinate in each bin.
{\it Top-left}: Miras used in the final fit. {\it Top-right:} The model of the bar (white contours) on top of the data. {Bottom-left:} The model of the bar. {\it Bottom-right:} the difference between the model and Mira distributions, i.e., residuals. The model shows the same number of stars as the number of analyzed Miras.}
\label{fig:data_and_model_YX}

\end{figure*}

\begin{figure*}
\centering
\includegraphics[scale=0.35]{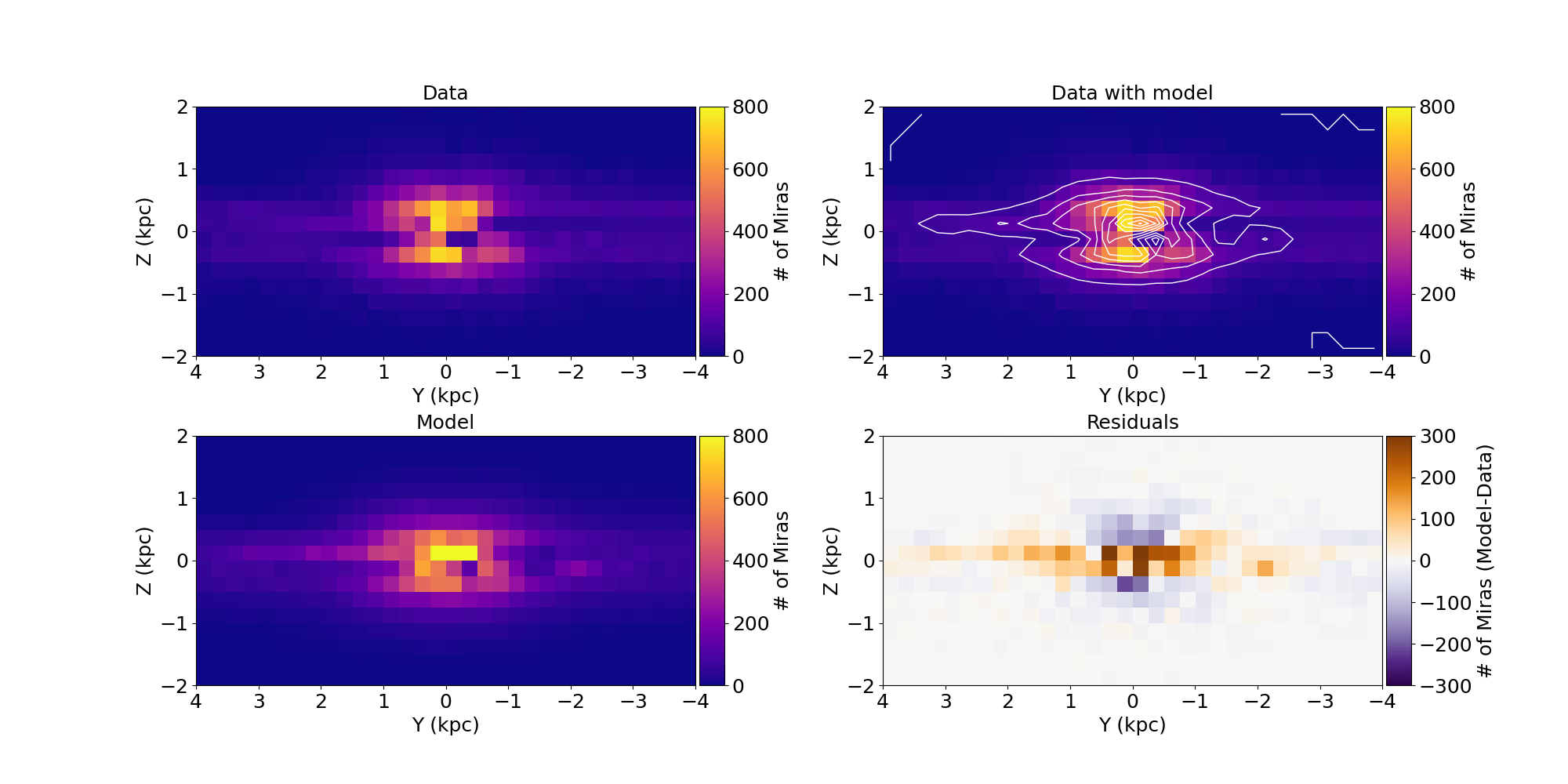}
\caption{The {\it Y--Z} projection of the three-dimensional distribution of Miras and the fitted model. Miras are summed along the $X$ coordinate in each bin.
{\it Top-left}: Miras used in the final fit. {\it Top-right:} The model of the bar (white contours) on top of the data. {Bottom-left:} The model of the bar. {\it Bottom-right:} the difference between the model and Mira distributions, i.e., residuals. The model shows the same number of stars as the number of analyzed Miras.}
\label{fig:data_and_model_YZ}
\end{figure*}

\begin{figure*}
\centering
\includegraphics[scale=0.35]{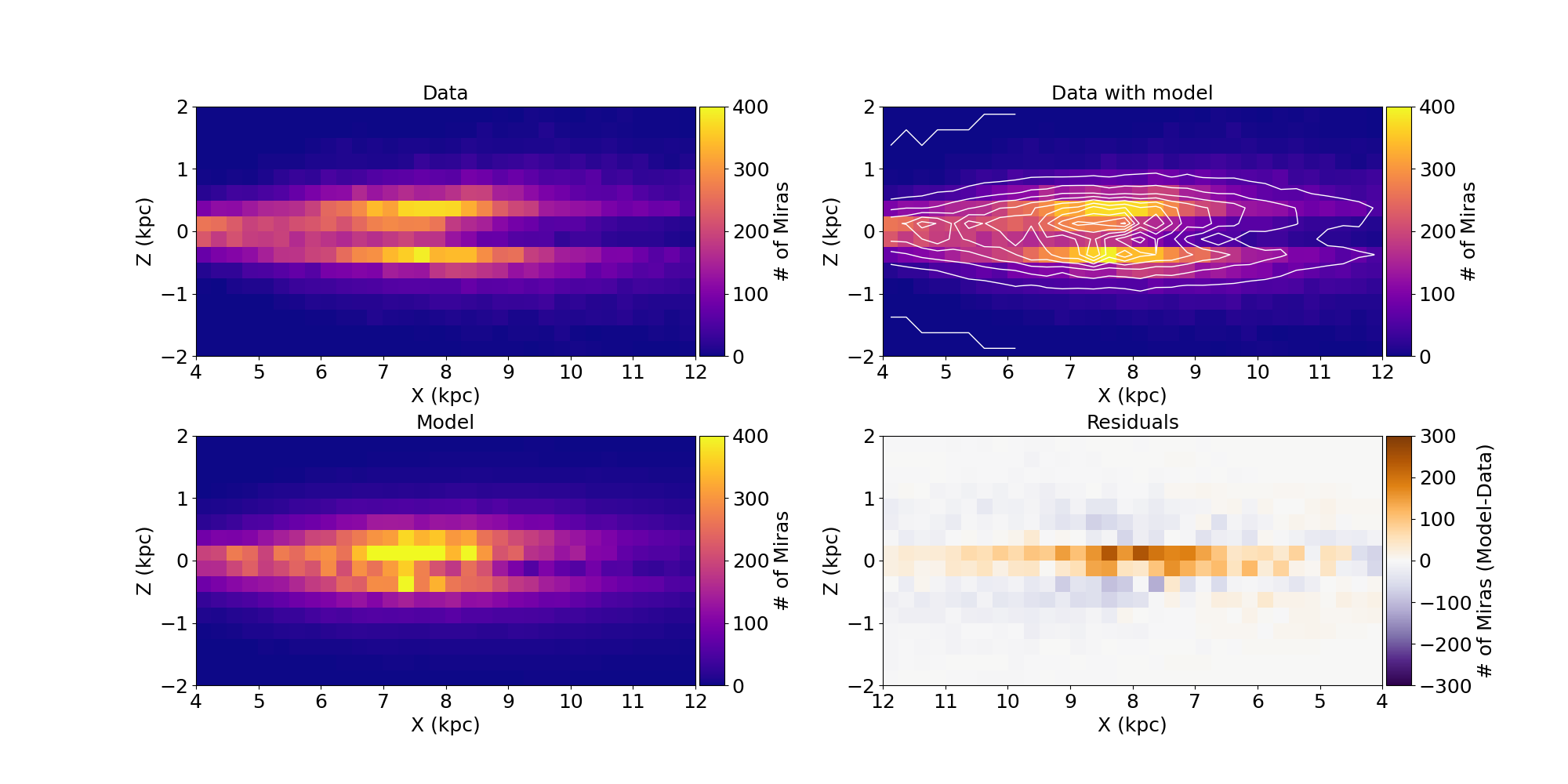}
\caption{The {\it X--Z}  projection of the three-dimensional distribution of Miras and the fitted model. Miras are summed along the $Y$ coordinate in each bin.
{\it Top-left}: Miras used in the final fit. {\it Top-right:} The model of the bar (white contours) on top of the data. {Bottom-left:} The model of the bar. {\it Bottom-right:} the difference between the model and Mira distributions, i.e., residuals. The model shows the same number of stars as the number of analyzed Miras.}
\label{fig:data_and_model_XZ}
\end{figure*}

\begin{figure*}[ht!]
\centering
\includegraphics[scale=0.7]{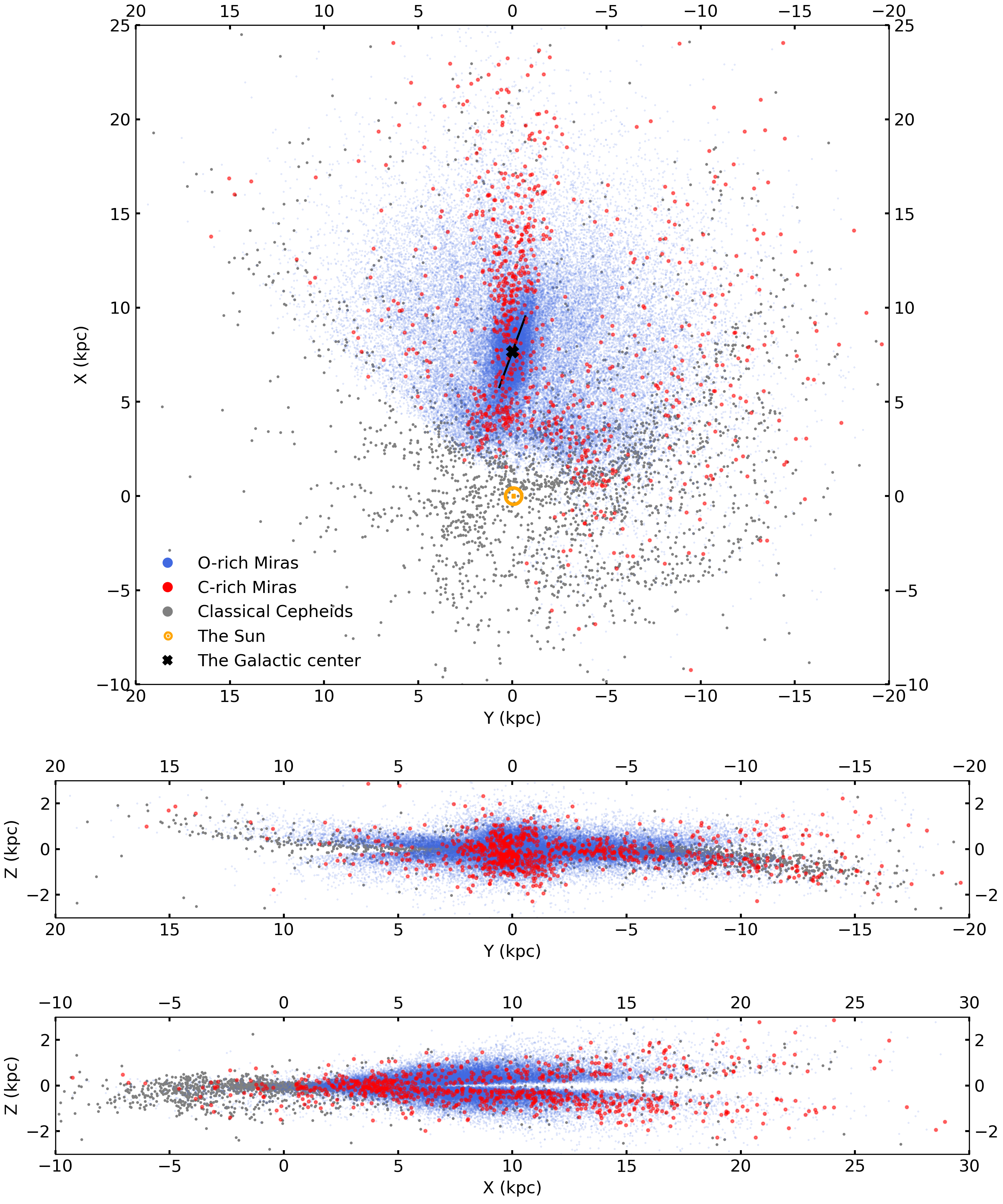}
\caption{Two-dimensional projections of the three-dimensional map of the Milky Way. With blue and red points, we show \mbox{O-rich}, and C-rich Miras respectively (whole sample with distance accuracy better than $20\%)$, while gray points represent Classical Cepheids analyzed by \citet{2019Sci...365..478S}. We marked the GC with a black cross ($X_{GC}$ = 7.66 kpc, \mbox{$Y_{GC}$ = $-$0.01 kpc}), while we marked the Solar System with an orange $\odot$ symbol. The black line along the bar shows the slope of the bar, i.e., $\theta$ angle ($\theta = 20.2^\circ$). {\it Top plot:} {\it Y--X} projection of the Milky Way. {\it Middle plot:} {\it Y--Z} projection of the Milky Way. {\it Bottom plot:} {\it X--Z} projection of the Milky Way.}
\label{fig:map}
\end{figure*}

In Figures \ref{fig:data_and_model_YX}, \ref{fig:data_and_model_YZ}, \ref{fig:data_and_model_XZ}, we present the data, model, and residuals in a two-dimensional projection of the analyzed cuboid. The corner plot, i.e., the two-dimensional projections of the multi-dimensional posterior parameter spaces fitted to the Galactic Mira distribution, is available through the OGLE website\footnote{\url{https://www.astrouw.edu.pl/ogle/ogle4/MILKY_WAY_3D_MAP/}} and Zenodo at \dataset[DOI: 10.5281/zenodo.7472598]{http://doi.org/10.5281/zenodo.7472598}. The corner plot was made using the \texttt{corner}\footnote{\url{https://corner.readthedocs.io/en/latest/}} Python library \citep{corner}. The two-dimensional projections of the three-dimensional map of the Milky Way are presented in Figure \ref{fig:map}.

\subsection{Systematic uncertainties} \label{subsec:systematic_errors}

The $\theta$ systematic uncertainty comes from the linear underestimation of $\theta$ described in Section \ref{subsec:bias_exploration}. During the estimation of the $\theta$ systematic uncertainty, we took into account uncertainties and scatter of Equation~\ref{eqn:theta_bias}. From the uncertainty propagation, when calculating $\theta_{\mathrm{TRUE}}$, we obtained the uncertainty of $0.6^\circ$, while the scatter of the relation is equal to $0.4^\circ$. Combining these two uncertainties in the quadrature gives the final $\theta$ systematic uncertainty, which is equal to $0.7^\circ$. This uncertainty is indicated in Table \ref{tab:final_solution} as $\sigma_{\mathrm{sys}}$.

The fitted model parameters vitally depend on the knowledge of the extinction, but also the zero-point of PLRs. First, the result could be affected by poorly measured extinction, which is highly variable toward the BLG. Second, the zero-point of the PLRs in the Milky Way may differ from the zero-point of the LMC PLRs, that we used. The zero-point may be affected by e.g., the higher metallicity of the Milky Way. Both of these factors may have a significant impact on the measured distances to the Mira variables, and thus on the final Mira distribution in the Milky Way.

We analyzed both above-mentioned effects. The first, extinction, was verified by measuring distances using different extinction laws. Using the same procedures as described in Section \ref{subsec:extinction} and Section \ref{sec:distances}, we measured the extinction in the {\it V}-band, then we transformed $A_V$ to the mid-IR extinctions $A_\lambda$ using the extinction laws derived by \citet{2018ApJ...859..137C} and \citet{2019ApJ...877..116W}: $A_{W1}/A_{V} = 0.039$, $A_{W2}/A_{V} = 0.026$, $A_{W3}/A_{V} = 0.040$, $A_{I1}/A_{V} = 0.037$, $A_{I2}/A_{V} = 0.026$, $A_{I3}/A_{V} = 0.019$, $A_{I4}/A_{V} = 0.025$, and we measured again the distance to each Mira star. As the different extinction law can cause a systematic shift of the whole Mira distribution in space, we estimated systematic uncertainties of $X_\mathrm{GC}$, $Y_\mathrm{GC}$, and $Z_\mathrm{GC}$. We calculated the mean distance difference between datasets with different extinction laws and found that on average the distances are shifted by about $4.5\%$. As the shift in distances can also affect two other Cartesian coordinates, i.e., $Y$, and $Z$, we transformed both sets of coordinates $(l,b,d)$ to $(X, Y, Z)$, and calculated the mean difference in each coordinate. Finally, we estimated the systematic uncertainties of $X_\mathrm{GC}$, $Y_\mathrm{GC}$, and $Z_\mathrm{GC}$ to be $0.38$ kpc, $0.01$ kpc, and $0.01$ kpc, respectively.

Another systematic shift in the distances may be caused by the zero-point of PLRs. To analyze this effect, we identified 41 OGLE-IV fields toward the BLG with the highest number of Miras and well-defined PLRs. We removed C-rich Miras as well as Miras with periods shorter than 100 days and longer than 400 days from the analysis. For each field, we calculated the PLR with a linear fit \citep[the same method as presented in][]{2021ApJ...919...99I}, calibrated at the distance to the Milky Way center $R_0$~=~8.178~kpc \citep[][]{2019A&A...625L..10G}, in each analyzed mid-IR band. Then we calculated the mean difference between the LMC and the Milky Way absolute zero-points in each band \citep[zero-points for LMC PLRs were taken from][]{2021ApJ...919...99I}. The mean zero-point offset between the LMC and Milky Way's PLRs is equal to $0.022$~mag, which translates into approximately $1\%$ in the distance offset. This means that the distance to each Mira star could be shifted by approximately $1\%$, depending on the line-of-sight. However, this uncertainty is much smaller than the typical Mira distance uncertainty, and adding them in the quadrature should not significantly affect the distance uncertainties reported in Table \ref{tab:all_information}. This zero-point offset could cause a shift of the entire Mira distribution by approximately $1\%$ toward larger distances. Therefore, we took this into account in the systematic uncertainty of the distance to the GC, i.e., $X_{\mathrm{GC}}$. The zero-point offset equal to $0.022$ mag means that \mbox{$X_{GC} = 7.66$ kpc} could be shifted by $0.08$ kpc. We added this systematic error in quadrature 
to the $X_{\mathrm{GC}}$ systematic error resulting from the change in the extinction law (which is equal to $0.38$ kpc). The difference in zero-point has no significant impact on $Y$ and $Z$ coordinates; therefore, we did not account for this in the systematic uncertainties of $Y_\mathrm{GC}$ and $Z_\mathrm{GC}$. The final systematic uncertainties are presented in Table \ref{tab:final_solution}.

\section{Discussion} \label{sec:discussion}

\subsection{Distance to the GC and the inclination of the bar}

We measured the distance to the GC $X_\mathrm{GC} = R_0 = 7.66 \pm 0.01 \mathrm{(stat.)} \pm 0.39 \mathrm{(sys.)}$ kpc. This result is consistent with other Milky Way studies based on various other indicators. For example, \citet{2005ApJ...628..246E} derived $R_0 = 7.62 \pm 0.32$ kpc based on orbits of young stars, while \citet{2008ApJ...689.1044G}, using the same method, but a different dataset, obtained $R_0 = 8.0 \pm 0.6$ kpc. \citet{2008A&A...492..419T} using kinematics of old stellar population got the value of $R_0 = 8.07 \pm 0.32 \mathrm{(stat.)} \pm 0.13 \mathrm{(sys.)} $ kpc, \citet{2006ApJ...647.1093N} using red clump stars found $R_0 = 7.52 \pm 0.10 \mathrm{(stat.)} \pm 0.35 \mathrm{(sys.)}$ kpc, while \citet{2014MNRAS.441.1105F} using globular clusters derived $R_0 = 7.4 \pm 0.2 \mathrm{(stat.)} \pm 0.2 \mathrm{(sys.)}$ kpc. The more recent analysis based on RR Lyrae stars \citep{2015ApJ...811..113P} gave $R_0 = 8.27 \pm 0.01 \mathrm{(stat.)} \pm 0.40 \mathrm{(sys.)}$ kpc. \citet{2009MNRAS.399.1709M} based on 100 Miras estimated the distance to GC to $R_0 = 8.24 \pm 0.08 \mathrm{(stat.)} \pm 0.42 \mathrm{(sys.)}$ kpc. This result slightly differs from ours, however, both results are consistent with the uncertainty limit. The difference may be caused by a few orders of magnitude smaller sample of Miras used by \citet{2009MNRAS.399.1709M}, less accurate PLRs based only on Spitzer data, a different method of extinction measurement, etc. One of the most accurate measurements of the distance to GC was made by the GRAVITY Collaboration and was based on astrometric and spectroscopic observations of star S2 orbiting the central black hole Sgr A*. This analysis provided \mbox{$R_0 = 8.178 \pm 0.013 \mathrm{(stat.)} \pm 0.022 \mathrm{(sys.)}$ kpc} \citep{2019A&A...625L..10G}. Comparing this result with ours, one may notice that our systematic uncertainty is one order of magnitude greater than the systematic uncertainty estimated by GRAVITY Collaboration. Since the GRAVITY Collaboration measurement was direct, the possible systematic uncertainty was dominated by instrumental effects, mostly in the measurement of the astrometry. This measurement was free of possible uncertainties in extinction measurement, PLRs calibration, differences in PLR zero-points between Milky Way and LMC, etc., which could potentially affect our result.

We obtained the inclination of the Galactic bar to the Sun-GC line-of-sight equal to \mbox{$\theta = 20.2^\circ \pm 0.6^\circ \mathrm{(stat.)} \pm 0.7^\circ \mathrm{(sys.)}$}. Our result is consistent with other previous studies that used e.g., red clump giants \citep{1997ApJ...477..163S} or RR Lyrae stars \citep{2015ApJ...811..113P}. The first analysis gave the range of angles between $20^\circ$ and $30^\circ$, while the second analysis derived inclination $\theta = 20^\circ \pm 3^\circ$. The other studies used a deep near-infrared wide-angle photometric analysis and measured the inclination of the bar to be $\theta = 22^\circ \pm 5.5^\circ$ \citep{2005MNRAS.358.1309B}, or red clump stars with a value of $\theta = 27^\circ \pm 2^\circ$ \citep{2013MNRAS.435.1874W}. The previous analysis of Mira stars distribution indicated, that the bar is inclined with the angle $\sim 21 ^\circ$ \citep{2020MNRAS.492.3128G}.

\subsection{Disk flaring and warping}

Studies of the Milky Way structure with different tracers showed that the disk is warped \citep[e.g.,][]{1998Natur.392..471S,2001ApJ...556..181D, 2006A&A...451..515M, 2006Sci...312.1773L, 2017A&A...602A..67A}. A recent analysis of the three-dimensional distribution of the classical Cepheids \citep{2019Sci...365..478S} revealed that the disk warp seen in these stars differs from previously proposed models. The authors concluded that the shape of the warp may differ between the young and older stellar populations.

When looking at the \mbox{Y-Z} projection of our Galaxy (middle plot of Figure \ref{fig:map}), the disk flaring is visible. The Mira disk population plausibly follows the warp seen in the young stellar population. Unfortunately, we are unable to confirm this thoroughly because our Mira sample is incomplete at the positive values of $Y$ and negative values of $Z$ -- this area is not visible from the southern hemisphere. The map presented in Figure \ref{fig:map} is the first such a detailed and accurate three-dimensional map of the Milky Way, showing both young- and intermediate-age stellar populations.

\subsection{The X-shaped structure}

Studies of the Milky Way bulge do not unequivocally prove the existence of an X-shape -- there exists evidence for and against the X-shaped bulge. The most promising evidence for the existence of the X-shaped structure is the split in the magnitude distribution of the red clump stars \citep{2010ApJ...721L..28N, 2010ApJ...724.1491M, 2013MNRAS.435.1874W, 2019MNRAS.489.3519C}. However, if this split was caused by an X-shaped bar, this would be evident in the Galaxy's three-dimensional structure. One group of stars should be closer to and the other farther away from the Sun along the same line of sight. \citet{2017ApJ...836..218L} analyzed the three-dimensional Milky Way structure based on the OGLE-III Mira sample \citep{2013AcA....63...21S} and concluded that there is no X-shaped structure. This analysis has several shortcomings. The authors used a small number of Miras compared to this analysis ($<10$\% of our sample) and analyzed a small range of Galactic latitudes not exceeding $|l| = 8 ^\circ$ \citep{2022MNRAS.517.6060S}. They also used a  simplified extinction and transformation of the {\it I}-band to the {\it K}-band mean magnitude, and did not account for distance uncertainties, which are significant as we showed in this paper. A similar analysis has been carried out by \citet{2022A&A...666L..13C} based on the recently published Mira sample in \citep{2022ApJS..260...46I}. This paper \citep{2022A&A...666L..13C} ignores distance uncertainties and uses simplified extinction and transformation of the {\it I}-band to the {\it K}-band mean magnitude. The authors came to a similar conclusion as \citet{2017ApJ...836..218L}, indicating that the boxy-bulge model is more favorable.

\citet{2020MNRAS.492.3128G} divided the Milky Way Miras into groups with different ages. The authors conclude that the distinction between the oldest and youngest Miras in the inner Galaxy is clearly visible. The old/metal-poor Miras dominate centrally, while the young/metal-rich Miras show the boxy/peanut like morphology with a characteristic X-shape.

Our results shed new light on the problem regarding the existence of the X-shape structure. It seems that the model proposed by \citet{2022MNRAS.514L...1S} describes quite well the three-dimensional distribution of Miras in the BLG, as shown in Figures \ref{fig:data_and_model_YX}, \ref{fig:data_and_model_YZ}, and \ref{fig:data_and_model_XZ}. The model parameters (Table~\ref{tab:final_solution}) that describe the X-shape, i.e., $\alpha$ -- the strength of the \mbox{X-shape}, and $c$ -- the slope of the \mbox{X-shape} in the $(X, Z)$ plane are statistically different from zero ($\alpha=0.780\pm0.065$ and $c = 3.598 \pm 0.204$) and are not biased (see Section \ref{subsec:bias_exploration}). Therefore, we can conclude that the X-shaped structure exists in the distribution represented by Mira variables.

\section{Conclusions} \label{sec:conclusions}

In this paper, we used the collection of 65,981 Miras \citep{2022ApJS..260...46I} to construct the three-dimensional map of the Milky Way and to study the structure of the BLG.
We crossmatched our Mira sample with the mid-IR data collected by the WISE \citep{2010AJ....140.1868W, 2011ApJ...731...53M, 2014ApJ...792...30M} and Spitzer \citep{2004ApJS..154....1W} space telescopes. We derived mean luminosities in up to seven mid-IR bands and used the mid-IR PLRs \citep{2021ApJ...919...99I} to determine distances to 65,385 Mira stars with a median accuracy of $6.6\%$. We fitted a 44-parameter model \citep{2022MNRAS.514L...1S} of the bulge to the Mira distribution taking into account the distance uncertainties by implementing Bayesian hierarchical inference. We established the distance to the GC equal to $R_0 = 7.66 \pm 0.01 \mathrm{(stat.)} \pm 0.39 \mathrm{(sys.)}$ kpc, while the inclination of the bulge to the Sun-GC line of sight equals $\theta = 20.2^\circ \pm 0.6^\circ \mathrm{(stat.)} \pm 0.7^\circ \mathrm{(sys.)}$. These results are consistent with previous results obtained based on different stellar tracers. The model we used contains parameters responsible for the X-shaped structure. Our result gives independent clues in favor of an X-shaped bulge, as the strength of the X-shape and its inclination are statistically different from zero ($\alpha=0.780\pm0.065$ and $c = 3.598 \pm 0.204$) and not biased. Further studies using different tracers or methods could reveal more accurately the nature of the X-shape.

\begin{acknowledgments}

We would like to thank the anonymous referee for suggestions that helped to improve this manuscript. PI is partially supported by Kartezjusz program No. POWR.03.02.00-00-I001/16-00, founded by the National Centre for Research and Development, Poland. SK acknowledges the support from the National Science Centre, Poland, via grant OPUS 2018/31/B/ST9/00334. 

This publication makes use of data products from WISE, which is a joint project of the University of California, Los Angeles, and the Jet Propulsion Laboratory/California Institute of Technology, funded by the National Aeronautics and Space Administration (NASA). This work is based in part on archival data obtained with the Spitzer Space Telescope, which was operated by the Jet Propulsion Laboratory, California Institute of Technology under a contract with NASA.

\end{acknowledgments}

\bibliography{paper}{}

\begin{thebibliography}{}
\expandafter\ifx\csname natexlab\endcsname\relax\def\natexlab#1{#1}\fi
\providecommand{\url}[1]{\href{#1}{#1}}
\providecommand{\dodoi}[1]{doi:~\href{http://doi.org/#1}{\nolinkurl{#1}}}
\providecommand{\doeprint}[1]{\href{http://ascl.net/#1}{\nolinkurl{http://ascl.net/#1}}}
\providecommand{\doarXiv}[1]{\href{https://arxiv.org/abs/#1}{\nolinkurl{https://arxiv.org/abs/#1}}}

\bibitem[{{Am{\^o}res} {et~al.}(2017){Am{\^o}res}, {Robin}, \&
  {Reyl{\'e}}}]{2017A&A...602A..67A}
{Am{\^o}res}, E.~B., {Robin}, A.~C., \& {Reyl{\'e}}, C. 2017, \aap, 602, A67,
  \dodoi{10.1051/0004-6361/201628461}

\bibitem[{{Babusiaux} \& {Gilmore}(2005)}]{2005MNRAS.358.1309B}
{Babusiaux}, C., \& {Gilmore}, G. 2005, \mnras, 358, 1309,
  \dodoi{10.1111/j.1365-2966.2005.08828.x}

\bibitem[{{Bellm} {et~al.}(2019){Bellm}, {Kulkarni}, {Graham}, {Dekany},
  {Smith}, {Riddle}, {Masci}, {Helou}, {Prince}, {Adams}, {Barbarino},
  {Barlow}, {Bauer}, {Beck}, {Belicki}, {Biswas}, {Blagorodnova}, {Bodewits},
  {Bolin}, {Brinnel}, {Brooke}, {Bue}, {Bulla}, {Burruss}, {Cenko}, {Chang},
  {Connolly}, {Coughlin}, {Cromer}, {Cunningham}, {De}, {Delacroix}, {Desai},
  {Duev}, {Eadie}, {Farnham}, {Feeney}, {Feindt}, {Flynn}, {Franckowiak},
  {Frederick}, {Fremling}, {Gal-Yam}, {Gezari}, {Giomi}, {Goldstein},
  {Golkhou}, {Goobar}, {Groom}, {Hacopians}, {Hale}, {Henning}, {Ho}, {Hover},
  {Howell}, {Hung}, {Huppenkothen}, {Imel}, {Ip}, {Ivezi{\'c}}, {Jackson},
  {Jones}, {Juric}, {Kasliwal}, {Kaspi}, {Kaye}, {Kelley}, {Kowalski},
  {Kramer}, {Kupfer}, {Landry}, {Laher}, {Lee}, {Lin}, {Lin}, {Lunnan},
  {Giomi}, {Mahabal}, {Mao}, {Miller}, {Monkewitz}, {Murphy}, {Ngeow},
  {Nordin}, {Nugent}, {Ofek}, {Patterson}, {Penprase}, {Porter}, {Rauch},
  {Rebbapragada}, {Reiley}, {Rigault}, {Rodriguez}, {van Roestel}, {Rusholme},
  {van Santen}, {Schulze}, {Shupe}, {Singer}, {Soumagnac}, {Stein}, {Surace},
  {Sollerman}, {Szkody}, {Taddia}, {Terek}, {Van Sistine}, {van Velzen},
  {Vestrand}, {Walters}, {Ward}, {Ye}, {Yu}, {Yan}, \&
  {Zolkower}}]{2019PASP..131a8002B}
{Bellm}, E.~C., {Kulkarni}, S.~R., {Graham}, M.~J., {et~al.} 2019, \pasp, 131,
  018002, \dodoi{10.1088/1538-3873/aaecbe}

\bibitem[{{Benjamin} {et~al.}(2003){Benjamin}, {Churchwell}, {Babler}, {Bania},
  {Clemens}, {Cohen}, {Dickey}, {Indebetouw}, {Jackson}, {Kobulnicky},
  {Lazarian}, {Marston}, {Mathis}, {Meade}, {Seager}, {Stolovy}, {Watson},
  {Whitney}, {Wolff}, \& {Wolfire}}]{2003PASP..115..953B}
{Benjamin}, R.~A., {Churchwell}, E., {Babler}, B.~L., {et~al.} 2003, \pasp,
  115, 953, \dodoi{10.1086/376696}

\bibitem[{{Bhardwaj} {et~al.}(2019){Bhardwaj}, {Kanbur}, {He}, {Rejkuba},
  {Matsunaga}, {de Grijs}, {Sharma}, {Singh}, {Baug}, {Ngeow}, \&
  {Ou}}]{2019ApJ...884...20B}
{Bhardwaj}, A., {Kanbur}, S., {He}, S., {et~al.} 2019, \apj, 884, 20,
  \dodoi{10.3847/1538-4357/ab38c2}

\bibitem[{{Bovy} {et~al.}(2016){Bovy}, {Rix}, {Green}, {Schlafly}, \&
  {Finkbeiner}}]{2016ApJ...818..130B}
{Bovy}, J., {Rix}, H.-W., {Green}, G.~M., {Schlafly}, E.~F., \& {Finkbeiner},
  D.~P. 2016, \apj, 818, 130, \dodoi{10.3847/0004-637X/818/2/130}

\bibitem[{{Burns} {et~al.}(2014){Burns}, {Nagayama}, {Handa}, {Omodaka},
  {Nakagawa}, {Nakanishi}, {Hayashi}, \& {Shizugami}}]{2014ApJ...797...39B}
{Burns}, R.~A., {Nagayama}, T., {Handa}, T., {et~al.} 2014, \apj, 797, 39,
  \dodoi{10.1088/0004-637X/797/1/39}

\bibitem[{{Burton}(1988)}]{1988gera.book..295B}
{Burton}, W.~B. 1988, in Galactic and Extragalactic Radio Astronomy, ed. K.~I.
  {Kellermann} \& G.~L. {Verschuur}, 295--358

\bibitem[{{Cabrera-Lavers} {et~al.}(2008){Cabrera-Lavers},
  {Gonz{\'a}lez-Fern{\'a}ndez}, {Garz{\'o}n}, {Hammersley}, \&
  {L{\'o}pez-Corredoira}}]{2008A&A...491..781C}
{Cabrera-Lavers}, A., {Gonz{\'a}lez-Fern{\'a}ndez}, C., {Garz{\'o}n}, F.,
  {Hammersley}, P.~L., \& {L{\'o}pez-Corredoira}, M. 2008, \aap, 491, 781,
  \dodoi{10.1051/0004-6361:200810720}

\bibitem[{{Calbet} {et~al.}(1995){Calbet}, {Mahoney}, {Garzon}, \&
  {Hammersley}}]{1995MNRAS.276..301C}
{Calbet}, X., {Mahoney}, T., {Garzon}, F., \& {Hammersley}, P.~L. 1995, \mnras,
  276, 301, \dodoi{10.1093/mnras/276.1.301}

\bibitem[{{Carey} {et~al.}(2008){Carey}, {Ali}, {Berriman}, {Boulanger},
  {Brunt}, {Cutri}, {Flagey}, {Gibson}, {Heyer}, {Hora}, {Indebetouw},
  {Kraemer}, {Kuchar}, {Latter}, {Marleau}, {Miville-Deschenes}, {Mizuno},
  {Molinari}, {Noriega-Crespo}, {Padgett}, {Paladini}, {Price}, {Rebull},
  {Rottler}, {Shenoy}, {Shipman}, \& {Testi}}]{2008sptz.prop50398C}
{Carey}, S., {Ali}, B., {Berriman}, B., {et~al.} 2008, {Spitzer Mapping of the
  Outer Galaxy (SMOG)}, Spitzer Proposal

\bibitem[{{Carollo} {et~al.}(2007){Carollo}, {Beers}, {Lee}, {Chiba}, {Norris},
  {Wilhelm}, {Sivarani}, {Marsteller}, {Munn}, {Bailer-Jones}, {Fiorentin}, \&
  {York}}]{2007Natur.450.1020C}
{Carollo}, D., {Beers}, T.~C., {Lee}, Y.~S., {et~al.} 2007, \nat, 450, 1020,
  \dodoi{10.1038/nature06460}

\bibitem[{{Catchpole} {et~al.}(2016){Catchpole}, {Whitelock}, {Feast},
  {Hughes}, {Irwin}, \& {Alard}}]{2016MNRAS.455.2216C}
{Catchpole}, R.~M., {Whitelock}, P.~A., {Feast}, M.~W., {et~al.} 2016, \mnras,
  455, 2216, \dodoi{10.1093/mnras/stv2372}

\bibitem[{{Chen} {et~al.}(2018){Chen}, {Wang}, {Deng}, \& {de
  Grijs}}]{2018ApJ...859..137C}
{Chen}, X., {Wang}, S., {Deng}, L., \& {de Grijs}, R. 2018, \apj, 859, 137,
  \dodoi{10.3847/1538-4357/aabfbc}

\bibitem[{{Chen} {et~al.}(2019){Chen}, {Wang}, {Deng}, {de Grijs}, {Liu}, \&
  {Tian}}]{2019NatAs...3..320C}
{Chen}, X., {Wang}, S., {Deng}, L., {et~al.} 2019, Nature Astronomy, 3, 320,
  \dodoi{10.1038/s41550-018-0686-7}

\bibitem[{{Chrob{\'a}kov{\'a}} {et~al.}(2022){Chrob{\'a}kov{\'a}},
  {L{\'o}pez-Corredoira}, \& {Garz{\'o}n}}]{2022A&A...666L..13C}
{Chrob{\'a}kov{\'a}}, {\v{Z}}., {L{\'o}pez-Corredoira}, M., \& {Garz{\'o}n}, F.
  2022, \aap, 666, L13, \dodoi{10.1051/0004-6361/202244810}

\bibitem[{{Churchwell} {et~al.}(2009){Churchwell}, {Babler}, {Meade},
  {Whitney}, {Benjamin}, {Indebetouw}, {Cyganowski}, {Robitaille}, {Povich},
  {Watson}, \& {Bracker}}]{2009PASP..121..213C}
{Churchwell}, E., {Babler}, B.~L., {Meade}, M.~R., {et~al.} 2009, \pasp, 121,
  213, \dodoi{10.1086/597811}

\bibitem[{{Ciambur} {et~al.}(2017){Ciambur}, {Graham}, \&
  {Bland-Hawthorn}}]{2017MNRAS.471.3988C}
{Ciambur}, B.~C., {Graham}, A.~W., \& {Bland-Hawthorn}, J. 2017, \mnras, 471,
  3988, \dodoi{10.1093/mnras/stx1823}

\bibitem[{{Clarke} {et~al.}(2019){Clarke}, {Wegg}, {Gerhard}, {Smith}, {Lucas},
  \& {Wylie}}]{2019MNRAS.489.3519C}
{Clarke}, J.~P., {Wegg}, C., {Gerhard}, O., {et~al.} 2019, \mnras, 489, 3519,
  \dodoi{10.1093/mnras/stz2382}

\bibitem[{{Dame} {et~al.}(2001){Dame}, {Hartmann}, \&
  {Thaddeus}}]{2001ApJ...547..792D}
{Dame}, T.~M., {Hartmann}, D., \& {Thaddeus}, P. 2001, \apj, 547, 792,
  \dodoi{10.1086/318388}

\bibitem[{{Delgado} {et~al.}(2019){Delgado}, {Sarro}, {Clementini}, {Muraveva},
  \& {Garofalo}}]{2019A&A...623A.156D}
{Delgado}, H.~E., {Sarro}, L.~M., {Clementini}, G., {Muraveva}, T., \&
  {Garofalo}, A. 2019, \aap, 623, A156, \dodoi{10.1051/0004-6361/201832945}

\bibitem[{{Drimmel} {et~al.}(2003){Drimmel}, {Cabrera-Lavers}, \&
  {L{\'o}pez-Corredoira}}]{2003A&A...409..205D}
{Drimmel}, R., {Cabrera-Lavers}, A., \& {L{\'o}pez-Corredoira}, M. 2003, \aap,
  409, 205, \dodoi{10.1051/0004-6361:20031070}

\bibitem[{{Drimmel} \& {Spergel}(2001)}]{2001ApJ...556..181D}
{Drimmel}, R., \& {Spergel}, D.~N. 2001, \apj, 556, 181, \dodoi{10.1086/321556}

\bibitem[{{Eisenhauer} {et~al.}(2005){Eisenhauer}, {Genzel}, {Alexander},
  {Abuter}, {Paumard}, {Ott}, {Gilbert}, {Gillessen}, {Horrobin}, {Trippe},
  {Bonnet}, {Dumas}, {Hubin}, {Kaufer}, {Kissler-Patig}, {Monnet},
  {Str{\"o}bele}, {Szeifert}, {Eckart}, {Sch{\"o}del}, \&
  {Zucker}}]{2005ApJ...628..246E}
{Eisenhauer}, F., {Genzel}, R., {Alexander}, T., {et~al.} 2005, \apj, 628, 246,
  \dodoi{10.1086/430667}

\bibitem[{{Feast} {et~al.}(1989){Feast}, {Glass}, {Whitelock}, \&
  {Catchpole}}]{1989MNRAS.241..375F}
{Feast}, M.~W., {Glass}, I.~S., {Whitelock}, P.~A., \& {Catchpole}, R.~M. 1989,
  \mnras, 241, 375, \dodoi{10.1093/mnras/241.3.375}

\bibitem[{{Feast} {et~al.}(1980){Feast}, {Robertson}, \&
  {Black}}]{1980MNRAS.190..227F}
{Feast}, M.~W., {Robertson}, B.~S.~C., \& {Black}, C. 1980, \mnras, 190, 227,
  \dodoi{10.1093/mnras/190.2.227}

\bibitem[{Foreman-Mackey(2016)}]{corner}
Foreman-Mackey, D. 2016, The Journal of Open Source Software, 1, 24,
  \dodoi{10.21105/joss.00024}

\bibitem[{{Foreman-Mackey} {et~al.}(2013){Foreman-Mackey}, {Hogg}, {Lang}, \&
  {Goodman}}]{2013PASP..125..306F}
{Foreman-Mackey}, D., {Hogg}, D.~W., {Lang}, D., \& {Goodman}, J. 2013, \pasp,
  125, 306, \dodoi{10.1086/670067}

\bibitem[{{Foreman-Mackey} {et~al.}(2014){Foreman-Mackey}, {Hogg}, \&
  {Morton}}]{2014ApJ...795...64F}
{Foreman-Mackey}, D., {Hogg}, D.~W., \& {Morton}, T.~D. 2014, \apj, 795, 64,
  \dodoi{10.1088/0004-637X/795/1/64}

\bibitem[{{Francis} \& {Anderson}(2014)}]{2014MNRAS.441.1105F}
{Francis}, C., \& {Anderson}, E. 2014, \mnras, 441, 1105,
  \dodoi{10.1093/mnras/stu631}

\bibitem[{{Gaia Collaboration} {et~al.}(2016){Gaia Collaboration}, {Prusti},
  {de Bruijne}, {Brown}, {Vallenari}, {Babusiaux}, {Bailer-Jones}, {Bastian},
  {Biermann}, {Evans}, {Eyer}, {Jansen}, {Jordi}, {Klioner}, {Lammers},
  {Lindegren}, {Luri}, {Mignard}, {Milligan}, {Panem}, {Poinsignon},
  {Pourbaix}, {Randich}, {Sarri}, {Sartoretti}, {Siddiqui}, {Soubiran},
  {Valette}, {van Leeuwen}, {Walton}, {Aerts}, {Arenou}, {Cropper}, {Drimmel},
  {H{\o}g}, {Katz}, {Lattanzi}, {O'Mullane}, {Grebel}, {Holland}, {Huc},
  {Passot}, {Bramante}, {Cacciari}, {Casta{\~n}eda}, {Chaoul}, {Cheek}, {De
  Angeli}, {Fabricius}, {Guerra}, {Hern{\'a}ndez}, {Jean-Antoine-Piccolo},
  {Masana}, {Messineo}, {Mowlavi}, {Nienartowicz}, {Ord{\'o}{\~n}ez-Blanco},
  {Panuzzo}, {Portell}, {Richards}, {Riello}, {Seabroke}, {Tanga},
  {Th{\'e}venin}, {Torra}, {Els}, {Gracia-Abril}, {Comoretto},
  {Garcia-Reinaldos}, {Lock}, {Mercier}, {Altmann}, {Andrae}, {Astraatmadja},
  {Bellas-Velidis}, {Benson}, {Berthier}, {Blomme}, {Busso}, {Carry},
  {Cellino}, {Clementini}, {Cowell}, {Creevey}, {Cuypers}, {Davidson}, {De
  Ridder}, {de Torres}, {Delchambre}, {Dell'Oro}, {Ducourant}, {Fr{\'e}mat},
  {Garc{\'\i}a-Torres}, {Gosset}, {Halbwachs}, {Hambly}, {Harrison}, {Hauser},
  {Hestroffer}, {Hodgkin}, {Huckle}, {Hutton}, {Jasniewicz}, {Jordan},
  {Kontizas}, {Korn}, {Lanzafame}, {Manteiga}, {Moitinho}, {Muinonen},
  {Osinde}, {Pancino}, {Pauwels}, {Petit}, {Recio-Blanco}, {Robin}, {Sarro},
  {Siopis}, {Smith}, {Smith}, {Sozzetti}, {Thuillot}, {van Reeven}, {Viala},
  {Abbas}, {Abreu Aramburu}, {Accart}, {Aguado}, {Allan}, {Allasia},
  {Altavilla}, {{\'A}lvarez}, {Alves}, {Anderson}, {Andrei}, {Anglada Varela},
  {Antiche}, {Antoja}, {Ant{\'o}n}, {Arcay}, {Atzei}, {Ayache}, {Bach},
  {Baker}, {Balaguer-N{\'u}{\~n}ez}, {Barache}, {Barata}, {Barbier}, {Barblan},
  {Baroni}, {Barrado y Navascu{\'e}s}, {Barros}, {Barstow}, {Becciani},
  {Bellazzini}, {Bellei}, {Bello Garc{\'\i}a}, {Belokurov}, {Bendjoya},
  {Berihuete}, {Bianchi}, {Bienaym{\'e}}, {Billebaud}, {Blagorodnova},
  {Blanco-Cuaresma}, {Boch}, {Bombrun}, {Borrachero}, {Bouquillon}, {Bourda},
  {Bouy}, {Bragaglia}, {Breddels}, {Brouillet}, {Br{\"u}semeister},
  {Bucciarelli}, {Budnik}, {Burgess}, {Burgon}, {Burlacu}, {Busonero}, {Buzzi},
  {Caffau}, {Cambras}, {Campbell}, {Cancelliere}, {Cantat-Gaudin}, {Carlucci},
  {Carrasco}, {Castellani}, {Charlot}, {Charnas}, {Charvet}, {Chassat},
  {Chiavassa}, {Clotet}, {Cocozza}, {Collins}, {Collins}, {Costigan}, {Crifo},
  {Cross}, {Crosta}, {Crowley}, {Dafonte}, {Damerdji}, {Dapergolas}, {David},
  {David}, {De Cat}, {de Felice}, {de Laverny}, {De Luise}, {De March}, {de
  Martino}, {de Souza}, {Debosscher}, {del Pozo}, {Delbo}, {Delgado},
  {Delgado}, {di Marco}, {Di Matteo}, {Diakite}, {Distefano}, {Dolding}, {Dos
  Anjos}, {Drazinos}, {Dur{\'a}n}, {Dzigan}, {Ecale}, {Edvardsson}, {Enke},
  {Erdmann}, {Escolar}, {Espina}, {Evans}, {Eynard Bontemps}, {Fabre},
  {Fabrizio}, {Faigler}, {Falc{\~a}o}, {Farr{\`a}s Casas}, {Faye}, {Federici},
  {Fedorets}, {Fern{\'a}ndez-Hern{\'a}ndez}, {Fernique}, {Fienga}, {Figueras},
  {Filippi}, {Findeisen}, {Fonti}, {Fouesneau}, {Fraile}, {Fraser}, {Fuchs},
  {Furnell}, {Gai}, {Galleti}, {Galluccio}, {Garabato}, {Garc{\'\i}a-Sedano},
  {Gar{\'e}}, {Garofalo}, {Garralda}, {Gavras}, {Gerssen}, {Geyer}, {Gilmore},
  {Girona}, {Giuffrida}, {Gomes}, {Gonz{\'a}lez-Marcos},
  {Gonz{\'a}lez-N{\'u}{\~n}ez}, {Gonz{\'a}lez-Vidal}, {Granvik}, {Guerrier},
  {Guillout}, {Guiraud}, {G{\'u}rpide}, {Guti{\'e}rrez-S{\'a}nchez}, {Guy},
  {Haigron}, {Hatzidimitriou}, {Haywood}, {Heiter}, {Helmi}, {Hobbs},
  {Hofmann}, {Holl}, {Holland }, {Hunt}, {Hypki}, {Icardi}, {Irwin}, {Jevardat
  de Fombelle}, {Jofr{\'e}}, {Jonker}, {Jorissen}, {Julbe}, {Karampelas},
  {Kochoska}, {Kohley}, {Kolenberg}, {Kontizas}, {Koposov}, {Kordopatis},
  {Koubsky}, {Kowalczyk}, {Krone-Martins}, {Kudryashova}, {Kull}, {Bachchan},
  {Lacoste-Seris}, {Lanza}, {Lavigne}, {Le Poncin-Lafitte}, {Lebreton},
  {Lebzelter}, {Leccia}, {Leclerc}, {Lecoeur-Taibi}, {Lemaitre}, {Lenhardt},
  {Leroux}, {Liao}, {Licata}, {Lindstr{\o}m}, {Lister}, {Livanou}, {Lobel},
  {L{\"o}ffler}, {L{\'o}pez}, {Lopez-Lozano}, {Lorenz}, {Loureiro},
  {MacDonald}, {Magalh{\~a}es Fernandes}, {Managau}, {Mann}, {Mantelet},
  {Marchal}, {Marchant}, {Marconi}, {Marie}, {Marinoni}, {Marrese},
  {Marschalk{\'o}}, {Marshall}, {Mart{\'\i}n-Fleitas}, {Martino}, {Mary},
  {Matijevi{\v{c}}}, {Mazeh}, {McMillan}, {Messina}, {Mestre}, {Michalik},
  {Millar}, {Miranda}, {Molina}, {Molinaro}, {Molinaro}, {Moln{\'a}r},
  {Moniez}, {Montegriffo}, {Monteiro}, {Mor}, {Mora}, {Morbidelli}, {Morel},
  {Morgenthaler}, {Morley}, {Morris}, {Mulone}, {Muraveva}, {Musella},
  {Narbonne}, {Nelemans}, {Nicastro}, {Noval}, {Ord{\'e}novic},
  {Ordieres-Mer{\'e}}, {Osborne}, {Pagani}, {Pagano}, {Pailler}, {Palacin},
  {Palaversa}, {Parsons}, {Paulsen}, {Pecoraro}, {Pedrosa}, {Pentik{\"a}inen},
  {Pereira}, {Pichon}, {Piersimoni}, {Pineau}, {Plachy}, {Plum}, {Poujoulet},
  {Pr{\v{s}}a}, {Pulone}, {Ragaini}, {Rago}, {Rambaux}, {Ramos-Lerate},
  {Ranalli}, {Rauw}, {Read}, {Regibo}, {Renk}, {Reyl{\'e}}, {Ribeiro},
  {Rimoldini}, {Ripepi}, {Riva}, {Rixon}, {Roelens}, {Romero-G{\'o}mez},
  {Rowell}, {Royer}, {Rudolph}, {Ruiz-Dern}, {Sadowski}, {Sagrist{\`a}
  Sell{\'e}s}, {Sahlmann}, {Salgado}, {Salguero}, {Sarasso}, {Savietto},
  {Schnorhk}, {Schultheis}, {Sciacca}, {Segol}, {Segovia}, {Segransan},
  {Serpell}, {Shih}, {Smareglia}, {Smart}, {Smith}, {Solano}, {Solitro},
  {Sordo}, {Soria Nieto}, {Souchay}, {Spagna}, {Spoto}, {Stampa}, {Steele},
  {Steidelm{\"u}ller}, {Stephenson}, {Stoev}, {Suess}, {S{\"u}veges}, {Surdej},
  {Szabados}, {Szegedi-Elek}, {Tapiador}, {Taris}, {Tauran}, {Taylor},
  {Teixeira}, {Terrett}, {Tingley}, {Trager}, {Turon}, {Ulla}, {Utrilla},
  {Valentini}, {van Elteren}, {Van Hemelryck}, {van Leeuwen}, {Varadi},
  {Vecchiato}, {Veljanoski}, {Via}, {Vicente}, {Vogt}, {Voss}, {Votruba},
  {Voutsinas}, {Walmsley}, {Weiler}, {Weingrill}, {Werner}, {Wevers},
  {Whitehead}, {Wyrzykowski}, {Yoldas}, {{\v{Z}}erjal}, {Zucker}, {Zurbach},
  {Zwitter}, {Alecu}, {Allen}, {Allende Prieto}, {Amorim},
  {Anglada-Escud{\'e}}, {Arsenijevic}, {Azaz}, {Balm}, {Beck}, {Bernstein},
  {Bigot}, {Bijaoui}, {Blasco}, {Bonfigli}, {Bono}, {Boudreault}, {Bressan},
  {Brown}, {Brunet}, {Bunclark}, {Buonanno}, {Butkevich}, {Carret}, {Carrion},
  {Chemin}, {Ch{\'e}reau}, {Corcione}, {Darmigny}, {de Boer}, {de Teodoro}, {de
  Zeeuw}, {Delle Luche}, {Domingues}, {Dubath}, {Fodor}, {Fr{\'e}zouls},
  {Fries}, {Fustes}, {Fyfe}, {Gallardo}, {Gallegos}, {Gardiol}, {Gebran},
  {Gomboc}, {G{\'o}mez}, {Grux}, {Gueguen}, {Heyrovsky}, {Hoar}, {Iannicola},
  {Isasi Parache}, {Janotto}, {Joliet}, {Jonckheere}, {Keil}, {Kim},
  {Klagyivik}, {Klar}, {Knude}, {Kochukhov}, {Kolka}, {Kos}, {Kutka}, {Lainey},
  {LeBouquin}, {Liu}, {Loreggia}, {Makarov}, {Marseille}, {Martayan},
  {Martinez-Rubi}, {Massart}, {Meynadier}, {Mignot}, {Munari}, {Nguyen},
  {Nordlander}, {Ocvirk}, {O'Flaherty}, {Olias Sanz}, {Ortiz}, {Osorio},
  {Oszkiewicz}, {Ouzounis}, {Palmer}, {Park}, {Pasquato}, {Peltzer}, {Peralta},
  {P{\'e}turaud}, {Pieniluoma}, {Pigozzi}, {Poels}, {Prat}, {Prod'homme},
  {Raison}, {Rebordao}, {Risquez}, {Rocca-Volmerange}, {Rosen}, {Ruiz-Fuertes},
  {Russo}, {Sembay}, {Serraller Vizcaino}, {Short}, {Siebert}, {Silva},
  {Sinachopoulos}, {Slezak}, {Soffel}, {Sosnowska}, {Strai{\v{z}}ys}, {ter
  Linden}, {Terrell}, {Theil}, {Tiede}, {Troisi}, {Tsalmantza}, {Tur},
  {Vaccari}, {Vachier}, {Valles}, {Van Hamme}, {Veltz}, {Virtanen}, {Wallut},
  {Wichmann}, {Wilkinson}, {Ziaeepour}, \& {Zschocke}}]{2016A&A...595A...1G}
{Gaia Collaboration}, {Prusti}, T., {de Bruijne}, J.~H.~J., {et~al.} 2016,
  \aap, 595, A1, \dodoi{10.1051/0004-6361/201629272}

\bibitem[{{Gaia Collaboration} {et~al.}(2018){Gaia Collaboration}, {Brown},
  {Vallenari}, {Prusti}, {de Bruijne}, {Babusiaux}, {Bailer-Jones}, {Biermann},
  {Evans}, {Eyer}, {Jansen}, {Jordi}, {Klioner}, {Lammers}, {Lindegren},
  {Luri}, {Mignard}, {Panem}, {Pourbaix}, {Randich}, {Sartoretti}, {Siddiqui},
  {Soubiran}, {van Leeuwen}, {Walton}, {Arenou}, {Bastian}, {Cropper},
  {Drimmel}, {Katz}, {Lattanzi}, {Bakker}, {Cacciari}, {Casta{\~n}eda},
  {Chaoul}, {Cheek}, {De Angeli}, {Fabricius}, {Guerra}, {Holl}, {Masana},
  {Messineo}, {Mowlavi}, {Nienartowicz}, {Panuzzo}, {Portell}, {Riello},
  {Seabroke}, {Tanga}, {Th{\'e}venin}, {Gracia-Abril}, {Comoretto},
  {Garcia-Reinaldos}, {Teyssier}, {Altmann}, {Andrae}, {Audard},
  {Bellas-Velidis}, {Benson}, {Berthier}, {Blomme}, {Burgess}, {Busso},
  {Carry}, {Cellino}, {Clementini}, {Clotet}, {Creevey}, {Davidson}, {De
  Ridder}, {Delchambre}, {Dell'Oro}, {Ducourant},
  {Fern{\'a}ndez-Hern{\'a}ndez}, {Fouesneau}, {Fr{\'e}mat}, {Galluccio},
  {Garc{\'\i}a-Torres}, {Gonz{\'a}lez-N{\'u}{\~n}ez}, {Gonz{\'a}lez-Vidal},
  {Gosset}, {Guy}, {Halbwachs}, {Hambly}, {Harrison}, {Hern{\'a}ndez},
  {Hestroffer}, {Hodgkin}, {Hutton}, {Jasniewicz}, {Jean-Antoine-Piccolo},
  {Jordan}, {Korn}, {Krone-Martins}, {Lanzafame}, {Lebzelter}, {L{\"o}ffler},
  {Manteiga}, {Marrese}, {Mart{\'\i}n-Fleitas}, {Moitinho}, {Mora}, {Muinonen},
  {Osinde}, {Pancino}, {Pauwels}, {Petit}, {Recio-Blanco}, {Richards},
  {Rimoldini}, {Robin}, {Sarro}, {Siopis}, {Smith}, {Sozzetti}, {S{\"u}veges},
  {Torra}, {van Reeven}, {Abbas}, {Abreu Aramburu}, {Accart}, {Aerts},
  {Altavilla}, {{\'A}lvarez}, {Alvarez}, {Alves}, {Anderson}, {Andrei},
  {Anglada Varela}, {Antiche}, {Antoja}, {Arcay}, {Astraatmadja}, {Bach},
  {Baker}, {Balaguer-N{\'u}{\~n}ez}, {Balm}, {Barache}, {Barata}, {Barbato},
  {Barblan}, {Barklem}, {Barrado}, {Barros}, {Barstow}, {Bartholom{\'e}
  Mu{\~n}oz}, {Bassilana}, {Becciani}, {Bellazzini}, {Berihuete}, {Bertone},
  {Bianchi}, {Bienaym{\'e}}, {Blanco-Cuaresma}, {Boch}, {Boeche}, {Bombrun},
  {Borrachero}, {Bossini}, {Bouquillon}, {Bourda}, {Bragaglia}, {Bramante},
  {Breddels}, {Bressan}, {Brouillet}, {Br{\"u}semeister}, {Brugaletta},
  {Bucciarelli}, {Burlacu}, {Busonero}, {Butkevich}, {Buzzi}, {Caffau},
  {Cancelliere}, {Cannizzaro}, {Cantat-Gaudin}, {Carballo}, {Carlucci},
  {Carrasco}, {Casamiquela}, {Castellani}, {Castro-Ginard}, {Charlot},
  {Chemin}, {Chiavassa}, {Cocozza}, {Costigan}, {Cowell}, {Crifo}, {Crosta},
  {Crowley}, {Cuypers}, {Dafonte}, {Damerdji}, {Dapergolas}, {David}, {David},
  {de Laverny}, {De Luise}, {De March}, {de Martino}, {de Souza}, {de Torres},
  {Debosscher}, {del Pozo}, {Delbo}, {Delgado}, {Delgado}, {Di Matteo},
  {Diakite}, {Diener}, {Distefano}, {Dolding}, {Drazinos}, {Dur{\'a}n},
  {Edvardsson}, {Enke}, {Eriksson}, {Esquej}, {Eynard Bontemps}, {Fabre},
  {Fabrizio}, {Faigler}, {Falc{\~a}o}, {Farr{\`a}s Casas}, {Federici},
  {Fedorets}, {Fernique}, {Figueras}, {Filippi}, {Findeisen}, {Fonti},
  {Fraile}, {Fraser}, {Fr{\'e}zouls}, {Gai}, {Galleti}, {Garabato},
  {Garc{\'\i}a-Sedano}, {Garofalo}, {Garralda}, {Gavel}, {Gavras}, {Gerssen},
  {Geyer}, {Giacobbe}, {Gilmore}, {Girona}, {Giuffrida}, {Glass}, {Gomes},
  {Granvik}, {Gueguen}, {Guerrier}, {Guiraud}, {Guti{\'e}rrez-S{\'a}nchez},
  {Haigron}, {Hatzidimitriou}, {Hauser}, {Haywood}, {Heiter}, {Helmi}, {Heu},
  {Hilger}, {Hobbs}, {Hofmann}, {Holland}, {Huckle}, {Hypki}, {Icardi},
  {Jan{\ss}en}, {Jevardat de Fombelle}, {Jonker}, {Juh{\'a}sz}, {Julbe},
  {Karampelas}, {Kewley}, {Klar}, {Kochoska}, {Kohley}, {Kolenberg},
  {Kontizas}, {Kontizas}, {Koposov}, {Kordopatis}, {Kostrzewa-Rutkowska},
  {Koubsky}, {Lambert}, {Lanza}, {Lasne}, {Lavigne}, {Le Fustec}, {Le
  Poncin-Lafitte}, {Lebreton}, {Leccia}, {Leclerc}, {Lecoeur-Taibi},
  {Lenhardt}, {Leroux}, {Liao}, {Licata}, {Lindstr{\o}m}, {Lister}, {Livanou},
  {Lobel}, {L{\'o}pez}, {Managau}, {Mann}, {Mantelet}, {Marchal}, {Marchant},
  {Marconi}, {Marinoni}, {Marschalk{\'o}}, {Marshall}, {Martino}, {Marton},
  {Mary}, {Massari}, {Matijevi{\v{c}}}, {Mazeh}, {McMillan}, {Messina},
  {Michalik}, {Millar}, {Molina}, {Molinaro}, {Moln{\'a}r}, {Montegriffo},
  {Mor}, {Morbidelli}, {Morel}, {Morris}, {Mulone}, {Muraveva}, {Musella},
  {Nelemans}, {Nicastro}, {Noval}, {O'Mullane}, {Ord{\'e}novic},
  {Ord{\'o}{\~n}ez-Blanco}, {Osborne}, {Pagani}, {Pagano}, {Pailler},
  {Palacin}, {Palaversa}, {Panahi}, {Pawlak}, {Piersimoni}, {Pineau}, {Plachy},
  {Plum}, {Poggio}, {Poujoulet}, {Pr{\v{s}}a}, {Pulone}, {Racero}, {Ragaini},
  {Rambaux}, {Ramos-Lerate}, {Regibo}, {Reyl{\'e}}, {Riclet}, {Ripepi}, {Riva},
  {Rivard}, {Rixon}, {Roegiers}, {Roelens}, {Romero-G{\'o}mez}, {Rowell},
  {Royer}, {Ruiz-Dern}, {Sadowski}, {Sagrist{\`a} Sell{\'e}s}, {Sahlmann},
  {Salgado}, {Salguero}, {Sanna}, {Santana-Ros}, {Sarasso}, {Savietto},
  {Schultheis}, {Sciacca}, {Segol}, {Segovia}, {S{\'e}gransan}, {Shih},
  {Siltala}, {Silva}, {Smart}, {Smith}, {Solano}, {Solitro}, {Sordo}, {Soria
  Nieto}, {Souchay}, {Spagna}, {Spoto}, {Stampa}, {Steele},
  {Steidelm{\"u}ller}, {Stephenson}, {Stoev}, {Suess}, {Surdej}, {Szabados},
  {Szegedi-Elek}, {Tapiador}, {Taris}, {Tauran}, {Taylor}, {Teixeira},
  {Terrett}, {Teyssand ier}, {Thuillot}, {Titarenko}, {Torra Clotet}, {Turon},
  {Ulla}, {Utrilla}, {Uzzi}, {Vaillant}, {Valentini}, {Valette}, {van Elteren},
  {Van Hemelryck}, {van Leeuwen}, {Vaschetto}, {Vecchiato}, {Veljanoski},
  {Viala}, {Vicente}, {Vogt}, {von Essen}, {Voss}, {Votruba}, {Voutsinas},
  {Walmsley}, {Weiler}, {Wertz}, {Wevers}, {Wyrzykowski}, {Yoldas},
  {{\v{Z}}erjal}, {Ziaeepour}, {Zorec}, {Zschocke}, {Zucker}, {Zurbach}, \&
  {Zwitter}}]{2018A&A...616A...1G}
{Gaia Collaboration}, {Brown}, A.~G.~A., {Vallenari}, A., {et~al.} 2018, \aap,
  616, A1, \dodoi{10.1051/0004-6361/201833051}

\bibitem[{Gelman {et~al.}(2004)Gelman, Carlin, Stern, \& Rubin}]{Gelman2004}
Gelman, A., Carlin, J.~B., Stern, H.~S., \& Rubin, D.~B. 2004, {B}ayesian data
  analysis, 2nd edn., Texts in Statistical Science Series (Chapman \& Hall/CRC,
  Boca Raton, FL)

\bibitem[{Gelman \& Hill(2007)}]{Gelman2007}
Gelman, A., \& Hill, J. 2007, Data analysis using regression and
  multilevel/hierarchical models, Vol. Analytical methods for social research
  (New York: Cambridge University Press)

\bibitem[{{Gerasimovic}(1928)}]{1928PNAS...14..963G}
{Gerasimovic}, B.~P. 1928, Proceedings of the National Academy of Science, 14,
  963, \dodoi{10.1073/pnas.14.12.963}

\bibitem[{{Ghez} {et~al.}(2008){Ghez}, {Salim}, {Weinberg}, {Lu}, {Do}, {Dunn},
  {Matthews}, {Morris}, {Yelda}, {Becklin}, {Kremenek}, {Milosavljevic}, \&
  {Naiman}}]{2008ApJ...689.1044G}
{Ghez}, A.~M., {Salim}, S., {Weinberg}, N.~N., {et~al.} 2008, \apj, 689, 1044,
  \dodoi{10.1086/592738}

\bibitem[{{Gingerich}(1999)}]{1999ApJ...525C.135G}
{Gingerich}, O. 1999, \apj, 525C, 135

\bibitem[{{Glass} \& {Evans}(1981)}]{1981Natur.291..303G}
{Glass}, I.~S., \& {Evans}, T.~L. 1981, \nat, 291, 303,
  \dodoi{10.1038/291303a0}

\bibitem[{{Glass} \& {Feast}(1982)}]{1982MNRAS.198..199G}
{Glass}, I.~S., \& {Feast}, M.~W. 1982, \mnras, 198, 199,
  \dodoi{10.1093/mnras/198.1.199}

\bibitem[{{GLIMPSE
  Team}(2020{\natexlab{a}})}]{https://doi.org/10.26131/irsa223}
{GLIMPSE Team}. 2020{\natexlab{a}}, GLIMPSE I Catalog,  IPAC,
  \dodoi{10.26131/IRSA223}

\bibitem[{{GLIMPSE
  Team}(2020{\natexlab{b}})}]{https://doi.org/10.26131/irsa200}
---. 2020{\natexlab{b}}, GLIMPSE II Catalog,  IPAC, \dodoi{10.26131/IRSA200}

\bibitem[{{GLIMPSE
  Team}(2020{\natexlab{c}})}]{https://doi.org/10.26131/irsa203}
---. 2020{\natexlab{c}}, GLIMPSE 3D Catalog,  IPAC, \dodoi{10.26131/IRSA203}

\bibitem[{{GLIMPSE
  Team}(2020{\natexlab{d}})}]{https://doi.org/10.26131/irsa214}
---. 2020{\natexlab{d}}, GLIMPSE 360 Catalog,  IPAC, \dodoi{10.26131/IRSA214}

\bibitem[{{GLIMPSE
  Team}(2020{\natexlab{e}})}]{https://doi.org/10.26131/irsa213}
---. 2020{\natexlab{e}}, Vela-Carina Catalog,  IPAC, \dodoi{10.26131/IRSA213}

\bibitem[{{GLIMPSE
  Team}(2020{\natexlab{f}})}]{https://doi.org/10.26131/irsa217}
---. 2020{\natexlab{f}}, Deep GLIMPSE Catalog,  IPAC, \dodoi{10.26131/IRSA217}

\bibitem[{{GLIMPSE
  Team}(2020{\natexlab{g}})}]{https://doi.org/10.26131/irsa226}
---. 2020{\natexlab{g}}, GLIMPSE SMOG Catalog,  IPAC, \dodoi{10.26131/IRSA226}

\bibitem[{{GLIMPSE
  Team}(2020{\natexlab{h}})}]{https://doi.org/10.26131/irsa225}
---. 2020{\natexlab{h}}, GLIMPSE Cygnus-X Catalog,  IPAC,
  \dodoi{10.26131/IRSA225}

\bibitem[{{Grady} {et~al.}(2020){Grady}, {Belokurov}, \&
  {Evans}}]{2020MNRAS.492.3128G}
{Grady}, J., {Belokurov}, V., \& {Evans}, N.~W. 2020, \mnras, 492, 3128,
  \dodoi{10.1093/mnras/stz3617}

\bibitem[{{GRAVITY Collaboration} {et~al.}(2019){GRAVITY Collaboration},
  {Abuter}, {Amorim}, {Baub{\"o}ck}, {Berger}, {Bonnet}, {Brandner},
  {Cl{\'e}net}, {Coud{\'e} Du Foresto}, {de Zeeuw}, {Dexter}, {Duvert},
  {Eckart}, {Eisenhauer}, {F{\"o}rster Schreiber}, {Garcia}, {Gao}, {Gendron},
  {Genzel}, {Gerhard}, {Gillessen}, {Habibi}, {Haubois}, {Henning}, {Hippler},
  {Horrobin}, {Jim{\'e}nez-Rosales}, {Jocou}, {Kervella}, {Lacour},
  {Lapeyr{\`e}re}, {Le Bouquin}, {L{\'e}na}, {Ott}, {Paumard}, {Perraut},
  {Perrin}, {Pfuhl}, {Rabien}, {Rodriguez Coira}, {Rousset}, {Scheithauer},
  {Sternberg}, {Straub}, {Straubmeier}, {Sturm}, {Tacconi}, {Vincent}, {von
  Fellenberg}, {Waisberg}, {Widmann}, {Wieprecht}, {Wiezorrek}, {Woillez}, \&
  {Yazici}}]{2019A&A...625L..10G}
{GRAVITY Collaboration}, {Abuter}, R., {Amorim}, A., {et~al.} 2019, \aap, 625,
  L10, \dodoi{10.1051/0004-6361/201935656}

\bibitem[{{Green} {et~al.}(2019){Green}, {Schlafly}, {Zucker}, {Speagle}, \&
  {Finkbeiner}}]{2019ApJ...887...93G}
{Green}, G.~M., {Schlafly}, E., {Zucker}, C., {Speagle}, J.~S., \&
  {Finkbeiner}, D. 2019, \apj, 887, 93, \dodoi{10.3847/1538-4357/ab5362}

\bibitem[{{Groenewegen} \& {Blommaert}(2005)}]{2005A&A...443..143G}
{Groenewegen}, M.~A.~T., \& {Blommaert}, J.~A.~D.~L. 2005, \aap, 443, 143,
  \dodoi{10.1051/0004-6361:20053131}

\bibitem[{{Herschel}(1785)}]{1785RSPT...75..213H}
{Herschel}, W. 1785, Philosophical Transactions of the Royal Society of London
  Series I, 75, 213

\bibitem[{{Hora} {et~al.}(2007){Hora}, {Bontemps}, {Megeath}, {Schneider},
  {Motte}, {Carey}, {Simon}, {Keto}, {Smith}, {Allen}, {Gutermuth}, {Fazio},
  {Kraemer}, {Mizuno}, {Price}, \& {Adams}}]{2007sptz.prop40184H}
{Hora}, J., {Bontemps}, S., {Megeath}, T., {et~al.} 2007, {A Spitzer Legacy
  Survey of the Cygnus-X Complex}, Spitzer Proposal

\bibitem[{{Ita} \& {Matsunaga}(2011)}]{2011MNRAS.412.2345I}
{Ita}, Y., \& {Matsunaga}, N. 2011, \mnras, 412, 2345,
  \dodoi{10.1111/j.1365-2966.2010.18056.x}

\bibitem[{{Ita} {et~al.}(2021){Ita}, {Menzies}, {Whitelock}, {Matsunaga},
  {Takayama}, {Nakada}, {Tanab{\'e}}, {Feast}, \&
  {Nagayama}}]{2021MNRAS.500...82I}
{Ita}, Y., {Menzies}, J.~W., {Whitelock}, P.~A., {et~al.} 2021, \mnras, 500,
  82, \dodoi{10.1093/mnras/staa3251}

\bibitem[{{Iwanek} {et~al.}(2021{\natexlab{a}}){Iwanek}, {Soszy{\'n}ski}, \&
  {Koz{\l}owski}}]{2021ApJ...919...99I}
{Iwanek}, P., {Soszy{\'n}ski}, I., \& {Koz{\l}owski}, S. 2021{\natexlab{a}},
  \apj, 919, 99, \dodoi{10.3847/1538-4357/ac10c5}

\bibitem[{{Iwanek} {et~al.}(2021{\natexlab{b}}){Iwanek}, {Koz{\l}owski},
  {Gromadzki}, {Soszy{\'n}ski}, {Wrona}, {Skowron}, {Ratajczak}, {Udalski},
  {Szyma{\'n}ski}, {Pietrukowicz}, {Ulaczyk}, {Poleski}, {Mr{\'o}z}, {Skowron},
  \& {Rybicki}}]{2021ApJS..257...23I}
{Iwanek}, P., {Koz{\l}owski}, S., {Gromadzki}, M., {et~al.} 2021{\natexlab{b}},
  \apjs, 257, 23, \dodoi{10.3847/1538-4365/ac1797}

\bibitem[{{Iwanek} {et~al.}(2022){Iwanek}, {Soszy{\'n}ski}, {Koz{\l}owski},
  {Poleski}, {Pietrukowicz}, {Skowron}, {Wrona}, {Mr{\'o}z}, {Udalski},
  {Szyma{\'n}ski}, {Skowron}, {Ulaczyk}, {Gromadzki}, {Rybicki}, \&
  {Ratajczak}}]{2022ApJS..260...46I}
{Iwanek}, P., {Soszy{\'n}ski}, I., {Koz{\l}owski}, S., {et~al.} 2022, \apjs,
  260, 46, \dodoi{10.3847/1538-4365/ac6676}

\bibitem[{{Kunder} {et~al.}(2020){Kunder}, {P{\'e}rez-Villegas}, {Rich},
  {Ogata}, {Murari}, {Boren}, {Johnson}, {Nataf}, {Walker}, {Bono}, {Koch},
  {Propris}, {Storm}, \& {Wojno}}]{2020AJ....159..270K}
{Kunder}, A., {P{\'e}rez-Villegas}, A., {Rich}, R.~M., {et~al.} 2020, \aj, 159,
  270, \dodoi{10.3847/1538-3881/ab8d35}

\bibitem[{{Leavitt} \& {Pickering}(1912)}]{1912HarCi.173....1L}
{Leavitt}, H.~S., \& {Pickering}, E.~C. 1912, Harvard College Observatory
  Circular, 173, 1

\bibitem[{{Levine} {et~al.}(2006){Levine}, {Blitz}, \&
  {Heiles}}]{2006Sci...312.1773L}
{Levine}, E.~S., {Blitz}, L., \& {Heiles}, C. 2006, Science, 312, 1773,
  \dodoi{10.1126/science.1128455}

\bibitem[{{L{\'o}pez-Corredoira}(2017)}]{2017ApJ...836..218L}
{L{\'o}pez-Corredoira}, M. 2017, \apj, 836, 218,
  \dodoi{10.3847/1538-4357/836/2/218}

\bibitem[{{L{\'o}pez-Corredoira} {et~al.}(2007){L{\'o}pez-Corredoira},
  {Cabrera-Lavers}, {Mahoney}, {Hammersley}, {Garz{\'o}n}, \&
  {Gonz{\'a}lez-Fern{\'a}ndez}}]{2007AJ....133..154L}
{L{\'o}pez-Corredoira}, M., {Cabrera-Lavers}, A., {Mahoney}, T.~J., {et~al.}
  2007, \aj, 133, 154, \dodoi{10.1086/509605}

\bibitem[{{Lopez-Corredoira} {et~al.}(1997){Lopez-Corredoira}, {Garzon},
  {Hammersley}, {Mahoney}, \& {Calbet}}]{1997MNRAS.292L..15L}
{Lopez-Corredoira}, M., {Garzon}, F., {Hammersley}, P., {Mahoney}, T., \&
  {Calbet}, X. 1997, \mnras, 292, L15, \dodoi{10.1093/mnras/292.1.L15}

\bibitem[{{L{\'o}pez-Corredoira} {et~al.}(2019){L{\'o}pez-Corredoira}, {Lee},
  {Garz{\'o}n}, \& {Lim}}]{2019A&A...627A...3L}
{L{\'o}pez-Corredoira}, M., {Lee}, Y.~W., {Garz{\'o}n}, F., \& {Lim}, D. 2019,
  \aap, 627, A3, \dodoi{10.1051/0004-6361/201935571}

\bibitem[{{Luri} {et~al.}(2018){Luri}, {Brown}, {Sarro}, {Arenou},
  {Bailer-Jones}, {Castro-Ginard}, {de Bruijne}, {Prusti}, {Babusiaux}, \&
  {Delgado}}]{2018A&A...616A...9L}
{Luri}, X., {Brown}, A.~G.~A., {Sarro}, L.~M., {et~al.} 2018, \aap, 616, A9,
  \dodoi{10.1051/0004-6361/201832964}

\bibitem[{{Mainzer} {et~al.}(2011){Mainzer}, {Bauer}, {Grav}, {Masiero},
  {Cutri}, {Dailey}, {Eisenhardt}, {McMillan}, {Wright}, {Walker}, {Jedicke},
  {Spahr}, {Tholen}, {Alles}, {Beck}, {Brand enburg}, {Conrow}, {Evans},
  {Fowler}, {Jarrett}, {Marsh}, {Masci}, {McCallon}, {Wheelock}, {Wittman},
  {Wyatt}, {DeBaun}, {Elliott}, {Elsbury}, {Gautier}, {Gomillion}, {Leisawitz},
  {Maleszewski}, {Micheli}, \& {Wilkins}}]{2011ApJ...731...53M}
{Mainzer}, A., {Bauer}, J., {Grav}, T., {et~al.} 2011, \apj, 731, 53,
  \dodoi{10.1088/0004-637X/731/1/53}

\bibitem[{{Mainzer} {et~al.}(2014){Mainzer}, {Bauer}, {Cutri}, {Grav},
  {Masiero}, {Beck}, {Clarkson}, {Conrow}, {Dailey}, {Eisenhardt}, {Fabinsky},
  {Fajardo-Acosta}, {Fowler}, {Gelino}, {Grillmair}, {Heinrichsen}, {Kendall},
  {Kirkpatrick}, {Liu}, {Masci}, {McCallon}, {Nugent}, {Papin}, {Rice},
  {Royer}, {Ryan}, {Sevilla}, {Sonnett}, {Stevenson}, {Thompson}, {Wheelock},
  {Wiemer}, {Wittman}, {Wright}, \& {Yan}}]{2014ApJ...792...30M}
{Mainzer}, A., {Bauer}, J., {Cutri}, R.~M., {et~al.} 2014, \apj, 792, 30,
  \dodoi{10.1088/0004-637X/792/1/30}

\bibitem[{{Marshall} {et~al.}(2006){Marshall}, {Robin}, {Reyl{\'e}},
  {Schultheis}, \& {Picaud}}]{2006A&A...453..635M}
{Marshall}, D.~J., {Robin}, A.~C., {Reyl{\'e}}, C., {Schultheis}, M., \&
  {Picaud}, S. 2006, \aap, 453, 635, \dodoi{10.1051/0004-6361:20053842}

\bibitem[{{Matsunaga} {et~al.}(2005){Matsunaga}, {Fukushi}, \&
  {Nakada}}]{2005MNRAS.364..117M}
{Matsunaga}, N., {Fukushi}, H., \& {Nakada}, Y. 2005, \mnras, 364, 117,
  \dodoi{10.1111/j.1365-2966.2005.09556.x}

\bibitem[{{Matsunaga} {et~al.}(2009){Matsunaga}, {Kawadu}, {Nishiyama},
  {Nagayama}, {Hatano}, {Tamura}, {Glass}, \& {Nagata}}]{2009MNRAS.399.1709M}
{Matsunaga}, N., {Kawadu}, T., {Nishiyama}, S., {et~al.} 2009, \mnras, 399,
  1709, \dodoi{10.1111/j.1365-2966.2009.15393.x}

\bibitem[{{Matsunaga} {et~al.}(2017){Matsunaga}, {Menzies}, {Feast},
  {Whitelock}, {Onozato}, {Barway}, \& {Aydi}}]{2017MNRAS.469.4949M}
{Matsunaga}, N., {Menzies}, J.~W., {Feast}, M.~W., {et~al.} 2017, \mnras, 469,
  4949, \dodoi{10.1093/mnras/stx1213}

\bibitem[{{McWilliam} \& {Zoccali}(2010)}]{2010ApJ...724.1491M}
{McWilliam}, A., \& {Zoccali}, M. 2010, \apj, 724, 1491,
  \dodoi{10.1088/0004-637X/724/2/1491}

\bibitem[{{Minniti} {et~al.}(2010){Minniti}, {Lucas}, {Emerson}, {Saito},
  {Hempel}, {Pietrukowicz}, {Ahumada}, {Alonso}, {Alonso-Garcia}, {Arias},
  {Bandyopadhyay}, {Barb{\'a}}, {Barbuy}, {Bedin}, {Bica}, {Borissova},
  {Bronfman}, {Carraro}, {Catelan}, {Clari{\'a}}, {Cross}, {de Grijs},
  {D{\'e}k{\'a}ny}, {Drew}, {Fari{\~n}a}, {Feinstein}, {Fern{\'a}ndez
  Laj{\'u}s}, {Gamen}, {Geisler}, {Gieren}, {Goldman}, {Gonzalez}, {Gunthardt},
  {Gurovich}, {Hambly}, {Irwin}, {Ivanov}, {Jord{\'a}n}, {Kerins}, {Kinemuchi},
  {Kurtev}, {L{\'o}pez-Corredoira}, {Maccarone}, {Masetti}, {Merlo},
  {Messineo}, {Mirabel}, {Monaco}, {Morelli}, {Padilla}, {Palma}, {Parisi},
  {Pignata}, {Rejkuba}, {Roman-Lopes}, {Sale}, {Schreiber}, {Schr{\"o}der},
  {Smith}, {}, {Soto}, {Tamura}, {Tappert}, {Thompson}, {Toledo}, {Zoccali}, \&
  {Pietrzynski}}]{2010NewA...15..433M}
{Minniti}, D., {Lucas}, P.~W., {Emerson}, J.~P., {et~al.} 2010, \na, 15, 433,
  \dodoi{10.1016/j.newast.2009.12.002}

\bibitem[{{Molina} {et~al.}(2019){Molina}, {Borissova}, {Catelan}, {Lucas},
  {Medina}, {Contreras Pe{\~n}a}, {Kurtev}, \& {Minniti}}]{2019MNRAS.482.5567M}
{Molina}, C.~N., {Borissova}, J., {Catelan}, M., {et~al.} 2019, \mnras, 482,
  5567, \dodoi{10.1093/mnras/sty3041}

\bibitem[{{Momany} {et~al.}(2006){Momany}, {Zaggia}, {Gilmore}, {Piotto},
  {Carraro}, {Bedin}, \& {de Angeli}}]{2006A&A...451..515M}
{Momany}, Y., {Zaggia}, S., {Gilmore}, G., {et~al.} 2006, \aap, 451, 515,
  \dodoi{10.1051/0004-6361:20054081}

\bibitem[{{Mr{\'o}z} {et~al.}(2019){Mr{\'o}z}, {Udalski}, {Skowron}, {Skowron},
  {Soszy{\'n}ski}, {Pietrukowicz}, {Szyma{\'n}ski}, {Poleski}, {Koz{\l}owski},
  \& {Ulaczyk}}]{2019ApJ...870L..10M}
{Mr{\'o}z}, P., {Udalski}, A., {Skowron}, D.~M., {et~al.} 2019, \apjl, 870,
  L10, \dodoi{10.3847/2041-8213/aaf73f}

\bibitem[{{Nakanishi} \& {Sofue}(2003)}]{2003PASJ...55..191N}
{Nakanishi}, H., \& {Sofue}, Y. 2003, \pasj, 55, 191,
  \dodoi{10.1093/pasj/55.1.191}

\bibitem[{{Nakanishi} \& {Sofue}(2006)}]{2006PASJ...58..847N}
---. 2006, \pasj, 58, 847, \dodoi{10.1093/pasj/58.5.847}

\bibitem[{{Nataf} {et~al.}(2010){Nataf}, {Udalski}, {Gould}, {Fouqu{\'e}}, \&
  {Stanek}}]{2010ApJ...721L..28N}
{Nataf}, D.~M., {Udalski}, A., {Gould}, A., {Fouqu{\'e}}, P., \& {Stanek},
  K.~Z. 2010, \apjl, 721, L28, \dodoi{10.1088/2041-8205/721/1/L28}

\bibitem[{{NEOWISE Team}(2020)}]{https://doi.org/10.26131/irsa144}
{NEOWISE Team}. 2020, NEOWISE-R Single Exposure (L1b) Source Table,  IPAC,
  \dodoi{10.26131/IRSA144}

\bibitem[{{Nishiyama} {et~al.}(2006){Nishiyama}, {Nagata}, {Sato}, {Kato},
  {Nagayama}, {Kusakabe}, {Matsunaga}, {Naoi}, {Sugitani}, \&
  {Tamura}}]{2006ApJ...647.1093N}
{Nishiyama}, S., {Nagata}, T., {Sato}, S., {et~al.} 2006, \apj, 647, 1093,
  \dodoi{10.1086/505529}

\bibitem[{{Palla} {et~al.}(2020){Palla}, {Matteucci}, {Spitoni}, {Vincenzo}, \&
  {Grisoni}}]{2020MNRAS.498.1710P}
{Palla}, M., {Matteucci}, F., {Spitoni}, E., {Vincenzo}, F., \& {Grisoni}, V.
  2020, \mnras, 498, 1710, \dodoi{10.1093/mnras/staa2437}

\bibitem[{{Pietrukowicz} {et~al.}(2013){Pietrukowicz}, {Dziembowski},
  {Mr{\'o}z}, {Soszy{\'n}ski}, {Udalski}, {Poleski}, {Szyma{\'n}ski}, {Kubiak},
  {Pietrzy{\'n}ski}, {Wyrzykowski}, {Ulaczyk}, {Koz{\l}owski}, \&
  {Skowron}}]{2013AcA....63..379P}
{Pietrukowicz}, P., {Dziembowski}, W.~A., {Mr{\'o}z}, P., {et~al.} 2013,
  \actaa, 63, 379.
\newblock \doarXiv{1311.5894}

\bibitem[{{Pietrukowicz} {et~al.}(2015){Pietrukowicz}, {Koz{\l}owski},
  {Skowron}, {Soszy{\'n}ski}, {Udalski}, {Poleski}, {Wyrzykowski},
  {Szyma{\'n}ski}, {Pietrzy{\'n}ski}, {Ulaczyk}, {Mr{\'o}z}, {Skowron}, \&
  {Kubiak}}]{2015ApJ...811..113P}
{Pietrukowicz}, P., {Koz{\l}owski}, S., {Skowron}, J., {et~al.} 2015, \apj,
  811, 113, \dodoi{10.1088/0004-637X/811/2/113}

\bibitem[{{Pietrukowicz} {et~al.}(2020){Pietrukowicz}, {Udalski},
  {Soszy{\'n}ski}, {Skowron}, {Wrona}, {Szyma{\'n}ski}, {Poleski}, {Ulaczyk},
  {Koz{\l}owski}, {Skowron}, {Mr{\'o}z}, {Rybicki}, {Iwanek}, \&
  {Gromadzki}}]{2020AcA....70..121P}
{Pietrukowicz}, P., {Udalski}, A., {Soszy{\'n}ski}, I., {et~al.} 2020, \actaa,
  70, 121, \dodoi{10.32023/0001-5237/70.2.3}

\bibitem[{{Poleski} {et~al.}(2021){Poleski}, {Skowron}, {Mr{\'o}z}, {Udalski},
  {Szyma{\'n}ski}, {Pietrukowicz}, {Ulaczyk}, {Rybicki}, {Iwanek}, {Wrona}, \&
  {Gromadzki}}]{2021AcA....71....1P}
{Poleski}, R., {Skowron}, J., {Mr{\'o}z}, P., {et~al.} 2021, \actaa, 71, 1,
  \dodoi{10.32023/0001-5237/71.1.1}

\bibitem[{{Portail} {et~al.}(2017){Portail}, {Gerhard}, {Wegg}, \&
  {Ness}}]{2017MNRAS.465.1621P}
{Portail}, M., {Gerhard}, O., {Wegg}, C., \& {Ness}, M. 2017, \mnras, 465,
  1621, \dodoi{10.1093/mnras/stw2819}

\bibitem[{{Reid} {et~al.}(2014){Reid}, {Menten}, {Brunthaler}, {Zheng}, {Dame},
  {Xu}, {Wu}, {Zhang}, {Sanna}, {Sato}, {Hachisuka}, {Choi}, {Immer},
  {Moscadelli}, {Rygl}, \& {Bartkiewicz}}]{2014ApJ...783..130R}
{Reid}, M.~J., {Menten}, K.~M., {Brunthaler}, A., {et~al.} 2014, \apj, 783,
  130, \dodoi{10.1088/0004-637X/783/2/130}

\bibitem[{{Riebel} {et~al.}(2010){Riebel}, {Meixner}, {Fraser}, {Srinivasan},
  {Cook}, \& {Vijh}}]{2010ApJ...723.1195R}
{Riebel}, D., {Meixner}, M., {Fraser}, O., {et~al.} 2010, \apj, 723, 1195,
  \dodoi{10.1088/0004-637X/723/2/1195}

\bibitem[{{Saito} {et~al.}(2012){Saito}, {Hempel}, {Minniti}, {Lucas},
  {Rejkuba}, {Toledo}, {Gonzalez}, {Alonso-Garc{\'\i}a}, {Irwin},
  {Gonzalez-Solares}, {Hodgkin}, {Lewis}, {Cross}, {Ivanov}, {Kerins},
  {Emerson}, {Soto}, {Am{\^o}res}, {Gurovich}, {D{\'e}k{\'a}ny}, {Angeloni},
  {Beamin}, {Catelan}, {Padilla}, {Zoccali}, {Pietrukowicz}, {Moni Bidin},
  {Mauro}, {Geisler}, {Folkes}, {Sale}, {Borissova}, {Kurtev}, {Ahumada},
  {Alonso}, {Adamson}, {Arias}, {Bandyopadhyay}, {Barb{\'a}}, {Barbuy},
  {Baume}, {Bedin}, {Bellini}, {Benjamin}, {Bica}, {Bonatto}, {Bronfman},
  {Carraro}, {Chen{\`e}}, {Clari{\'a}}, {Clarke}, {Contreras}, {Corvill{\'o}n},
  {de Grijs}, {Dias}, {Drew}, {Fari{\~n}a}, {Feinstein},
  {Fern{\'a}ndez-Laj{\'u}s}, {Gamen}, {Gieren}, {Goldman},
  {Gonz{\'a}lez-Fern{\'a}ndez}, {Grand}, {Gunthardt}, {Hambly}, {Hanson},
  {He{\l}miniak}, {Hoare}, {Huckvale}, {Jord{\'a}n}, {Kinemuchi}, {Longmore},
  {L{\'o}pez-Corredoira}, {Maccarone}, {Majaess}, {Mart{\'\i}n}, {Masetti},
  {Mennickent}, {Mirabel}, {Monaco}, {Morelli}, {Motta}, {Palma}, {Parisi},
  {Parker}, {Pe{\~n}aloza}, {Pietrzy{\'n}ski}, {Pignata}, {Popescu}, {Read},
  {Rojas}, {Roman-Lopes}, {Ruiz}, {Saviane}, {Schreiber}, {Schr{\"o}der},
  {Sharma}, {Smith}, {Sodr{\'e}}, {Stead}, {Stephens}, {Tamura}, {Tappert},
  {Thompson}, {Valenti}, {Vanzi}, {Walton}, {Weidmann}, \&
  {Zijlstra}}]{2012A&A...537A.107S}
{Saito}, R.~K., {Hempel}, M., {Minniti}, D., {et~al.} 2012, \aap, 537, A107,
  \dodoi{10.1051/0004-6361/201118407}

\bibitem[{{Sanders} {et~al.}(2022){Sanders}, {Matsunaga}, {Kawata}, {Smith},
  {Minniti}, \& {Lucas}}]{2022MNRAS.517..257S}
{Sanders}, J.~L., {Matsunaga}, N., {Kawata}, D., {et~al.} 2022, \mnras, 517,
  257, \dodoi{10.1093/mnras/stac2274}

\bibitem[{{Sato} \& {Chiba}(2022)}]{2022ApJ...927..145S}
{Sato}, G., \& {Chiba}, M. 2022, \apj, 927, 145,
  \dodoi{10.3847/1538-4357/ac47fb}

\bibitem[{{Savino} {et~al.}(2020){Savino}, {Koch}, {Prudil}, {Kunder}, \&
  {Smolec}}]{2020A&A...641A..96S}
{Savino}, A., {Koch}, A., {Prudil}, Z., {Kunder}, A., \& {Smolec}, R. 2020,
  \aap, 641, A96, \dodoi{10.1051/0004-6361/202038305}

\bibitem[{{Semczuk} {et~al.}(2022{\natexlab{a}}){Semczuk}, {Dehnen},
  {Sch{\"o}nrich}, \& {Athanassoula}}]{2022MNRAS.509.4532S}
{Semczuk}, M., {Dehnen}, W., {Sch{\"o}nrich}, R., \& {Athanassoula}, E.
  2022{\natexlab{a}}, \mnras, 509, 4532, \dodoi{10.1093/mnras/stab3294}

\bibitem[{{Semczuk} {et~al.}(2022{\natexlab{b}}){Semczuk}, {Dehnen},
  {Sch{\"o}nrich}, \& {Athanassoula}}]{2022MNRAS.517.6060S}
---. 2022{\natexlab{b}}, \mnras, 517, 6060, \dodoi{10.1093/mnras/stac3085}

\bibitem[{{Shapley}(1918{\natexlab{a}})}]{1918ApJ....48...89S}
{Shapley}, H. 1918{\natexlab{a}}, \apj, 48, 89, \dodoi{10.1086/142419}

\bibitem[{{Shapley}(1918{\natexlab{b}})}]{1918ApJ....48..154S}
---. 1918{\natexlab{b}}, \apj, 48, 154, \dodoi{10.1086/142423}

\bibitem[{{Shapley}(1918{\natexlab{c}})}]{1918PASP...30...42S}
---. 1918{\natexlab{c}}, \pasp, 30, 42, \dodoi{10.1086/122686}

\bibitem[{{Shappee} {et~al.}(2014){Shappee}, {Prieto}, {Grupe}, {Kochanek},
  {Stanek}, {De Rosa}, {Mathur}, {Zu}, {Peterson}, {Pogge}, {Komossa}, {Im},
  {Jencson}, {Holoien}, {Basu}, {Beacom}, {Szczygie{\l}}, {Brimacombe},
  {Adams}, {Campillay}, {Choi}, {Contreras}, {Dietrich}, {Dubberley},
  {Elphick}, {Foale}, {Giustini}, {Gonzalez}, {Hawkins}, {Howell}, {Hsiao},
  {Koss}, {Leighly}, {Morrell}, {Mudd}, {Mullins}, {Nugent}, {Parrent},
  {Phillips}, {Pojmanski}, {Rosing}, {Ross}, {Sand}, {Terndrup}, {Valenti},
  {Walker}, \& {Yoon}}]{2014ApJ...788...48S}
{Shappee}, B.~J., {Prieto}, J.~L., {Grupe}, D., {et~al.} 2014, \apj, 788, 48,
  \dodoi{10.1088/0004-637X/788/1/48}

\bibitem[{{Skowron} {et~al.}(2019{\natexlab{a}}){Skowron}, {Skowron},
  {Mr{\'o}z}, {Udalski}, {Pietrukowicz}, {Soszy{\'n}ski}, {Szyma{\'n}ski},
  {Poleski}, {Koz{\l}owski}, {Ulaczyk}, {Rybicki}, \&
  {Iwanek}}]{2019Sci...365..478S}
{Skowron}, D.~M., {Skowron}, J., {Mr{\'o}z}, P., {et~al.} 2019{\natexlab{a}},
  Science, 365, 478, \dodoi{10.1126/science.aau3181}

\bibitem[{{Skowron} {et~al.}(2019{\natexlab{b}}){Skowron}, {Skowron},
  {Mr{\'o}z}, {Udalski}, {Pietrukowicz}, {Soszy{\'n}ski}, {Szyma{\'n}ski},
  {Poleski}, {Koz{\l}owski}, {Ulaczyk}, {Rybicki}, {Iwanek}, {. Wrona}, \&
  {Gromadzki}}]{2019AcA....69..305S}
---. 2019{\natexlab{b}}, \actaa, 69, 305, \dodoi{10.32023/0001-5237/69.4.1}

\bibitem[{{Smart} {et~al.}(1998){Smart}, {Drimmel}, {Lattanzi}, \&
  {Binney}}]{1998Natur.392..471S}
{Smart}, R.~L., {Drimmel}, R., {Lattanzi}, M.~G., \& {Binney}, J.~J. 1998,
  \nat, 392, 471, \dodoi{10.1038/33096}

\bibitem[{{Sormani} {et~al.}(2022){Sormani}, {Gerhard}, {Portail}, {Vasiliev},
  \& {Clarke}}]{2022MNRAS.514L...1S}
{Sormani}, M.~C., {Gerhard}, O., {Portail}, M., {Vasiliev}, E., \& {Clarke}, J.
  2022, \mnras, 514, L1, \dodoi{10.1093/mnrasl/slac046}

\bibitem[{{Soszy{\'n}ski} {et~al.}(2005){Soszy{\'n}ski}, {Udalski}, {Kubiak},
  {Szyma{\'n}ski}, {Pietrzy{\'n}ski}, {{\.Z}ebru{\'n}}, {Szewczyk},
  {Wyrzykowski}, \& {Ulaczyk}}]{2005AcA....55..331S}
{Soszy{\'n}ski}, I., {Udalski}, A., {Kubiak}, M., {et~al.} 2005, \actaa, 55,
  331.
\newblock \doarXiv{astro-ph/0512578}

\bibitem[{{Soszy{\'n}ski} {et~al.}(2009){Soszy{\'n}ski}, {Udalski},
  {Szyma{\'n}ski}, {Kubiak}, {Pietrzy{\'n}ski}, {Wyrzykowski}, {Szewczyk},
  {Ulaczyk}, \& {Poleski}}]{2009AcA....59..239S}
{Soszy{\'n}ski}, I., {Udalski}, A., {Szyma{\'n}ski}, M.~K., {et~al.} 2009,
  \actaa, 59, 239.
\newblock \doarXiv{0910.1354}

\bibitem[{{Soszy{\'n}ski} {et~al.}(2013){Soszy{\'n}ski}, {Udalski},
  {Szyma{\'n}ski}, {Kubiak}, {Pietrzy{\'n}ski}, {Wyrzykowski}, {Ulaczyk},
  {Poleski}, {Koz{\l}owski}, {Pietrukowicz}, \&
  {Skowron}}]{2013AcA....63...21S}
---. 2013, \actaa, 63, 21.
\newblock \doarXiv{1304.2787}

\bibitem[{{Stanek} {et~al.}(1994){Stanek}, {Mateo}, {Udalski}, {Szymanski},
  {Kaluzny}, \& {Kubiak}}]{1994ApJ...429L..73S}
{Stanek}, K.~Z., {Mateo}, M., {Udalski}, A., {et~al.} 1994, \apjl, 429, L73,
  \dodoi{10.1086/187416}

\bibitem[{{Stanek} {et~al.}(1997){Stanek}, {Udalski}, {Szyma{\'N}ski},
  {Ka{\L}u{\.Z}ny}, {Kubiak}, {Mateo}, \&
  {Krzemi{\'N}ski}}]{1997ApJ...477..163S}
{Stanek}, K.~Z., {Udalski}, A., {Szyma{\'N}ski}, M., {et~al.} 1997, \apj, 477,
  163, \dodoi{10.1086/303702}

\bibitem[{{Tian} {et~al.}(2020){Tian}, {Liu}, {Wang}, {Xu}, {Yang}, {Zhang}, \&
  {Xue}}]{2020ApJ...899..110T}
{Tian}, H., {Liu}, C., {Wang}, Y., {et~al.} 2020, \apj, 899, 110,
  \dodoi{10.3847/1538-4357/aba1ec}

\bibitem[{{Timberlake}(2011)}]{2011arXiv1112.3635T}
{Timberlake}, T. 2011, arXiv e-prints, arXiv:1112.3635.
\newblock \doarXiv{1112.3635}

\bibitem[{{Trippe} {et~al.}(2008){Trippe}, {Gillessen}, {Gerhard}, {Bartko},
  {Fritz}, {Maness}, {Eisenhauer}, {Martins}, {Ott}, {Dodds-Eden}, \&
  {Genzel}}]{2008A&A...492..419T}
{Trippe}, S., {Gillessen}, S., {Gerhard}, O.~E., {et~al.} 2008, \aap, 492, 419,
  \dodoi{10.1051/0004-6361:200810191}

\bibitem[{{Udalski}(2003)}]{2003AcA....53..291U}
{Udalski}, A. 2003, \actaa, 53, 291.
\newblock \doarXiv{astro-ph/0401123}

\bibitem[{{Udalski} {et~al.}(2015){Udalski}, {Szyma{\'n}ski}, \&
  {Szyma{\'n}ski}}]{2015AcA....65....1U}
{Udalski}, A., {Szyma{\'n}ski}, M.~K., \& {Szyma{\'n}ski}, G. 2015, \actaa, 65,
  1.
\newblock \doarXiv{1504.05966}

\bibitem[{{Urago} {et~al.}(2020){Urago}, {Omodaka}, {Nagayama}, {Watabe},
  {Miyanosita}, {Matsunaga}, \& {Burns}}]{2020ApJ...891...50U}
{Urago}, R., {Omodaka}, T., {Nagayama}, T., {et~al.} 2020, \apj, 891, 50,
  \dodoi{10.3847/1538-4357/ab70b1}

\bibitem[{{Wang} {et~al.}(2020){Wang}, {L{\'o}pez-Corredoira}, {Huang},
  {Chang}, {Zhang}, {Carlin}, {Chen}, {Chrob{\'a}kov{\'a}}, \&
  {Chen}}]{2020ApJ...897..119W}
{Wang}, H.~F., {L{\'o}pez-Corredoira}, M., {Huang}, Y., {et~al.} 2020, \apj,
  897, 119, \dodoi{10.3847/1538-4357/ab93ad}

\bibitem[{{Wang} \& {Chen}(2019)}]{2019ApJ...877..116W}
{Wang}, S., \& {Chen}, X. 2019, \apj, 877, 116,
  \dodoi{10.3847/1538-4357/ab1c61}

\bibitem[{{Wegg} \& {Gerhard}(2013)}]{2013MNRAS.435.1874W}
{Wegg}, C., \& {Gerhard}, O. 2013, \mnras, 435, 1874,
  \dodoi{10.1093/mnras/stt1376}

\bibitem[{{Weiland} {et~al.}(1994){Weiland}, {Arendt}, {Berriman}, {Dwek},
  {Freudenreich}, {Hauser}, {Kelsall}, {Lisse}, {Mitra}, {Moseley}, {Odegard},
  {Silverberg}, {Sodroski}, {Spiesman}, \& {Stemwedel}}]{1994ApJ...425L..81W}
{Weiland}, J.~L., {Arendt}, R.~G., {Berriman}, G.~B., {et~al.} 1994, \apjl,
  425, L81, \dodoi{10.1086/187315}

\bibitem[{{Werner} {et~al.}(2004){Werner}, {Roellig}, {Low}, {Rieke}, {Rieke},
  {Hoffmann}, {Young}, {Houck}, {Brandl}, {Fazio}, {Hora}, {Gehrz}, {Helou},
  {Soifer}, {Stauffer}, {Keene}, {Eisenhardt}, {Gallagher}, {Gautier}, {Irace},
  {Lawrence}, {Simmons}, {Van Cleve}, {Jura}, {Wright}, \&
  {Cruikshank}}]{2004ApJS..154....1W}
{Werner}, M.~W., {Roellig}, T.~L., {Low}, F.~J., {et~al.} 2004, \apjs, 154, 1,
  \dodoi{10.1086/422992}

\bibitem[{{Whitelock} {et~al.}(2006){Whitelock}, {Feast}, {Marang}, \&
  {Groenewegen}}]{2006MNRAS.369..751W}
{Whitelock}, P.~A., {Feast}, M.~W., {Marang}, F., \& {Groenewegen}, M.~A.~T.
  2006, \mnras, 369, 751, \dodoi{10.1111/j.1365-2966.2006.10322.x}

\bibitem[{{Whitelock} {et~al.}(2008){Whitelock}, {Feast}, \& {Van
  Leeuwen}}]{2008MNRAS.386..313W}
{Whitelock}, P.~A., {Feast}, M.~W., \& {Van Leeuwen}, F. 2008, \mnras, 386,
  313, \dodoi{10.1111/j.1365-2966.2008.13032.x}

\bibitem[{{Whitney} {et~al.}(2008){Whitney}, {Arendt}, {Babler}, {Benjamin},
  {Beuther}, {Bhattacharya}, {Blum}, {Bracker}, {Brunt}, {Carey}, {Churchwell},
  {Clemens}, {Cohen}, {Elmegreen}, {Frinchaboy}, {Heitsch}, {Hoare}, {Hora},
  {Indebetouw}, {Jackson}, {Jarrett}, {Kerton}, {Kobulnicky}, {Kraemer},
  {Lucas}, {Majewski}, {Marengo}, {Meade}, {Meixner}, {Mizuno}, {Molinari},
  {Povich}, {Price}, {Rathborne}, {Reach}, {Reid}, {Rho}, {Robitaille},
  {Sewilo}, {Shenoy}, {Smith}, {Smith}, {Stauffer}, {Stolovy}, {Ubeda}, {Van
  Dyk}, {van Loon}, {Volk}, {Watson}, {Wolff}, {Yusef-Zadeh}, \&
  {Zasowski}}]{2008sptz.prop60020W}
{Whitney}, B., {Arendt}, R., {Babler}, B., {et~al.} 2008, {GLIMPSE360:
  Completing the Spitzer Galactic Plane Survey}, Spitzer Proposal

\bibitem[{{Whitney} {et~al.}(2011){Whitney}, {Benjamin}, {Churchwell}, {Meade},
  {Babler}, {Allen}, {Anderson}, {Balser}, {Bania}, {Blitz}, {Boyer}, {Brunt},
  {Chakrabarti}, {Chambers}, {Clemens}, {Cohen}, {Cotera}, {Cyganowski},
  {Davis}, {Elmegreen}, {Frinchaboy}, {Froebrich}, {Hora}, {Indebetouw},
  {Ioannidis}, {Jarrett}, {Kerton}, {Kolbulnicky}, {Kraemer}, {Kumar}, {Liu},
  {Lucas}, {Majewski}, {Mauerhan}, {Marengo}, {Megeath}, {Minniti}, {Mottram},
  {Povich}, {Robitaille}, {Rood}, {Sewilo}, {Smith}, {Smith}, {Stanke},
  {Stauffer}, {Van Dyk}, {van Loon}, {Volk}, {Watson}, {Wolf-Chase}, \&
  {Zasowski}}]{2011sptz.prop80074W}
{Whitney}, B., {Benjamin}, R., {Churchwell}, E., {et~al.} 2011, {Deep GLIMPSE:
  Exploring the Far Side of the Galaxy}, Spitzer Proposal

\bibitem[{{Wilson} \& {Merrill}(1942)}]{1942ApJ....95..248W}
{Wilson}, R.~E., \& {Merrill}, P.~W. 1942, \apj, 95, 248,
  \dodoi{10.1086/144391}

\bibitem[{{WISE Team}(2020)}]{https://doi.org/10.26131/irsa134}
{WISE Team}. 2020, AllWISE Multiepoch Photometry Table,  IPAC,
  \dodoi{10.26131/IRSA134}

\bibitem[{{Wood}(2000)}]{2000PASA...17...18W}
{Wood}, P.~R. 2000, \pasa, 17, 18, \dodoi{10.1071/AS00018}

\bibitem[{{Wright} {et~al.}(2010){Wright}, {Eisenhardt}, {Mainzer}, {Ressler},
  {Cutri}, {Jarrett}, {Kirkpatrick}, {Padgett}, {McMillan}, {Skrutskie},
  {Stanford}, {Cohen}, {Walker}, {Mather}, {Leisawitz}, {Gautier}, {McLean},
  {Benford}, {Lonsdale}, {Blain}, {Mendez}, {Irace}, {Duval}, {Liu}, {Royer},
  {Heinrichsen}, {Howard}, {Shannon}, {Kendall}, {Walsh}, {Larsen}, {Cardon},
  {Schick}, {Schwalm}, {Abid}, {Fabinsky}, {Naes}, \&
  {Tsai}}]{2010AJ....140.1868W}
{Wright}, E.~L., {Eisenhardt}, P. R.~M., {Mainzer}, A.~K., {et~al.} 2010, \aj,
  140, 1868, \dodoi{10.1088/0004-6256/140/6/1868}

\bibitem[{{Xue} {et~al.}(2016){Xue}, {Jiang}, {Gao}, {Liu}, {Wang}, \&
  {Li}}]{2016ApJS..224...23X}
{Xue}, M., {Jiang}, B.~W., {Gao}, J., {et~al.} 2016, \apjs, 224, 23,
  \dodoi{10.3847/0067-0049/224/2/23}

\bibitem[{{Yuan} {et~al.}(2017{\natexlab{a}}){Yuan}, {He}, {Macri}, {Long}, \&
  {Huang}}]{2017AJ....153..170Y}
{Yuan}, W., {He}, S., {Macri}, L.~M., {Long}, J., \& {Huang}, J.~Z.
  2017{\natexlab{a}}, \aj, 153, 170, \dodoi{10.3847/1538-3881/aa63f1}

\bibitem[{{Yuan} {et~al.}(2017{\natexlab{b}}){Yuan}, {Macri}, {He}, {Huang},
  {Kanbur}, \& {Ngeow}}]{2017AJ....154..149Y}
{Yuan}, W., {Macri}, L.~M., {He}, S., {et~al.} 2017{\natexlab{b}}, \aj, 154,
  149, \dodoi{10.3847/1538-3881/aa86f1}

\bibitem[{{Yuan} {et~al.}(2018){Yuan}, {Macri}, {Javadi}, {Lin}, \&
  {Huang}}]{2018AJ....156..112Y}
{Yuan}, W., {Macri}, L.~M., {Javadi}, A., {Lin}, Z., \& {Huang}, J.~Z. 2018,
  \aj, 156, 112, \dodoi{10.3847/1538-3881/aad330}

\bibitem[{{Zasowski} {et~al.}(2009){Zasowski}, {Majewski}, {Indebetouw},
  {Meade}, {Nidever}, {Patterson}, {Babler}, {Skrutskie}, {Watson}, {Whitney},
  \& {Churchwell}}]{2009ApJ...707..510Z}
{Zasowski}, G., {Majewski}, S.~R., {Indebetouw}, R., {et~al.} 2009, \apj, 707,
  510, \dodoi{10.1088/0004-637X/707/1/510}

\end{thebibliography}
\bibliographystyle{aasjournal}

\end{document}